\documentclass[journal, comsoc]{IEEEtran}
\IEEEoverridecommandlockouts

\IEEEoverridecommandlockouts
\usepackage{cite}
\usepackage{amsmath,amssymb,amsfonts}
\usepackage{graphicx}
\usepackage{textcomp}
\usepackage{xcolor}
\usepackage{authblk}
\usepackage{url}
\usepackage{booktabs,makecell,tabularx}

\usepackage{soul}

\usepackage{threeparttable}
\usepackage{subfigure}
\usepackage{svg}
\usepackage{multirow}

\usepackage{array}

\usepackage[ruled,vlined,linesnumbered]{algorithm2e}
\let\oldnl\nl
\newcommand{\nonl}{\renewcommand{\nl}{\let\nl\oldnl}}

\def\BibTeX{{\rm B\kern-.05em{\sc i\kern-.025em b}\kern-.08em
    T\kern-.1667em\lower.7ex\hbox{E}\kern-.125emX}}

\usepackage{hyperref}
\hypersetup{
     colorlinks = true,
     linkcolor = blue,
     anchorcolor = red,
     citecolor = magenta,
     filecolor = blue,
     urlcolor = black,
}

\usepackage{siunitx}

\begin{document}

\title{CAN-Trace Attack: Exploit CAN Messages to Uncover Driving Trajectories}

\author{Xiaojie~Lin,
        Baihe~Ma,
        Xu~Wang,~\IEEEmembership{Member,~IEEE},
        Guangsheng~Yu,
        Ying~He,~\IEEEmembership{Senior Member,~IEEE},
        Wei~Ni,~\IEEEmembership{Fellow,~IEEE},
        and~Ren~Ping~Liu,~\IEEEmembership{Senior Member,~IEEE}
    \IEEEcompsocitemizethanks{
        \IEEEcompsocthanksitem X. Lin, B. Ma, X. Wang, Y. He, G. Yu, W. Ni, and R. P. Liu are with the Global Big Data Technologies Centre, University of Technology Sydney, Australia. \protect
        E-mail: \{xiaojie.lin, baihe.ma, xu.wang, ying.he, guangsheng.yu, wei.ni, renping.liu\}@uts.edu.au
    }
}

\maketitle

\begin{abstract}
Driving trajectory data remains vulnerable to privacy breaches despite existing mitigation measures. Traditional methods for detecting driving trajectories typically rely on map-matching the path using Global Positioning System (GPS) data, which is susceptible to GPS data outage.
This paper introduces CAN-Trace, a novel privacy attack mechanism that leverages Controller Area Network (CAN) messages to uncover driving trajectories, posing a significant risk to drivers' long-term privacy. 
A new trajectory reconstruction algorithm is proposed to transform the CAN messages, specifically vehicle speed and accelerator pedal position, into weighted graphs accommodating various driving statuses. 
CAN-Trace identifies driving trajectories using graph-matching algorithms applied to the created graphs in comparison to road networks.
We also design a new metric to evaluate matched candidates, which allows for potential data gaps and matching inaccuracies.
Empirical validation under various real-world conditions, encompassing different vehicles and driving regions, demonstrates the efficacy of CAN-Trace: it achieves an attack success rate of up to $90.59\%$ in the urban region, and $99.41\%$ in the suburban region.

\end{abstract}


\section{Introduction}
\label{intro}

Location privacy is a serious topic in the rapidly developing field of Intelligent Transportation Systems (ITS), attracting significant attention from both industry and academia~\cite{krumm2009survey, wang2018user}.
Sensitive location information, such as a driver's home or work address, is increasingly at risk due to privacy breaches in the ITS ecosystem.
In the ITS ecosystem, vehicles can share location and motion data among themselves and with broader in-vehicle networks~\cite{ma2009location, huang2020recent, ma2023location}.
The misuse of driver location data can reveal driving trajectories and the geographical locations of drivers, thereby compromising the personal privacy of drivers~\cite{corser2016evaluating, ma2022personalized}.

Map-matching process, which identifies a vehicle's location on a road network, is pivotal in these potential privacy breaches~\cite{krakiwsky1988kalman, assam2013private, carter2019using}. 
These attacks allow adversaries to accurately track and pinpoint vehicle locations using transmitted data~\cite{kong2018security}. 
Traditional methods to detect the driving trajectory have largely been built around Global Positioning System (GPS)~\cite{jianjun2011improved} or a combination of GPS and additional sensors like the on-board camera~\cite{roncella2005photogrammetric, liu2021vehicle}, Inertial Measurement Units (IMU)~\cite{georgy2011enhanced, sasani2016improving} or On-Board Diagnostics (OBD) devices~\cite{chen2019trajcompressor}. 
However, the traditional methods suffer from weaknesses such as data loss, GPS outages, and restricted access, thus offering loopholes for privacy attacks~\cite{iqbal2010privacy, lim2016land, xiao2018goi}.

Advanced techniques have evolved to refine the map-matching process by incorporating mobile magnetometers~\cite{li2018location}. 
The mobile magnetometers can provide granular vehicle tracking by identifying unique vehicle movements like stops, turns, and lane changes. However, the method using mobile phones faces the challenge of performance degradation when the position is shifted by the vehicle's occupants, whether intentionally or unintentionally. The variability in phone positioning further complicates the map-matching process and decreases the robustness of these methods against privacy attacks~\cite{li2018location}.

Controller Area Network (CAN) is a vehicle communication protocol, which defines the transmission of vehicle motion and control messages, such as vehicle speed, acceleration, and steering angle, allowing vehicular Electronic Control Units (ECUs) to communicate with each other within the in-vehicle network~\cite{hpl2002introduction, voss2008comprehensible,johansson2005vehicle}.
CAN messages offer a wealth of real-time and precise vehicle motion data, effectively mitigating the risks associated with data outage and integrity issues~\cite{szijj2015hacking, yan2015two}. 
CAN messages can be easily accessed through interfaces such as the OBD-II port~\cite{klinedinst2016board, wen2020plug, lin2022multi}, commonly exploited by third-party car diagnostic software, OBD-II applications, and OEM assistance applications.

In this paper, we pioneer a CAN-centric driving trajectory tracking strategy that serves as an alternative to the map-matching methods reliant on GPS or magnetometers. We use basic CAN messages as an innovative data source to decipher driving trajectories in the form of weighted line graphs. Subsequently, we reframe the problem of trajectory identification as the task of pinpointing the corresponding subgraph, derived from CAN messages, in a targeted road network. 
{\color{black}The proposed CAN-Trace attack directly addresses critical privacy challenges, such as uncovering private home addresses, frequently visited locations, and travel patterns, by exploiting unencrypted CAN messages. These messages, which can be easily accessed through standard interfaces like the OBD-II port, represent an overlooked but significant attack vector embedded within the vehicle itself. Unlike traditional privacy risks associated with GPS data or external sensors, CAN-Trace highlights the inherent vulnerabilities of vehicles, emphasizing the urgent need for robust security measures and secure data management practices to safeguard sensitive driving data.}
The key contributions are as follows:

\begin{itemize}

\item We propose a novel trajectory reconstruction algorithm that composes weighted graphs to depict driving trajectories. The algorithm discerns key vehicle movements, i.e., stops, turns, and driving lengths, using only speed and pedal CAN messages.

\item We are the first to introduce the graph-based matching technique for driving trajectory identification from road networks. We also design a new metric to evaluate matching results, accounting discrepancies between the CAN perspective and the actual road network.

\item We validate the proposed CAN-Trace attack via extensive real-world experiments across diverse road networks and vehicles, demonstrating success rates up to 90.59\% in the urban region and 99.41\% in the suburban region.

\end{itemize}

The rest of the paper is organized as follows: 
Section~\ref{threat_model} gives an overview of the system and attack models and Section~\ref{data} describes each process of the CAN-Trace attack in detail. 
Section~\ref{performance} presents the evaluation of the CAN-Trace attack in the real-world environment.
Section~\ref{related} summarizes the related work.
Section~\ref{conclusion} concludes this paper.

\begin{figure}[t]
\centering
    \includegraphics[width=1\linewidth]{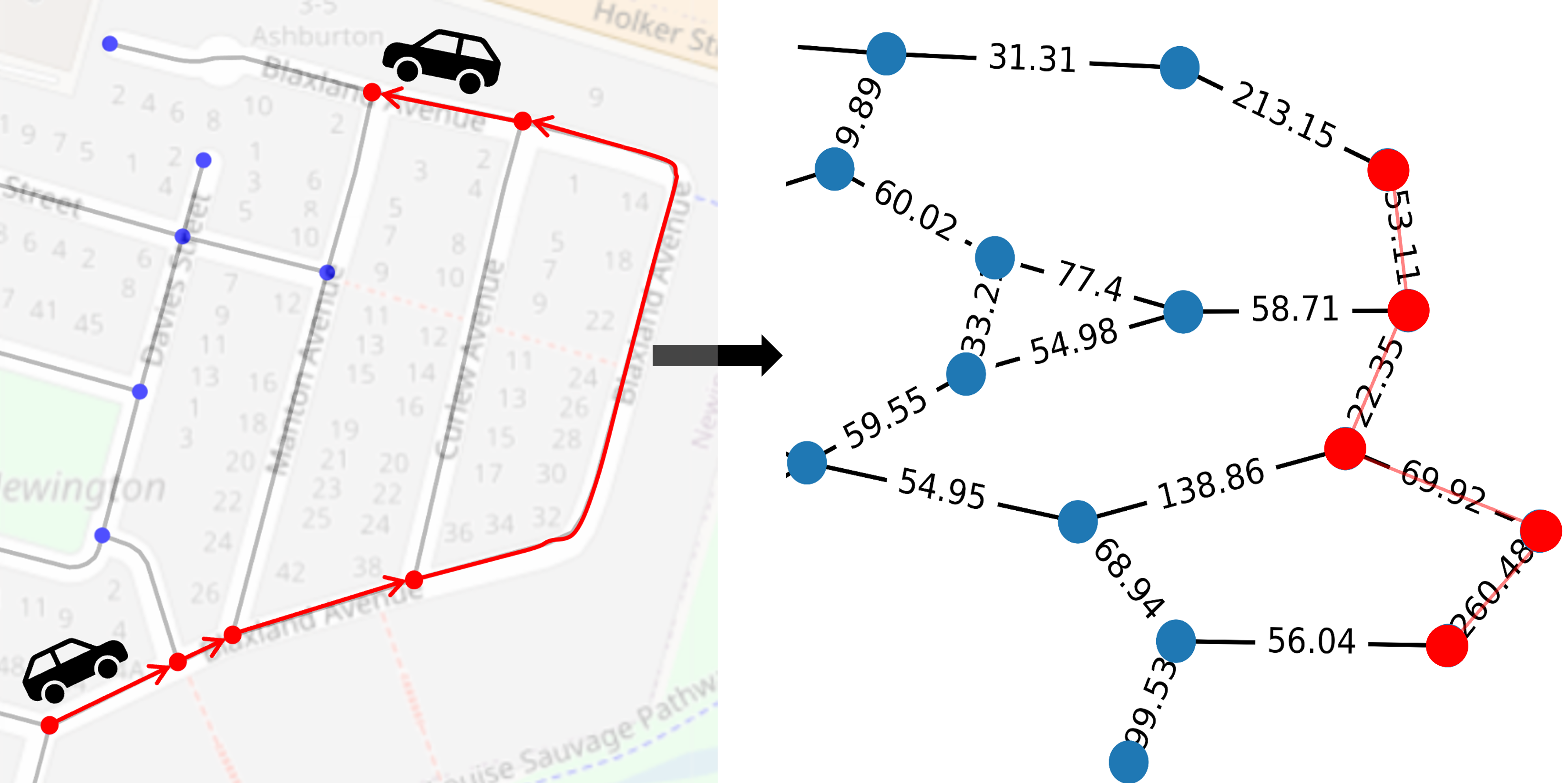}
    \caption{Road network to graph: a road network is converted from the 2D street map into a graph $G=(V, E)$, where edges represent road segments and the weight of an edge indicates the length of the corresponding road segment. The actual driving trajectory can be represented as a subgraph $G^*=(V^*, E^*)$. Blue nodes indicate road intersections not covered by the driving trajectory, while red nodes denote intersections in the driving trajectory.}
    \label{fig:certainArea}
\end{figure}

\begin{figure}[t]
\centering
    \includegraphics[width=0.8\linewidth]{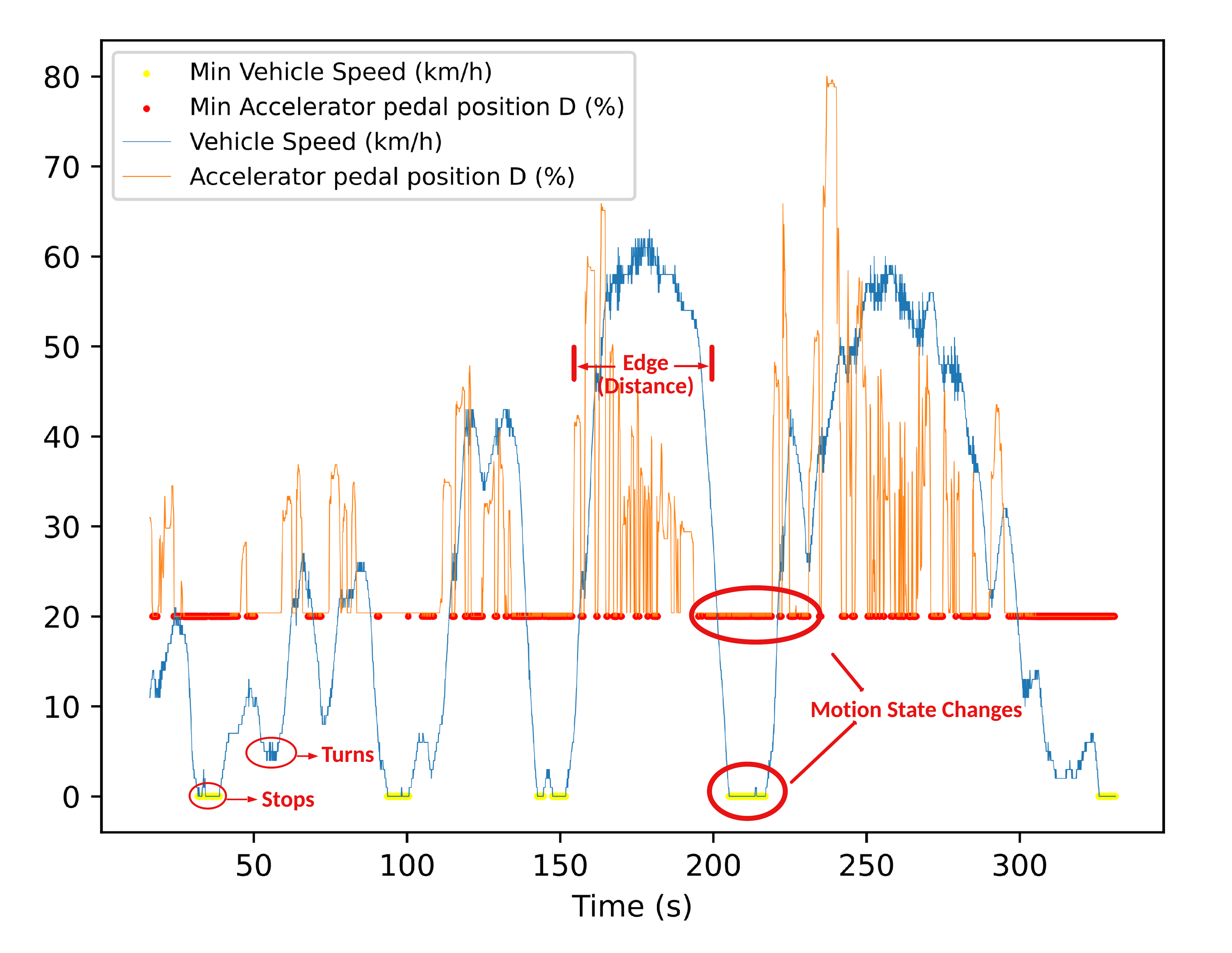}
    \caption{CAN to driving trajectory: vehicle stops or turns can be identified as nodes by analyzing the vehicle speed and pedal position data in OBD response, and the length of the road segments can be calculated by the distance between two adjacent nodes.}
    \label{fig:pidValueRelation}
\end{figure}

\section{System and Attack Models}
\label{threat_model}

The proposed CAN-Trace attack considers road networks and deduces the driving trajectory of the on-road vehicle by collecting and analyzing CAN messages through a driving period.
Table~\ref{denotion} summarizes the notation used throughout this paper.
The system model is described as follows:

\begin{itemize}
  \item \textbf{Attack Target}: The attack target is to expose the geolocation of the driver and identify the driving trajectories.

  \item \textbf{Adversary Capability}: Adversaries have full prior knowledge of the road networks, i.e., where the target vehicle is located within a city or a suburb size.
  Adversaries can collect vehicular motion data by analyzing the CAN message to infer vehicle activities on road networks, e.g., stops or turns.

\end{itemize}

\begin{table}[t]
  \centering
  \caption{Notation and Definitions}
  \begin{tabularx}{\linewidth}{ l X l }
    \toprule
    \thead{\textbf{Notation}} & \thead{\textbf{Description}}\\
    \midrule
    \makecell[lt]{$G$  } & Data graph of the road network\\
    \addlinespace[2pt]
    \makecell[lt]{$G^*$  } & Graph of the actual driving trajectory\\
    \addlinespace[2pt]
    \makecell[lt]{ $\bar R$  } & Graph of the constructed driving trajectory\\
    \addlinespace[2pt]
    \makecell[lt]{ $\hat R = \{r_k\}$  } & Set of matched subgraphs $r_k$\\
    \addlinespace[2pt]
    \makecell[lt]{ $E$ / $E^*$/ $E^R$ / $E^{r_k}$ } & Set of edges in $G$ /  $G^*$ / $\bar R$ /$r_k$\\
    \addlinespace[2pt]
    \makecell[lt]{$V$ / $V^*$/ $V^R$ / $V^{r_k}$ } & Set of nodes in $G$ /  $G^*$ / $\bar R$ /$r_k$\\
    \addlinespace[2pt]
    \makecell[lt]{ $w_{i,j}$ / $w_{i,j}^*$ / $w_{i,j}^R$ / $w_{i,j}^k$ } & Weight of $e_{i,j}$ / $e_{i,j}^*$ / $e_{i,j}^R$ / $e_{i,j}^{r_k}$ \\
    \addlinespace[2pt]
    \makecell[lt]{  $\mathcal{M} = \{M^s, M^p\}$ } &  Set of CAN messages of OBD response data\\
    \addlinespace[2pt]
    \makecell[lt]{ $M^s = \{m_i^s,t_i^s\}$} &   Set of OBD response of vehicle speed value $m_i^s$ and the timestamp $t_i^s$\\
    \addlinespace[2pt]
    \makecell[lt]{$M^p = \{m_i^p,t_j^p\}$} &   Set of OBD response of accelerator pedal position value $m_i^p$ and the timestamp $t_j^p$\\
    \addlinespace[2pt]
    \makecell[lt]{ $N_T^s$ / $N_T^p$  } &  Number of collected OBD messages of vehicle speed/pedal position\\
     \addlinespace[2pt]
    \makecell[lt]{ $N'_k$  } &  Number of correctly inferred nodes in $r_k$\\
    \addlinespace[2pt]
    \makecell[lt]{ $Q^*$  } &  Number of nodes in the graph of actual driving trajectory\\
    \addlinespace[2pt]
    \makecell[lt]{$\Delta^s$ / $\Delta^p$} & Time difference threshold to determine the nodes of vehicle stops/turns \\
    \addlinespace[2pt]
    \makecell[lt]{$\sigma$} &  Relative tolerance of edge weight used in the matching process\\
    \addlinespace[2pt]
    \makecell[lt]{$\theta$} &  Distance difference of the $e_{i,j}^R$ and $e_{i,j}^k$ pair of $\bar R$ and $r_k$\\
    \addlinespace[2pt]
    \makecell[lt]{$P$ } & Attack precision\\
    \addlinespace[2pt]
    \makecell[lt]{$\Psi$ } & Attack success rate \\
    \addlinespace[2pt]
    \makecell[lt]{$D$} &  Spatial distance offset between the deduced and the actual driving trajectory\\
    \makecell[lt]{$\mathbb{F}$} &  False negative rate\\
    \bottomrule
  \end{tabularx}
  \label{denotion}
\end{table}

\subsection{Prior Knowledge}

\subsubsection{Road Network}

A road network can be extracted from a Two-Dimensional (2D) street map and converted into an undirected weighted graph $G=(V, E)$, which contains the geospatial information of streets.
The set of nodes, $V$, represents the road intersections in the road network, while the set of edges, $E$, are the segments that connect adjacent nodes of $V$ within the road network.  
The weights of edges are the lengths of the segments in the road network.

As shown in Fig.~\ref{fig:certainArea}, the actual driving trajectory is regarded as a graph $G^* = (V^*, E^*)$  that is a subgraph of $G$.
The road intersections passed by the car are denoted as $V^*$, and the road segments passed are denoted as $E^*$.
The problem of detecting the actual driving trajectory is then converted into that of finding the corresponding subgraph $G^*$ within $G$ by matching the graph constructed from CAN messages.

\subsubsection{CAN Messages}

The adversaries obtain OBD response $\mathcal{M} = \{M^s, M^p\}$ to construct the line graph of the driving path $\bar R$ as $\mathcal{M} \rightarrow \bar R = (V^R, E^R)$.
Analyzing the motion data embedded in $\mathcal{M}$, the adversaries identify the vehicle stops or turns as the nodes $V^R$, and the path between adjacent nodes as edge $E^R$.
CAN-Trace attack leverages the CAN messages of vehicle speed $M^s = \{(m_i^s,t_i^s)\}$ (where $i\leq N_T^s$) and those of the accelerator pedal position $M^p = \{(m_j^p,t_j^p)\}$ (where $j\leq N_T^p$) to construct the line graph of the driving path.
Here, $(m_i^s,t_i^s)$ is the $i$-th vehicle speed message of the vehicle speed value $m_i^s$ at the time $t_i^s$, and $(m_j^p,t_j^p)$ is the $j$-th vehicle pedal position message of the pedal position value $m_i^p$ at the time $t_j^p$.
$N_T^s$ and $N_T^p$ are the number of messages in $M^s$ and $M^p$, respectively.
As demonstrated in Fig.~\ref{fig:pidValueRelation}, vehicle activities such as stops and turns can be identified from the OBD response statistics.

\begin{figure}[t]
        \centering
        \includegraphics[width=.85\columnwidth]{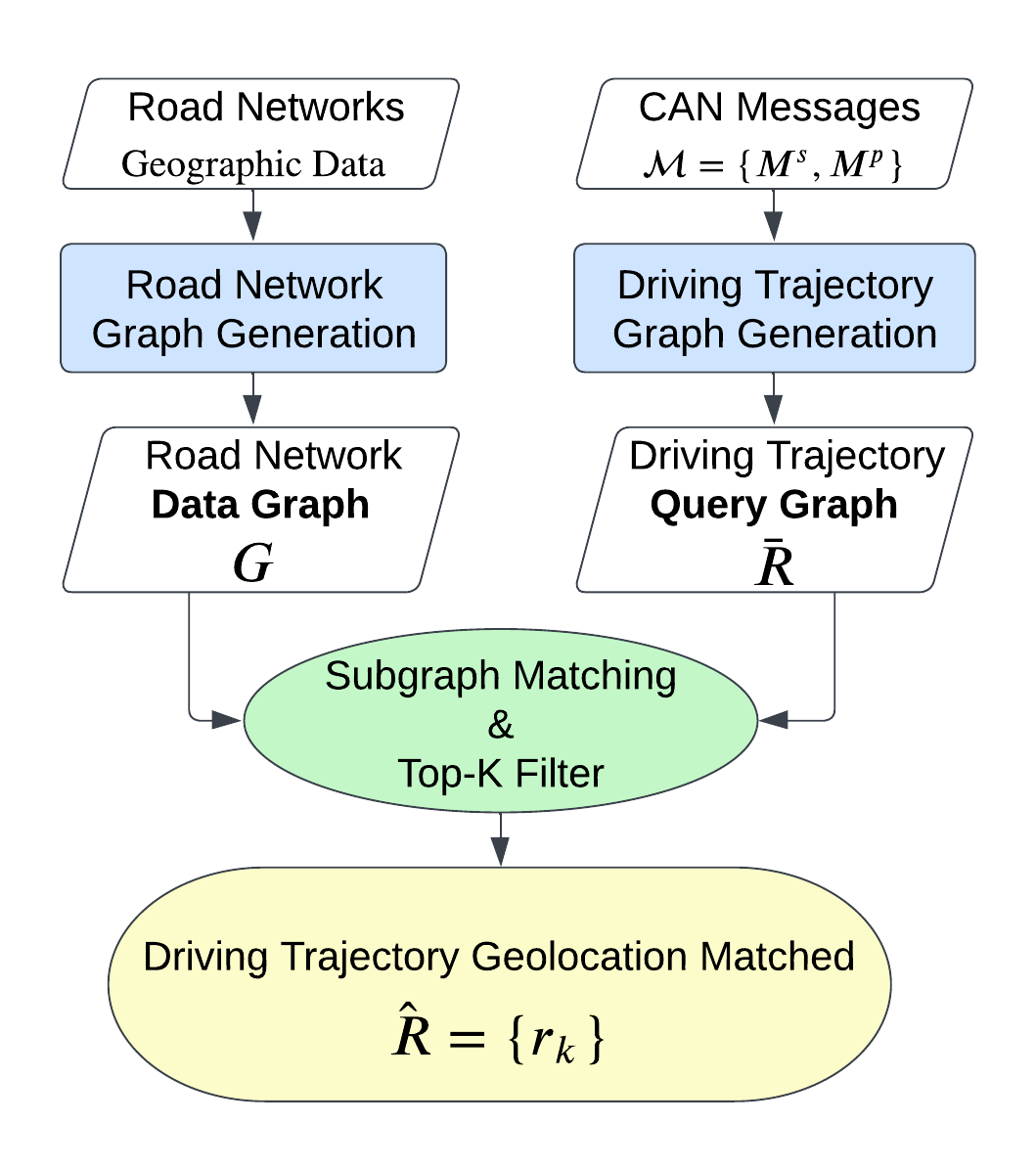}
        \caption{Attack model: the CAN-Trace attack identifies the driving trajectory by matching the driving path graph $\bar R$ with the road network graph $G$. 
        The road network graph generation converts geographic data of road networks into $G$ using OpenStreetMap.
        The driving trajectory graph generation leverages the CAN messages to transform vehicle motion data into $\bar R$.
        The generated $\bar R$ is further matched with $G$ using a subgraph matching algorithm to find subgraphs $r_k$, i.e., driving trajectory candidates.
         Top-$K$ ranking rule is utilized to select a specified number of deduced subgraph candidates as the attack result $\hat{R}$.
        }
        \label{fig:systemModel}
\end{figure}

\section{CAN-Trace Attack}
\label{data}

To the best of our knowledge, the CAN-Trace attack is the first to propose the OBD-II port as a new attack interface and the CAN messages as the data source.
As demonstrated in Fig.~\ref{fig:systemModel}, the CAN-Trace attack consists of the driving trajectory graph generation, the road network graph generation, the subgraph matching, and the Top-$K$ ranking.
The CAN-Trace attack constructs the driving trajectory graph $\bar R$ based on vehicle stops and turns from CAN messages.
Then, it applies subgraph matching algorithms to match $\bar R$ with the road network graph $G$.
Since more than one subgraph candidate $r_k$ can be matched by the CAN-Trace attack, ranking rules are applied to identify the Top-$K$ subgraphs.
Algorithms~\ref{alg:can2graphALG} and~\ref{alg:merged} illustrate the attack method.

\begin{algorithm*}[htbp]
\footnotesize
\caption{CAN to Graph Conversion}
\label{alg:can2graphALG}
\nonl{
\begin{minipage}{.48\linewidth}
\SetAlgoLined
\SetKwInOut{Input}{Input}
\SetKwInOut{Output}{Output}
\Input{CAN messages $\mathcal{M} = \{M^s, M^p\}$}
\Output{Graph of constructed driving trajectory $\bar R = (V^R, E^R)$}
\nonl{$\triangleright$ \textbf{Initialization}}

\nonl{\tcc{\scriptsize Put data points whose $m^s=0$ or $m^p = \min(m^p)$ into the candidate sets $P^s$ and $P^p$.}}

\nl $P^s \gets \{m^{s=0}\}$, $P^p \gets \{m^{p = \min(m^p)}\}$\; \label{lst:line:algo-1-1}

\nonl{\tcc{\scriptsize Use the time gap between two consecutive candidates for clustering. }}

\nl $T^s = \{d(P^s_i, P^s_{i+1})\}$, $T^p = \{d(P^p_j, P^p_{j+1})\}$;

\nonl{\tcc{\scriptsize Cluster $T^s$ and $T^p$ for thresholds $\Delta^s$ and $\Delta^p$.}}

\nl \textbf{Cluster}($T^s$), \textbf{Cluster}($T^p$)\; \label{lst:line:algo-1-3}
$\Delta^s \gets$ \textbf{UpperBoundOfStoppingClusters($T^s$)}\;

$\Delta^p \gets$ \textbf{UpperBoundOfTurningClusters($T^p$)}\; \label{lst:line:algo-1-5}

\BlankLine
{\nonl$ \triangleright$ \textbf{Graph Construction}}\\
\tcc{\scriptsize Initial index $i$ and $j$ for the loop on $M^s$.}
$i,j=1$\; \label{lst:line:algo-1-6}
\For{$j<N^s_T$}{
\If{$m^s_j=0$}{
\If{$t_j^s - t_i^s \geq \Delta^s$}{
$V^R$.\textbf{append}\big($(m^s_j, t^s_j)$\big)\;
}
$i=j$
}
$j=j+1$
}\label{lst:line:algo-1-15}
\LinesNotNumbered
\end{minipage}\hfill
\begin{minipage}{.4\linewidth}
\LinesNumbered
\SetAlgoLined
\tcc{\scriptsize Initial index $i$ for the loop to parse $M^p$.}

\nl $i,j=1$\;\label{lst:line:algo-1-16}
\For{$j<N^p_T$}{
\If{$m^p_j = \min(m^p)$}{
\If{$t_j^p - t_i^p \geq \Delta^p$}{
$V^R$.\textbf{append}\big($(m^p_j, t^p_j)$\big)\;}
$i=j$\;
}
$j=j+1$\;
}\label{lst:line:algo-1-25}
\tcc{\scriptsize Sort $V^R$ based on the message timestamp.}
$V^R$.\textbf{sort()}\; \label{lst:line:algo-1-26}
\tcc{\scriptsize Merge the duplicated nodes.}
\For{$v^R_k$ and $v^R_{k+1}$ in $V^R$}{ \label{lst:line:algo-1-27}
Calculate the weight of edge $w_{k,k+1}^R$; \qquad  $\triangleright$ with~\eqref{equ:route_dis}\\
\tcc{\scriptsize Select the shortest road network segment in $G$ denoted as $\min(w_{i,j})$}
\If{$w^R_{k,k+1} < \min(w_{i,j})$}{
\tcc{\scriptsize Remove nodes whose edges less than $\min(w_{i,j})$.}
$V^R$.delete($v^R_k$)\;
}
} \label{lst:line:algo-1-32}
\tcc{\scriptsize Construct $\bar R$ with the merged $V^R$}
$E^R \gets \{e^R_{k, k+1}\}$ where $v_k, v_{k+1} \in V^R$\;
\Return $\bar R = (V^R, E^R)$
\LinesNotNumbered
\end{minipage}
}
\end{algorithm*}

\begin{algorithm}[htbp]
\footnotesize
\caption{Driving Trajectory Detection: Matching and Filtering}
\label{alg:merged}
\SetAlgoLined
\SetKwInOut{Input}{Input}
\SetKwInOut{Output}{Output}
\Input{Road network graph $G$, graph of constructed driving trajectory $\bar R$, relative tolerance $\sigma$, Top-$K$ value $K$
}

\Output{A set of detected driving trajectory $\hat R = \{r_k\}$}

\tcc{\scriptsize Set different tolerance $\sigma$ in matching process to find $l$ subgraphs until $l  \geq K$ with $\max(\sigma)$}
Set $\sigma$ and $K$ in matching $\bar R$ with $G$ to identify subgraph candidate $r_l$\; \label{lst:line:algo-2-1}
Calculate the difference of edge pair $\varpi^l_{i,j}= |w^l_{i,j} - w_{i,j}|$\;
\If{ $\forall$ $\varpi^l_{i,j} \leq \sigma \times w_{i,j}$}{
\tcc{\scriptsize Append matched nodes and edges to $r_l$}
$V^{r_l} \gets v_i, v_j$\;
$E^{r_l} \gets e_{i,j}^{r_l}$\;
Calculate the weight of $e_{i,j}^{r_l}$ as $w_{i,j}^{l}$; \qquad  $\triangleright$ with~\eqref{equ:route_dis}\\
} \label{lst:line:algo-2-7}
\tcc{\scriptsize Calculate the average distance difference between the edge pair of line graph $\bar R$ and $r_l$ as $\theta_l$}
$\theta_l \gets \sum_{i=1}^{Q^*-1} \frac{|{w_{i,i+1}^R}-{w_{i,i+1}^l}|}{Q^*-1}$; \qquad  $\triangleright$ with~\eqref{equ:topK}\\ \label{lst:line:algo-2-8}
\tcc{\scriptsize Sort $r_l$ by $\theta_l$ as $\{r_k\}$ }
$\{r_k\} \gets \textbf{Sort}(r_l, \theta_l)$\;
\tcc{Select $K$ subgraphs as $\hat{R}$}
\Return $\hat R = \{r_k\}, k \leq K$ \label{lst:line:algo-2-10}
\end{algorithm}

\subsection{Graph Generation}
\label{subsec:can2graph}

To the best of our knowledge, we are the first to generate a road network graph by using the CAN messages to infer the driving trajectory.
The detailed steps are discussed as follows.

\smallskip
\subsubsection{\textbf{Road Network Data Graph}}

Vehicles are subject to traffic rules and are constrained to drivable road segments in the road network $G=(V,E)$.
Nodes $V$ and edges $E$ are constructed to generate a purified $G$ as given by:
\begin{itemize}
  \item \textbf{Nodes}: The $i$-th node $v_i$ stands for the $i$-th entry or exit of a road segment in the road network. 
  A node can have multiple entries or exits, e.g., a roundabout or intersection. 
 
  \item \textbf{Edges}: An edge $e_{i,j}$ exists if there is a route between $v_i$ and $v_j$. 
  The value of $e_{i,j}$ is the route length of the road segment between $v_i$ and $v_j$.
 
\end{itemize}

The actual driving trajectory is located within the road network, which means $G^*$ is the subgraph of $G$ satisfying $V^* \subseteq V$ and $E^* \subseteq E$.

\smallskip
\subsubsection{\textbf{Graph of Constructed Driving Trajectory}}

Adversaries generate the graph of the constructed driving trajectory $\bar R = (V^R, E^R)$ using the OBD response set $\mathcal{M} = \{M^s, M^p\}$ of a driving period $T$. 
During the period, $N_T^s$ speed messages and $N_T^p$ pedal position messages are collected, so we have $i \leq N_T^s$ and $j \leq N_T^p$ for any $m_i^s$ and $m_j^p$ in $\mathcal{M}$.

Nodes $V^R$ refer to the vehicle stops and turns embedded in the OBD response $\mathcal{M} = \{M^s, M^p\}$.
Vehicle status switches between driving and stopping or turning are determined as follows:
\begin{itemize}
  \item \textbf{Stops at traffic signals or signs}: $m_i^s = 0$.
  \item \textbf{Turns at the road intersections}: $m_j^p = \min(m^p)$.

\end{itemize}

The minimum value of the pedal position indicates the driver releases the throttle.
Note that the default value of the pedal position is not equal to zero.
We assume that the vehicle turns at the road intersection when the minimum pedal position lasts for a period.

\begin{figure}[!htbp]
\centering
     \subfigure[Trajectory graph construction from vehicle speed to stops]{
    \begin{minipage}[t]{0.5\textwidth}
    \centering
    \includegraphics[width=3.4in]{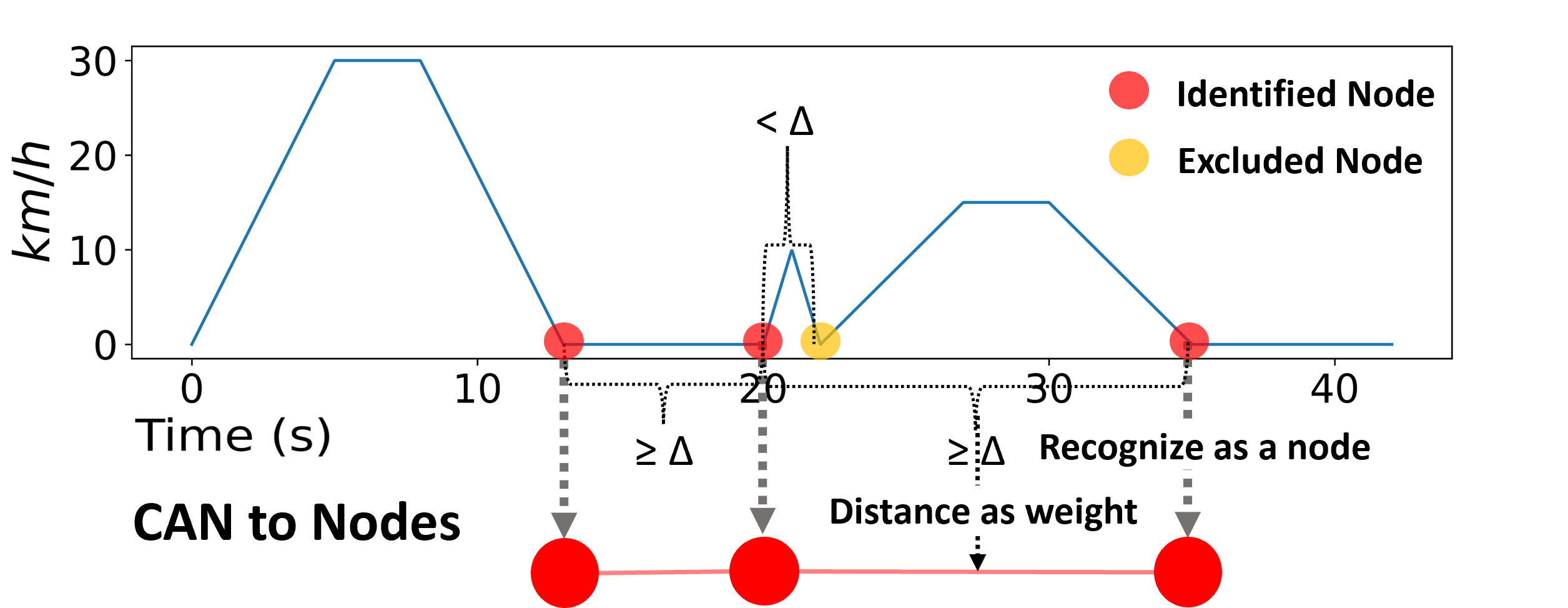}
    \end{minipage}
    \label{fig:can2graphDetails}
    }
    \subfigure[Algorithm flowchart of graph construction from CAN messages]{
    \begin{minipage}[t]{0.5\textwidth}
    \centering
    \includegraphics[width=3.4in]{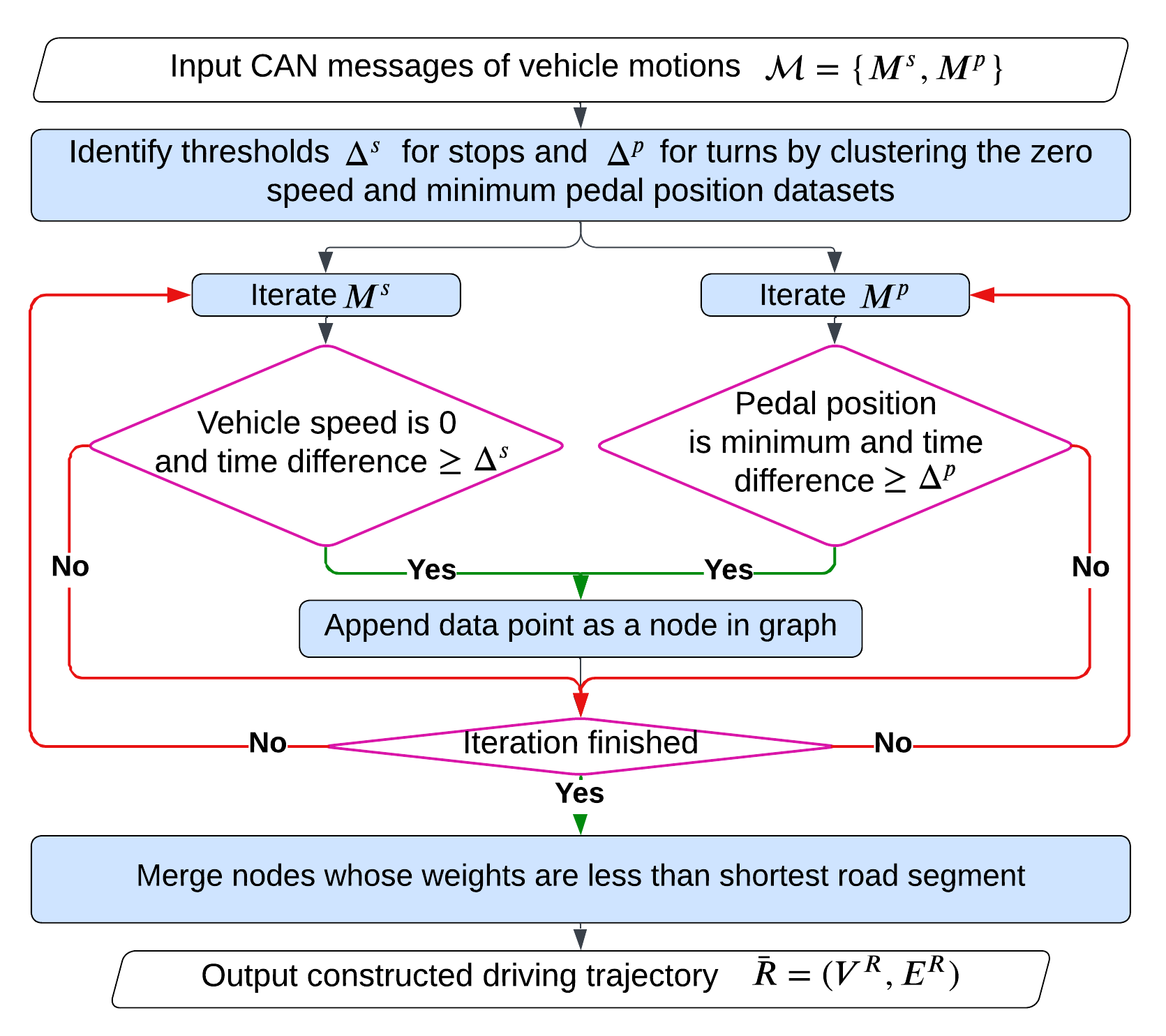}
    \end{minipage}
    \label{fig:can2graphFlow}
    }
    \caption{Graph construction: nodes are parsed and extracted from CAN messages. CAN-Trace attack calculates the time differences between two data points that are possible node candidates. Nodes are identified when the time differences are greater than the examined thresholds. 
    The distance between two nodes is calculated as the edge weight. The $x$-axis in (a) represents the elapsed driving time in seconds, while the $y$-axis indicates the vehicle speed in kilometers per hour. The algorithm flowchart in (b) represents the workflow of constructing graphs by parsing CAN messages and identifying nodes of stops and turns.}
    \label{fig:can2graph}
\end{figure}

As illustrated in Fig.~\ref{fig:can2graphDetails}, nodes $V^R$ can be identified from CAN messages by analyzing the data points and factors like the time difference.
Details of the node selection in constructing the graph are found in Lines~\ref{lst:line:algo-1-1}--\ref{lst:line:algo-1-5} of Algorithm~\ref{alg:can2graphALG}.
The CAN-Trace attack feeds the set of all $m^{s=0}$ as $P^s$ and $m^{p={\min(p)}}$ as $P^p$.
The time differences are calculated by $d(P_i^s, P_{i+1}^s) = t_{i+1}^s - t_i^s$ for $P^s$ and $d(P_j^p, P_{j+1}^p) = t_{j+1}^p - t_j^p$ for $P^p$, respectively.
{\color{black}These time difference data can be processed with machine learning algorithms to differentiate meaningful stops and turns from transient statuses. Specifically, we use the classic and efficient K-means clustering algorithm to partition the time differences of $P^s$ and $P^p$ into clusters, minimizing intra-cluster variance. The clustering process groups similar time differences together, with each cluster reflecting a distinct driving status. Clusters with small time differences typically represent transient statuses, such as brief pauses or lane changes, while clusters with large time differences correspond to significant stops or turns. Then, the biggest time difference in the cluster with the smallest time differences are used as the delimiters, denoted by $\Delta^s$ and $\Delta^p$, acting as filters to exclude irrelevant movements.} For example, $\Delta^s$ eliminates short-term stops like brief pauses in traffic, while $\Delta^p$ removes lane changes.

The method to construct the nodes in the line graph of the driving trajectory $\bar R$ is described in Algorithm~\ref{alg:can2graphALG} from Lines~\ref{lst:line:algo-1-6}--\ref{lst:line:algo-1-26}.
The node of a vehicle stop can be identified when the time difference between two data points in $M^s$ is greater than or equal to $\Delta^s$ and the two points satisfy $m_i^s=0$ and $m_j^s=0$.
The node of a vehicle turning can be determined when the time difference between two data points in $M^p$ is greater than or equal to $\Delta^p$, and the two points satisfy $m_i^p=\min(m^p)$ and $m_j^p=\min(m^p)$.
The nodes in $V^R$ are filtered out and built. Then, all the nodes are sorted by the message timestamp $t^R$.

The redundant nodes in $V^R$ need to be merged before constructing the edges $E^R$ since the same node can be generated twice from the vehicle speed and pedal position separately.
The duplicated nodes in $V^R$ have the following situations:
\begin{itemize}
  \item \textbf{Short-term start and stop before red lights:} fully stop and short-term start when the driver quickly pushes and releases the throttle can lead to repeated nodes.
  \item \textbf{Pedal releasing before stop signs:} the driver may release the throttle before the vehicle fully stops, i.e., $t_j^{\min(p)} < t_i^{s=0}$ of the duplicated nodes $v_i^R$ and $v_j^R$.
\end{itemize}

As given by Lines~\ref{lst:line:algo-1-27}--\ref{lst:line:algo-1-32} in Algorithm~\ref{alg:can2graphALG}, all duplicated nodes are merged by checking the edge weight between two adjacent nodes $v^R_k$ and $v^R_{k+1}$ in $V^{R}$.
The edge weight $w_{k,{k+1}}^R$ stands for the identified driving path connected by the two consecutive nodes, i.e., the road segment between $v^R_k$ and $v^R_{k+1}$.
As the driving trajectory is located in a certain road network area, any edge weight should be no less than the length of the shortest road segments in the road network area, i.e., $\forall w_{k,{k+1}}^R \geq \min(w_{i,j})$.
Once the edge weight of $w_{k,{k+1}}^R$ does not satisfy the condition, the CAN-Trace attack will merge the node by deleting the node $v_k^R$.

Edges $E^R$, representing the driving path between two nodes, are constructed by connecting the consecutive nodes in $V^R$ after merging nodes.
Since the constructed graph of the driving trajectory from CAN messages is a line graph, each node is connected with only one or two nodes in the graph.
The weight $w_{i,i+1}^R$ of the edge $e_{i,i+1}^R$ is the road length between nodes $v_i$ and $v_{i+1}$, where $v_i$ and $v_{i+1}$ are the $i$-th and $(i+1)$-th intersections, respectively.
CAN-Trace attack calculates the distance traveled by the vehicle using discrete data points of vehicle speed and corresponding timestamps.
The weight $w_{i,i+1}^R$ can be calculated by using the rectangle method to estimate the driving distance, which is as given by
\begin{equation}
\label{equ:route_dis}
    w_{i,i+1}^R=\sum_{ t_a \leq t_l < t_{b}}m_l^s\times(t_{l+1}-t_{l}),
\end{equation}
where $t_a$ and $t_b$ are the timestamps when the vehicle leaves the node $v_i$ and arrives the node $v_{i+1}$, respectively. 
The timestamp of the CAN messages between $v_i$ and $v_{i+1}$ when the vehicle is driving during the period from $t_a$ to $t_b$ is denoted as $t_l$.
The vehicle speed at a given timestamp $t_l$ is denoted as $m^s_l$.
The time difference between two consecutive timestamps $t_l$ and $t_{l+1}$ is represented as $(t_{l+1} - t_l)$.
The driving length between two consecutive nodes $v_i$ and $v_{i+1}$, i.e., $w_{i,i+1}^R$, is calculated by multiplying the vehicle speed $m^s_l$ with the time difference $(t_{l+1} - t_l)$.

The driving trajectory construction from CAN messages is detailed in Algorithm~\ref{alg:can2graphALG} and illustrated in Fig.~\ref{fig:can2graphFlow}. The CAN-Trace attack reconstructs the driving trajectory by leveraging vehicle motion data from CAN messages, specifically vehicle speed and pedal position. CAN-Trace begins by determining speed and pedal thresholds, i.e., $\Delta^s$ and $\Delta^p$, using the K-means clustering algorithm on the respective CAN messages. Next, CAN-Trace identifies node candidates by selecting zero-speed CAN messages and minimal-pedal CAN messages when the time differences exceed the identified thresholds, as shown in Fig.~\ref{fig:can2graphFlow}. The zero-speed and minimal-pedal CAN messages are verified, as they imply that the vehicle is either stopping at or crossing intersections.
The distance between node candidates is calculated using the speed data and serves as the edge weight between them. CAN-Trace then removes edges whose weights are shorter than the shortest road segment in the road network, merging the end nodes of these edges accordingly.

\subsection{Proposed Trajectory Matching and Detection}
\label{subsec:filteringRule}

By using the generated graph, the CAN-Trace attack employs a subgraph matching algorithm to find the isomorphic subgraph of $G$.
The matching process can be formulated as given by
\begin{equation}
\label{equ:match}
    \hat {\mathcal R}=F(G,\bar R)=\{r_k\},
\end{equation}
where $\hat {\mathcal R}$ is the set of the matched subgraphs. $r_k$ is the $k$-th subgraph in $\hat {\mathcal R}$ and $F$ is the matching function, e.g., Vento-Foggia 2 (VF2)~\cite{cordella2004sub}.
VF2 is an improved matching method that employs a depth-first search strategy and utilizes feasibility rules to prune the search space when finding subgraph isomorphism.
With less memory requirements, VF2 is capable of effectively matching graphs of thousands of nodes and edges.
In the proposed attack, the VF2 is employed due to its efficiency.
By matching the road network graph $G$ with the constructed graph from CAN messages $\bar R$, the matched subgraphs $r_k$ can infer the driving trajectory with the geolocation data.

As given by Lines~\ref{lst:line:algo-2-1}--\ref{lst:line:algo-2-7} in Algorithm~\ref{alg:merged}, the subgraph $\hat{R}$ is identified by matching and sorting the subgraph candidates $r_l$.
Note that the lengths of road segments in the network graph $G$ and the constructed graph $\bar R$ can differ due to various factors, such as inaccurate map data and different driving behaviors.

\smallskip
\noindent\textbf{Relative Tolerance.}
The relative tolerance $\sigma$ bounds the tolerated value of the edge weight when matching the $\bar R$ with $G$ to identify the subgraph candidates $r_l$.
The tolerated length is calculated as $\sigma \times w_{i,{i+1}}$.
The weight difference between the edge pair of $e^{r_l}_{i,j}$ and $e_{i,j}$ is examined as $\varpi^l_{i,j}$.
The CAN-Trace attack finds the subgraph candidates $r_l$ when $\forall$ $\varpi^l_{i,{i+1}} \leq \sigma \times w_{i,{i+1}}$.

\smallskip
\noindent\textbf{Top-$K$ ranking.}
The Top-$K$ ranking method is used to sort and filter the identified subgraphs $r_l$ into the set $\hat{R} = \{r_k\}$, as referencing to Lines~\ref{lst:line:algo-2-8}--\ref{lst:line:algo-2-10} in Algorithm~\ref{alg:merged}.
The CAN-Trace attack calculates the distance difference between the edge pair of the driving path of $\bar R$ and the matched subgraph of $r_l$, i.e., $w_{i,{i+1}}^R-w_{i,{i+1}}^l$.
The mean of the differences for the entire driving trajectory, denoted as $\theta_l$, is assessed to rank the matched subgraphs, as given by
\begin{equation}
\label{equ:topK}
    \theta_l=\sum_{i=1}^{Q^*-1} \frac{|{w_{i,i+1}^R}-{w_{i,i+1}^l}|}{Q^*-1},
\end{equation}
where $w_{i,{i+1}}^R$ is the weight of the driving path $e_{i,{i+1}}^R$, and $w_{i,{i+1}}^l$ is the weight of the matched path $e_{i,{i+1}}^{r_l}$. 
The number of nodes in the actual driving trajectory graph is denoted as $Q^*$.
With the setting of $K$, CAN-Trace attack ranks $r_l$ by $\theta_l$ to filter out $K$ subgraphs $r_k$ as the attack results $\hat R = \{r_k\}$, where $k \leq K$.

\subsection{Complexity Analysis}
The proposed attack includes the sequential execution of CAN to graph conversion (i.e., Algorithm~\ref{alg:can2graphALG}) and trajectory detection (i.e., Algorithm~\ref{alg:merged}).
In Algorithm~\ref{alg:can2graphALG} Line~\ref{lst:line:algo-1-3}, the K-means clustering step for finding the thresholds $\Delta^s$ and $\Delta^p$ operates at $O(N_T^s\log N_T^s + N_T^p\log N_T^p)$~\cite{10056862}. 
The complexity of the graph construction steps in Algorithm~\ref{alg:can2graphALG} Lines~\ref{lst:line:algo-1-6}--\ref{lst:line:algo-1-15}, Lines~\ref{lst:line:algo-1-16}--\ref{lst:line:algo-1-25}, and Lines~\ref{lst:line:algo-1-27}--\ref{lst:line:algo-1-32}, are $O(N_T^s)$, $O(N_T^p)$, and $O(V^R)$, respectively, where $N_T^s$ and $N_T^p$ are the numbers of collected CAN messages of vehicle speed and pedal, respectively.
The sorting step of Algorithm~\ref{alg:can2graphALG} Line~\ref{lst:line:algo-1-26} has the complexity of $O(V^R \log V^R)$, where $V^R$ is the number of nodes in the constructed driving trajectory graph $\bar R$.
Thus, the computational and time complexity of Algorithm~\ref{alg:can2graphALG} is $O((\log N_T^s+1)N_T^s + (\log N_T^p+1)N_T^p + V^R + V^R \log V^R)$.
In practice, the vehicle speed and pedal position messages can be collected at the same rate, and the number of the vehicle messages is much greater than the number of stops in the driving trajectory, i.e., $N_T^s = N_T^p \gg V^R$. Thus, the complexity of Algorithm~\ref{alg:can2graphALG} can be written as $O(N_T^s\log N_T^s)$.

The most time-consuming step of Algorithm~\ref{alg:merged} is the subgraph matching. 
The VF2-based implementation exhibits a complexity of $O(N_V^2)$ in the best case and $O(N_V! N_V)$ in the worst case~\cite{cordella2004sub}, where $N_V$ is the number of nodes in the graph of the road network. 
As a result, the overall computational and time complexity of the proposed attack is between $O(N_T^s\log N_T^s + N_V^2)$ and $O(N_T^s\log N_T^s + N_V! N_V)$. 
We note that the proposed CAN-Trace attack can be conducted offline, with CAN messages collected during driving and processed afterward. Therefore, the attack is not time-critical. The above-mentioned complexity is tolerable.

\subsection{Assessment Criteria}

The node in $r_k$ is determined as true positive when the geolocation is the same as the corresponding node in $G^*$.
The attack performance is assessed by the attack success rate $\Psi$, attack precision $P$, and the spatial distance offset $D$.

The attack success rate $\Psi$ refers to the attack coverage of the CAN-Trace attack,
evaluated by
\begin{equation}
\label{equ:asr}
\Psi = \frac{\left|\bigcup_{k=1}^{K} (V^{r_k} \cap V^*)\right|}{Q^*},
\end{equation}
where $V^{r_k}$ is the inferred nodes of the subgraph $r_k$, $K$ is the number of matched subgraphs, and  $V^*$ is the set of nodes in the actual driving trajectory.
The size of the distinct set containing all correctly inferred nodes in $r_k$ is calculated as $\left|\bigcup_{k=1}^{K} (V^{r_k} \cap V^*)\right|$.
The number of nodes in the graph of the actual driving trajectory  $G^*$ is denoted as $Q^*$.

The attack precision $P$ refers to the positive predictive value, i.e., the True Positive (TP) divided by the sum of the true positive and False Positive (FP), which indicates the effectiveness of the attack.
In our experiments, $P$ represents the average percentage of the correctly matched nodes of the attack result and is as given by
\begin{equation}
\begin{split}
\label{equ:accuracy}
P &= \frac{TP}{TP+FP} =\frac{1}{K} \sum_{k=1}^{K} \frac{N'_{k}}{Q^*},
\end{split}
\end{equation}
where $N'_k$ is the number of correctly inferred nodes in $r_k$.

The spatial distance offset $D$ quantifies the geospatial discrepancy between the inferred driving trajectory $r_k$ and the actual driving trajectory $G^*$, which is given in~\eqref{equ:sd}.

\begin{equation}
\label{equ:sd}
D = \frac{1}{K} \sum_{k=1}^{K} \frac{d(V^*, V^{r_k})}{Q^*},
\end{equation}
where $V^{r_k}$ is the set of nodes in the selected subgraph $r_k$, and $d(V^*, V^{r_k})$ is the average of the Euclidean distance between each node pair of the selected subgraph and that of the actual driving trajectory.

The false negative rate $\mathbb{F}$ is the average ratio of nodes in the actual driving trajectory $G^*$ that are not correctly identified in the matched subgraph $r_k$, as given by
\begin{equation}
\begin{split}
\label{equ:FPR}
\mathbb{F} =\frac{1}{K} \sum_{k=1}^{K} \frac{N^{\mathbb{F}}_{k}}{Q^*},
\end{split}
\end{equation}
where $N_k^{\mathbb{F}}$ represents the number of nodes that are in the actual driving trajectory graph $G^*$ but not identified in the matched subgraph $r_k$, and $Q^*$ is the number of nodes in the graph of the actual driving trajectory.

\begin{figure*}[!htbp]
    \centering
    \includegraphics[width=0.8\linewidth]{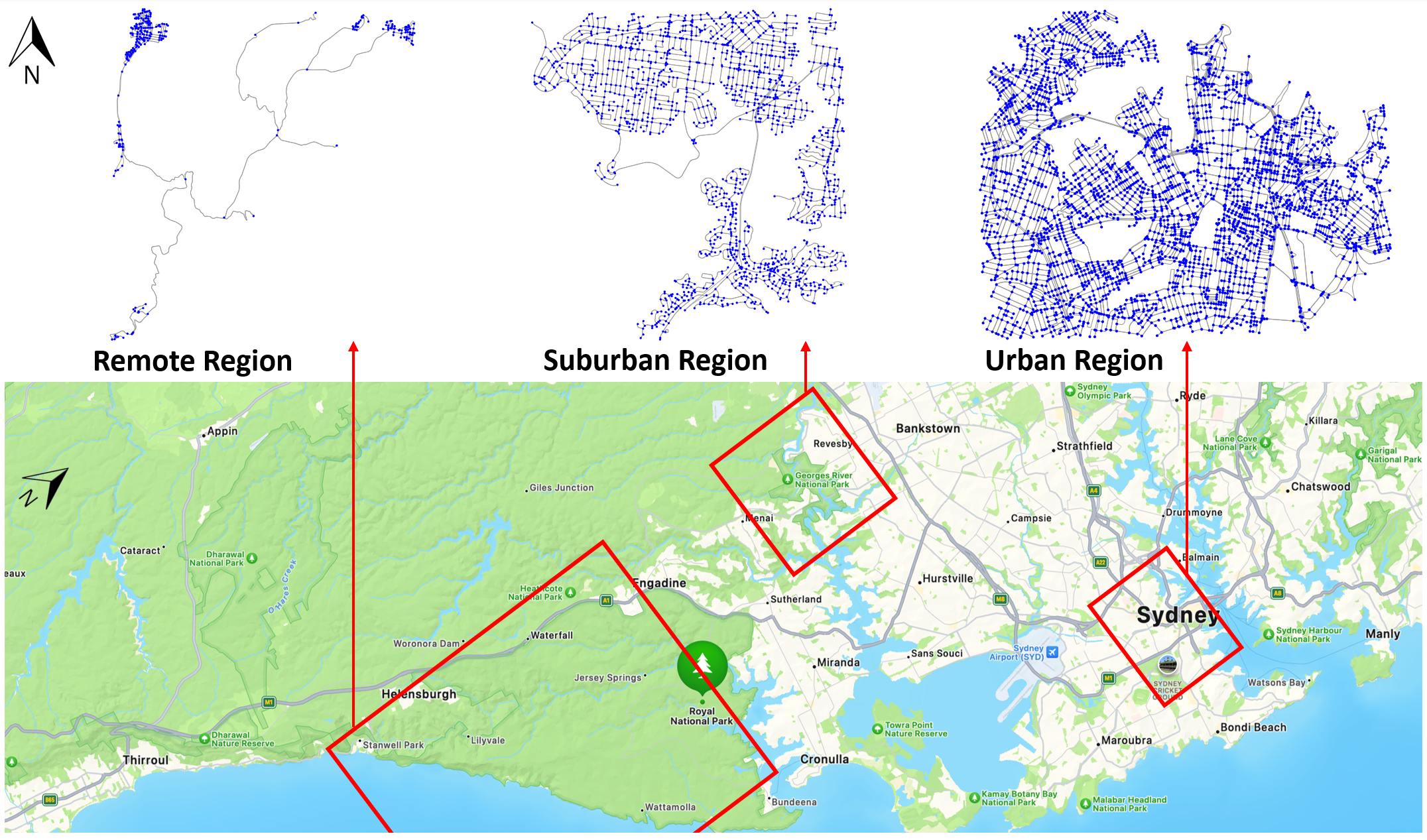}
    \caption{Topology graphs of used road networks in remote, suburban, and urban regions, showing road network density from low to high. The road networks are from Sydney city, extracted using OpenStreetMap. Blue dots represent road network intersections, and edges represent the road segments connecting the intersections. }
    \label{fig:RoadNetworkSetting}
\end{figure*}

\section{Performance Evaluation}
\label{performance}

This section evaluates the proposed CAN-Trace attack in a real road network under different experimental settings and subjects, including the type and area size of the road network, the models and years of experimental cars, and the length of the actual driving trajectory.

\subsection{Experiment Setup}

We collect driving data from three different cars (i.e., a 2013 petrol car, a 2015 petrol car, and a 2022 hybrid car). All three cars can provide the required vehicle motion data, i.e., speed and pedal data, via the standard OBD-II protocol. A laptop is connected to each vehicle's OBD-II port using the same CAN analyzer (specifically, the PEAK PCAN-USB Pro FD adapter\footnote{The PEAK PCAN-USB Pro FD adapter allows a computer to connect the vehicle CAN bus via the OBD-II port. \url{https://www.peak-system.com/PCAN-USB-Pro-FD.366.0.html?&L=1}}). The CAN analyzer reads vehicle speed and pedal data at 0.1-second intervals and logs the motion data on the laptop with the complementary software
(i.e., PCAN-Explorer\footnote{PCAN-Explorer is a professional Windows software that equips with PEAK devices to view, transmit, and record real-time CAN messages. \url{https://www.peak-system.com/PCAN-Explorer-6.415.0.html?&L=1}}). The collected CAN messages are converted from the raw trace format (.trc) into CSV format, with decoded numerical vehicle speed and pedal data, along with the timestamp of each data entry. The consistent data collection process and settings help minimize discrepancies attributable to different vehicles. While the current data collection process requires physical devices to be connected to the CAN bus, typically through the OBD-II port, car manufacturers and vehicle applications may have direct access to CAN messages.

The proposed CAN-Trace method is tested across 279 driving trajectories in three types of regions within Sydney: remote, suburban, and urban, as illustrated in Fig.~\ref{fig:RoadNetworkSetting}. The remote regions feature low density, long road segments, and distinct street blocks. The suburban regions have medium density, a mix of short and long road segments, and partially similar street blocks. The urban regions are characterized by high density, short road segments, and similar street blocks.

The matching process of the CAN-Trace attack is performed on a Windows 10 laptop with Intel(R) Core(TM) i7-1185G7 CPU@3GHz and 16GB RAM.
The server runs in an Anaconda environment with Python 3.6.8, NetworkX 2.5.1, OSMnx 1.2.1 and Numpy 1.19.5.

\subsection{Experimental Setting}
\label{performance_variable}

The experimental vehicle is driven in different types of road networks, i.e., urban and suburban, to evaluate the attack generalizability.
The evaluation is also performed across diverse configurations, i.e., the road network area, the value $K$ for Top-$K$ ranking, and the number of nodes in the driving trajectory graph $Q^*$.

\smallskip
\noindent\textbf{Road Network Area.}
The road network area represents the size of the road network where the constructed trajectories from CAN messages are matched.
A large road network area stands for a bigger graph and more nodes and edges.
In this experiment, the road network area is set to various values, i.e., \qtyproduct{5 x 5}{\km}, \qtyproduct{10 x 10}{\km},  \qtyproduct{15 x 15}{\km}, \qtyproduct{20 x 20}{\km},  \qtyproduct{25 x 25}{\km}, and \qtyproduct{30 x 30}{\km}.
A \qtyproduct{5 x 5}{\km} road network area crosses two or more suburbs of the urban region and is only located in one suburb of the suburban region.
The majority of the metro city of Sydney can be covered in a \qtyproduct{30 x 30}{\km} road network area, which means our experimental settings can prove the attack performance regarding the victim's daily range of driving.

\smallskip
\noindent\textbf{Driving Trajectory Length.}
The length of an actual driving trajectory can be captured by the number of nodes in the related graph, denoted by $Q^*$. 
The experiments have different settings of $Q^*$ value: \qtyproduct{5}{}, \qtyproduct{7}{}, \qtyproduct{10}{}, \qtyproduct{12}{} and \qtyproduct{15}{}.
In our experiments, the driving trajectory length is approximately \qtyproduct{100}{m} long for the edge between two nodes.
Similar road network patterns are intensive in the urban region but sparse in the suburban region.

\smallskip
\noindent\textbf{Value of $K$.}
Top-$K$ ranking, which determines the size of $\hat{R}$, is designed to select $K$ detected subgraph candidates $\hat{R} = \{r_k\}$ as the attack results.
The attack results in $\hat{R}$ occasionally contain outliers (i.e., a subgraph candidate with zero nodes matched correctly) at smaller $K$ values within a large road network area, which leads to a lower attack success rate $\Psi$.
A higher $K$ value considers a broader range of matched candidates to enhance attack robustness, but the trade-off is a lower attack precision $P$.

\subsection{Experimental Results}
\label{performance_attack}

The overall trend of all the attack success rate $\Psi$, attack precision $P$, and the spatial distance offset $D$ is consistent in the urban and suburban regions, as shown in Figs.~\ref{fig:asr}--\ref{fig:Distance}.

\smallskip
\subsubsection{\textbf{Attack Success Rate}}

\begin{figure*}[!htbp]
    \centering
    \subfigure[Attack within urban region grouped by road network area]{
    \begin{minipage}[t]{0.22\textwidth}
    \centering
    \includegraphics[width=1.7in]{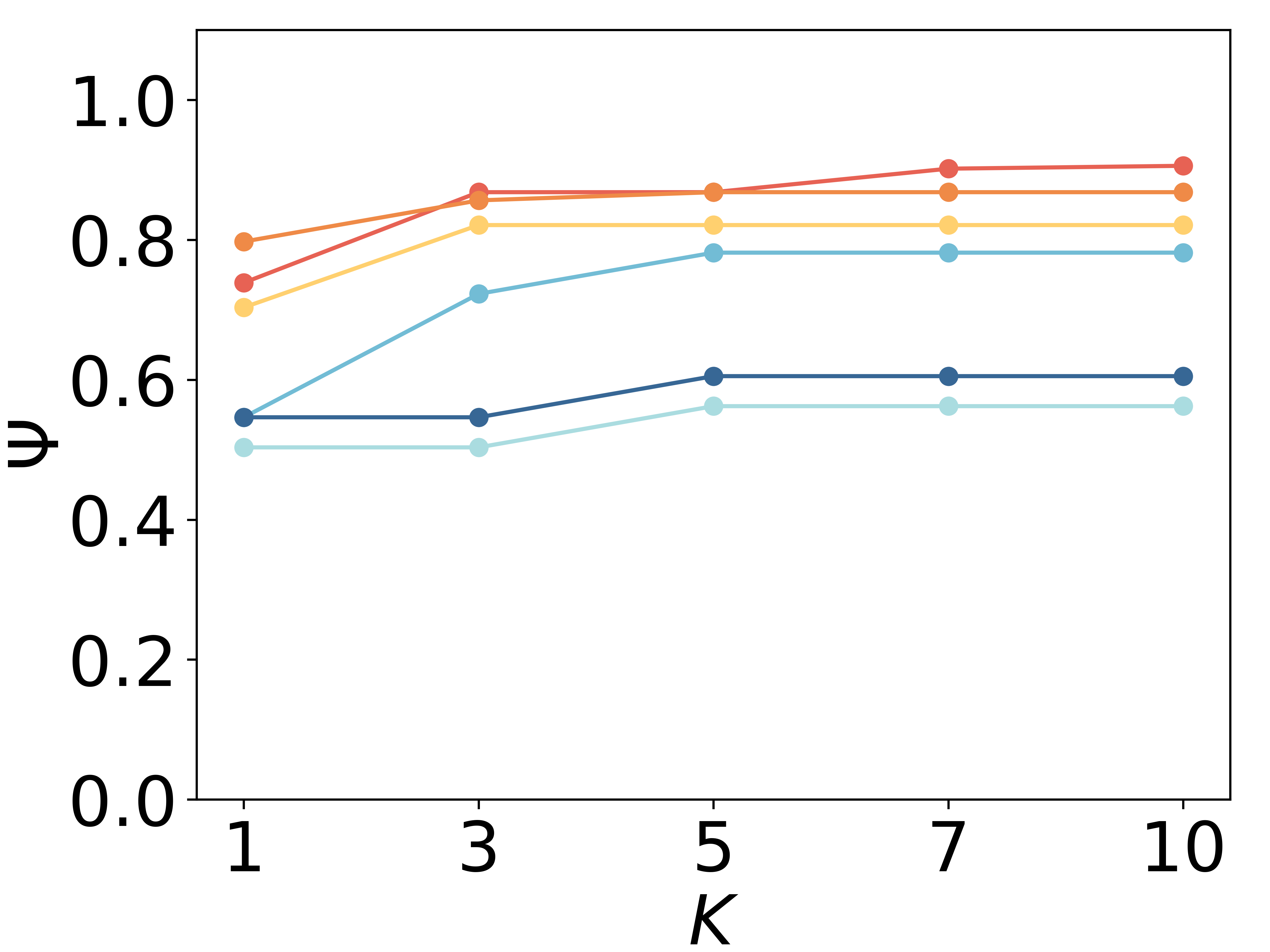}
    \end{minipage}
    \label{fig:asr_urban_area}
    }
    \subfigure[Attack within suburban region grouped by road network area]{
    \begin{minipage}[t]{0.22\textwidth}
    \centering
    \includegraphics[width=1.7in]{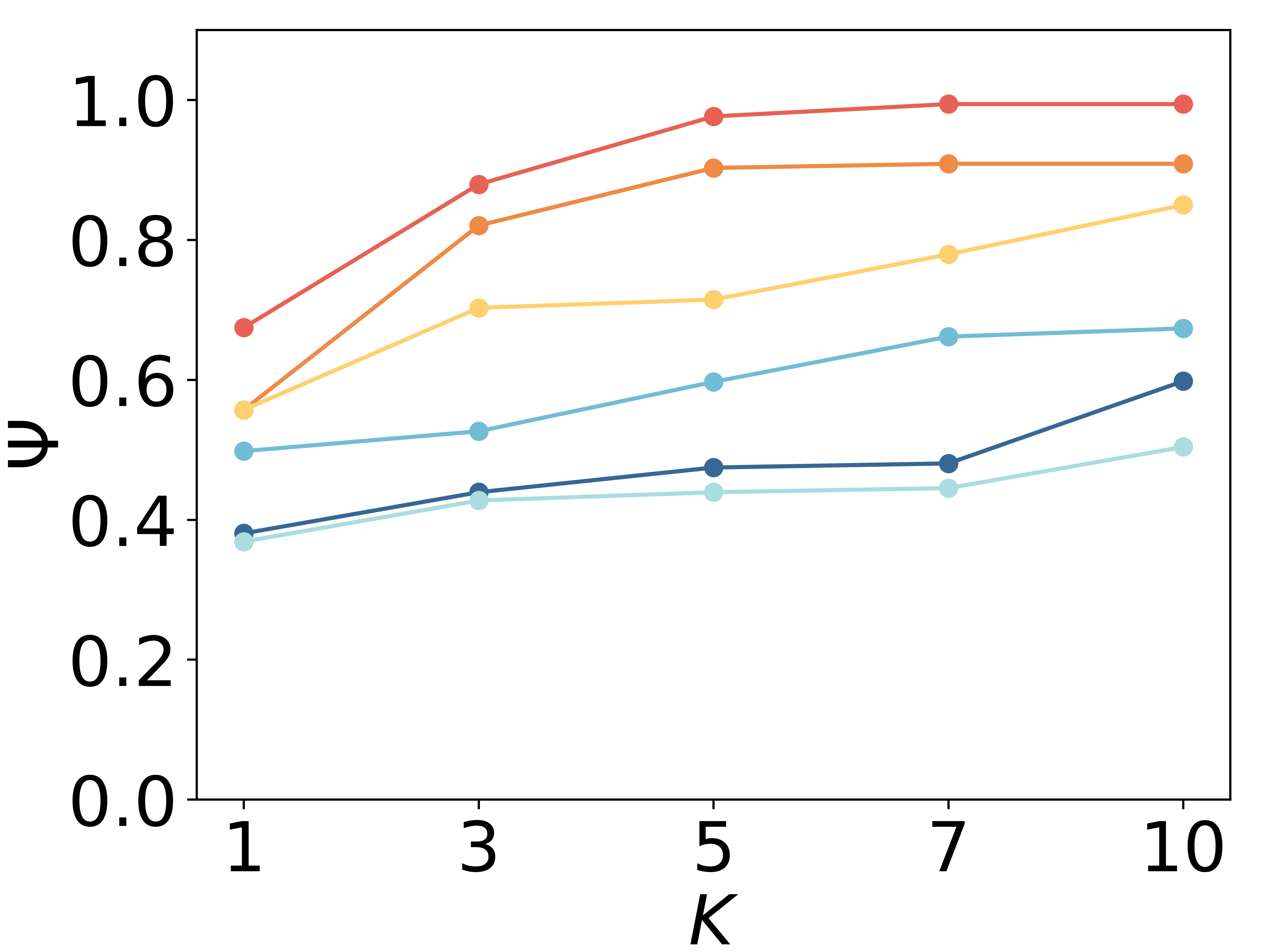}
    \end{minipage}
    \label{fig:asr_suburban_area}
    }
    \subfigure[Attack within urban region grouped by $Q^*$]{
    \begin{minipage}[t]{0.22\textwidth}
    \centering
    \includegraphics[width=1.7in]{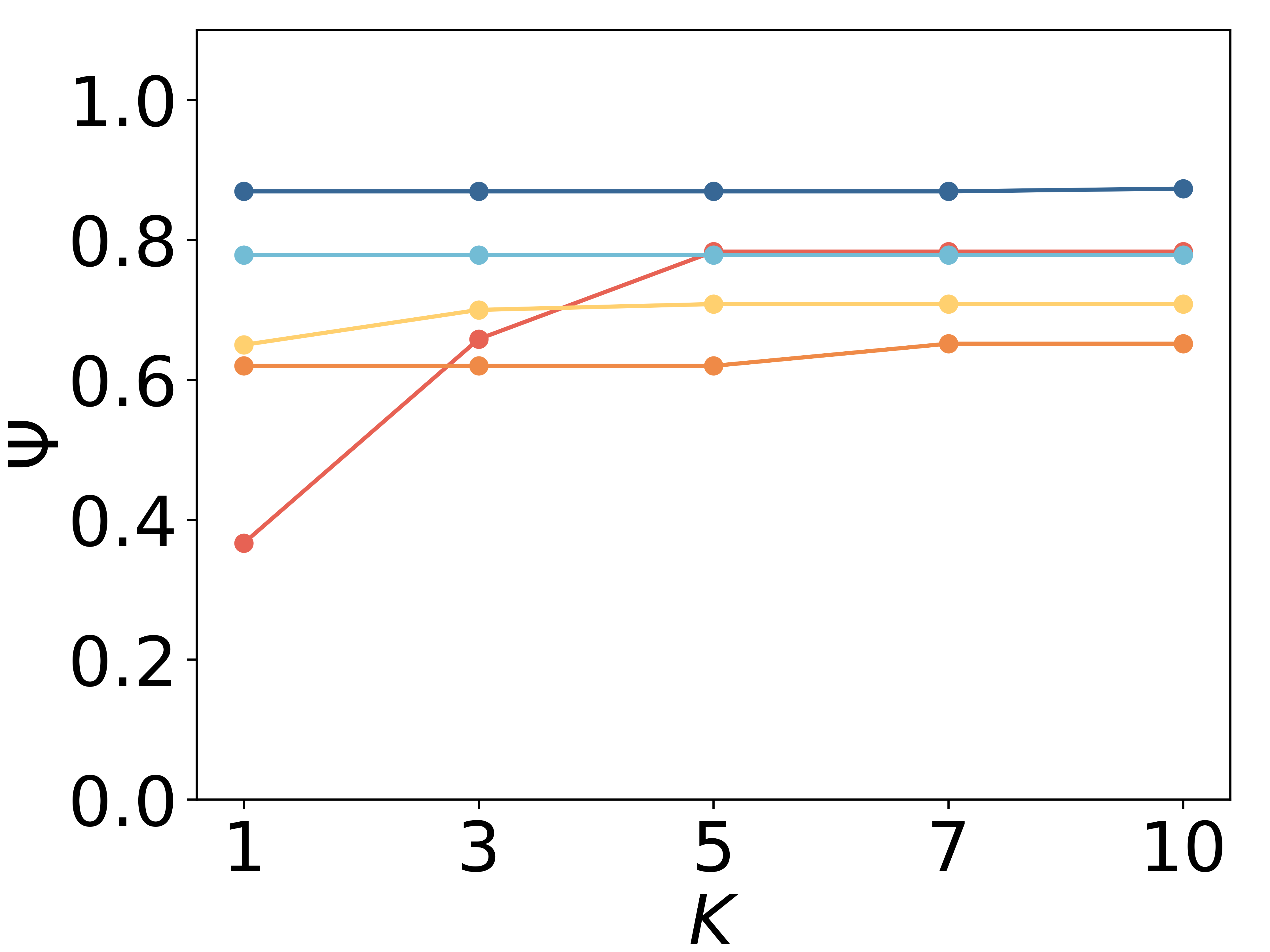}
    \end{minipage}
    \label{fig:asr_urban_node}
    }
    \subfigure[Attack within suburban region grouped by $Q^*$]{
    \begin{minipage}[t]{0.22\textwidth}
    \centering
    \includegraphics[width=1.7in]{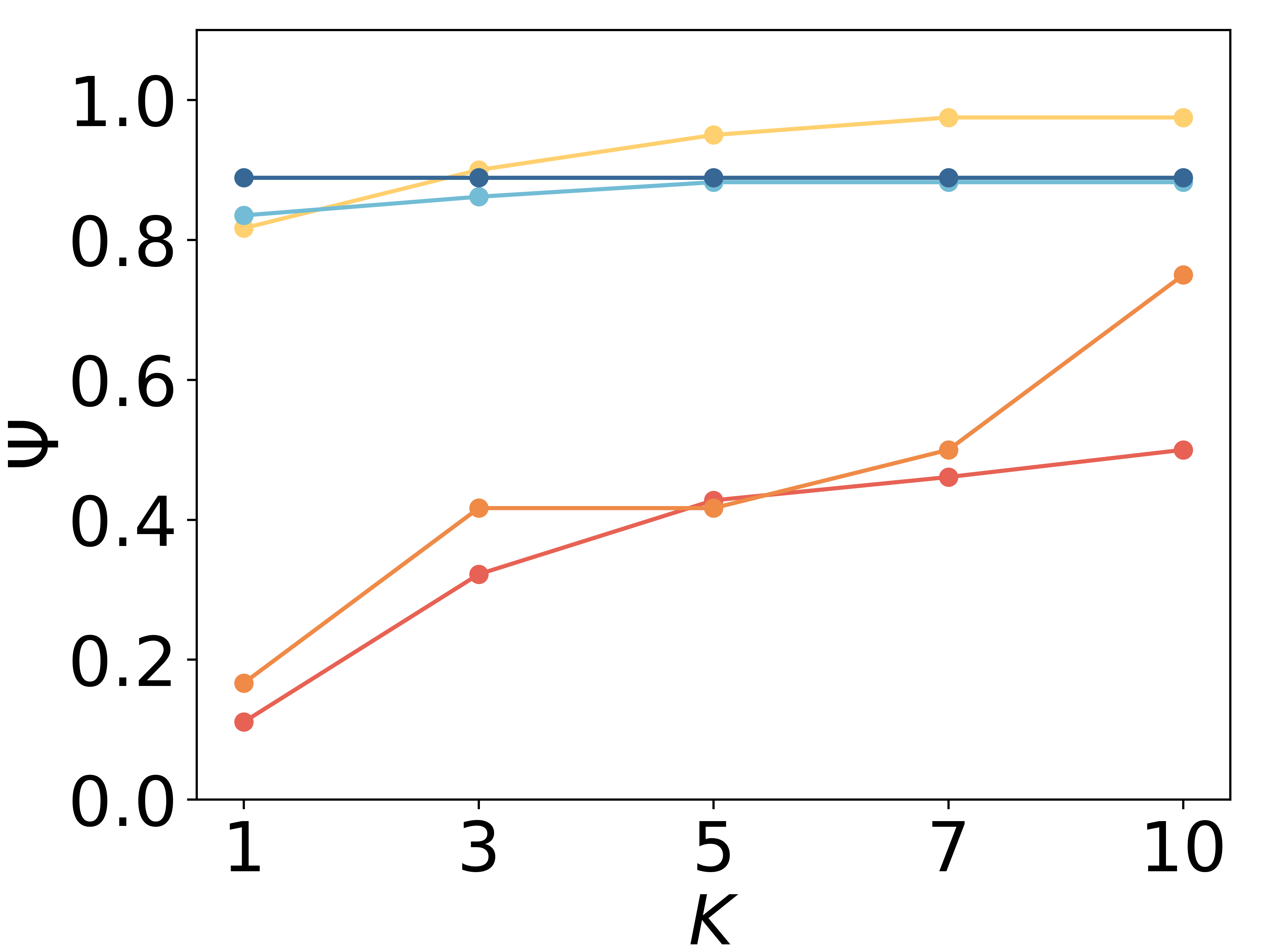}
    \end{minipage}
    \label{fig:asr_suburban_node}
    }
\caption{Comparison of attack success rate $\Psi$ within urban and suburban road networks: $y$-axes are the average values of $\Psi$ among different $Q^*$ in subplots (a) and (b), and different road network areas in subplots (c) and (d).
The bar colors in subplots (a) and (b) represent the size of road network areas as follows: \textcolor{red}{$\bullet$}~\qtyproduct{5 x 5}{\km}, \textcolor{orange}{$\bullet$}~\qtyproduct{10 x 10}{\km}, \textcolor{yellow}{$\bullet$}~\qtyproduct{15 x 15}{\km}, \textcolor{blue}{$\bullet$}~\qtyproduct{20 x 20}{\km}, \textcolor{black}{$\bullet$}~\qtyproduct{25 x 25}{\km}, and \textcolor{cyan}{$\bullet$}~\qtyproduct{30 x 30}{\km}.
The bar colors in subplots (c) and (d) represent the numbers of nodes in the graph of actual driving trajectory as follows: \textcolor{red}{$\bullet$}~$Q^* = 5$, \textcolor{orange}{$\bullet$}~$Q^* = 7$, \textcolor{yellow}{$\bullet$}~$Q^* = 10$, \textcolor{blue}{$\bullet$}~$Q^* = 12$, and \textcolor{black}{$\bullet$}~$Q^* = 15$.}
\label{fig:asr}
\end{figure*}

\begin{figure*}[!htbp]
    \centering
    \subfigure[Attack within urban region grouped by $K$]{
    \begin{minipage}[t]{0.22\textwidth}
    \centering
    \includegraphics[width=1.7in]{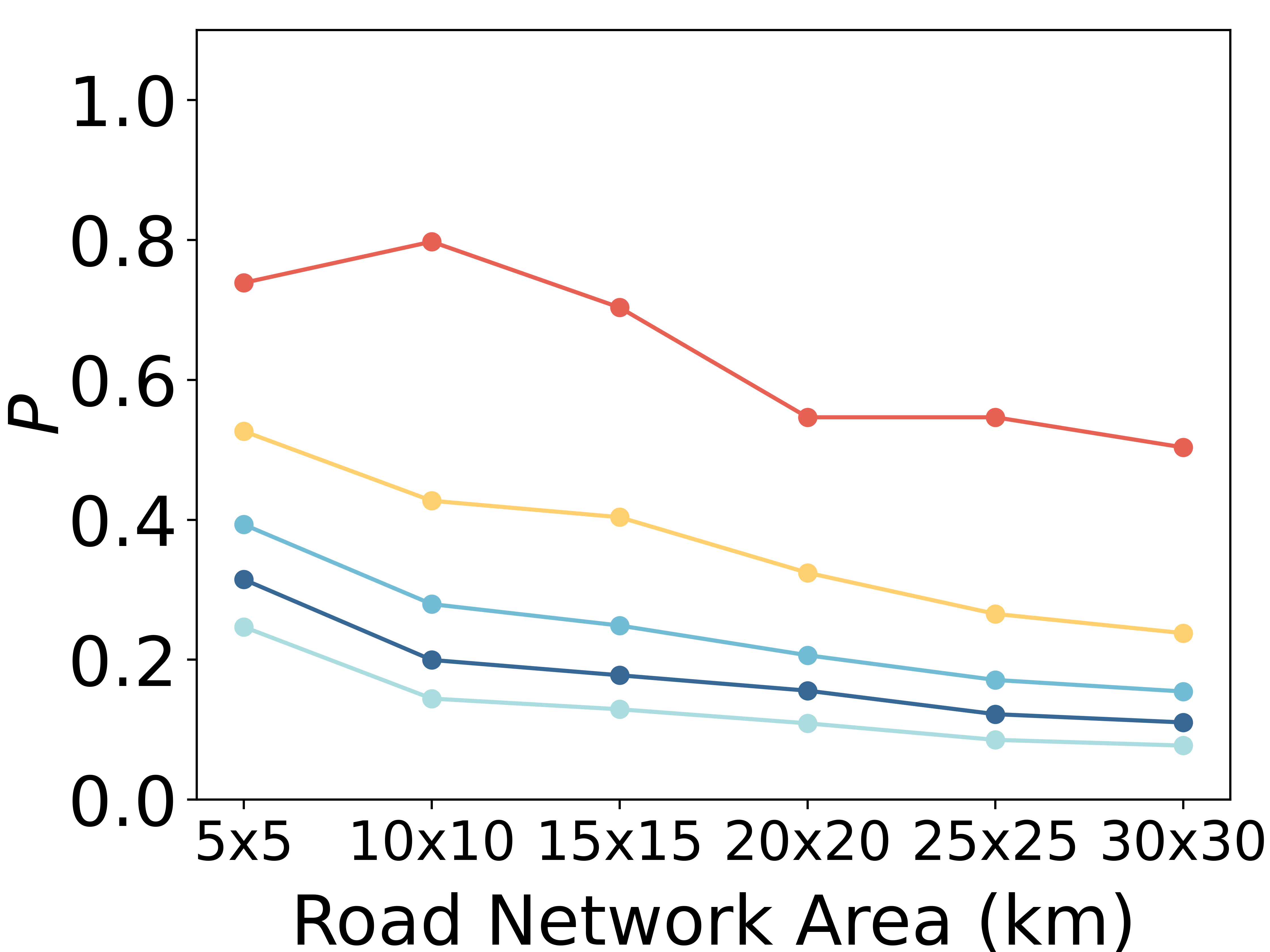}
    \end{minipage}
    \label{fig:precision_urban_area}
    }
    \subfigure[Attack within suburban region grouped by $K$]{
    \begin{minipage}[t]{0.22\textwidth}
    \centering
    \includegraphics[width=1.7in]{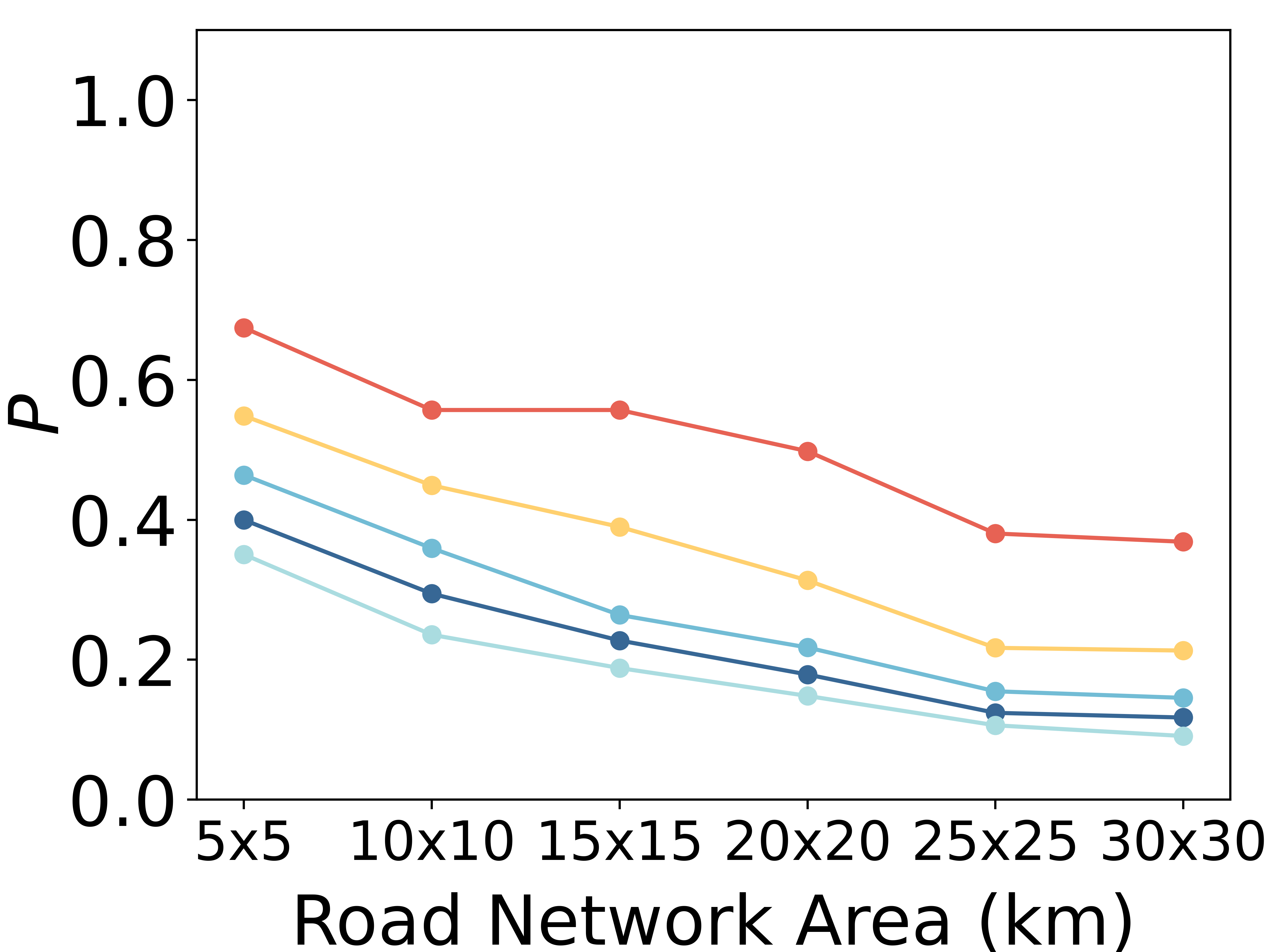}
    \end{minipage}
    \label{fig:precision_suburban_area}
    }
    \subfigure[Attack within urban region grouped by $K$]{
    \begin{minipage}[t]{0.22\textwidth}
    \centering
    \includegraphics[width=1.7in]{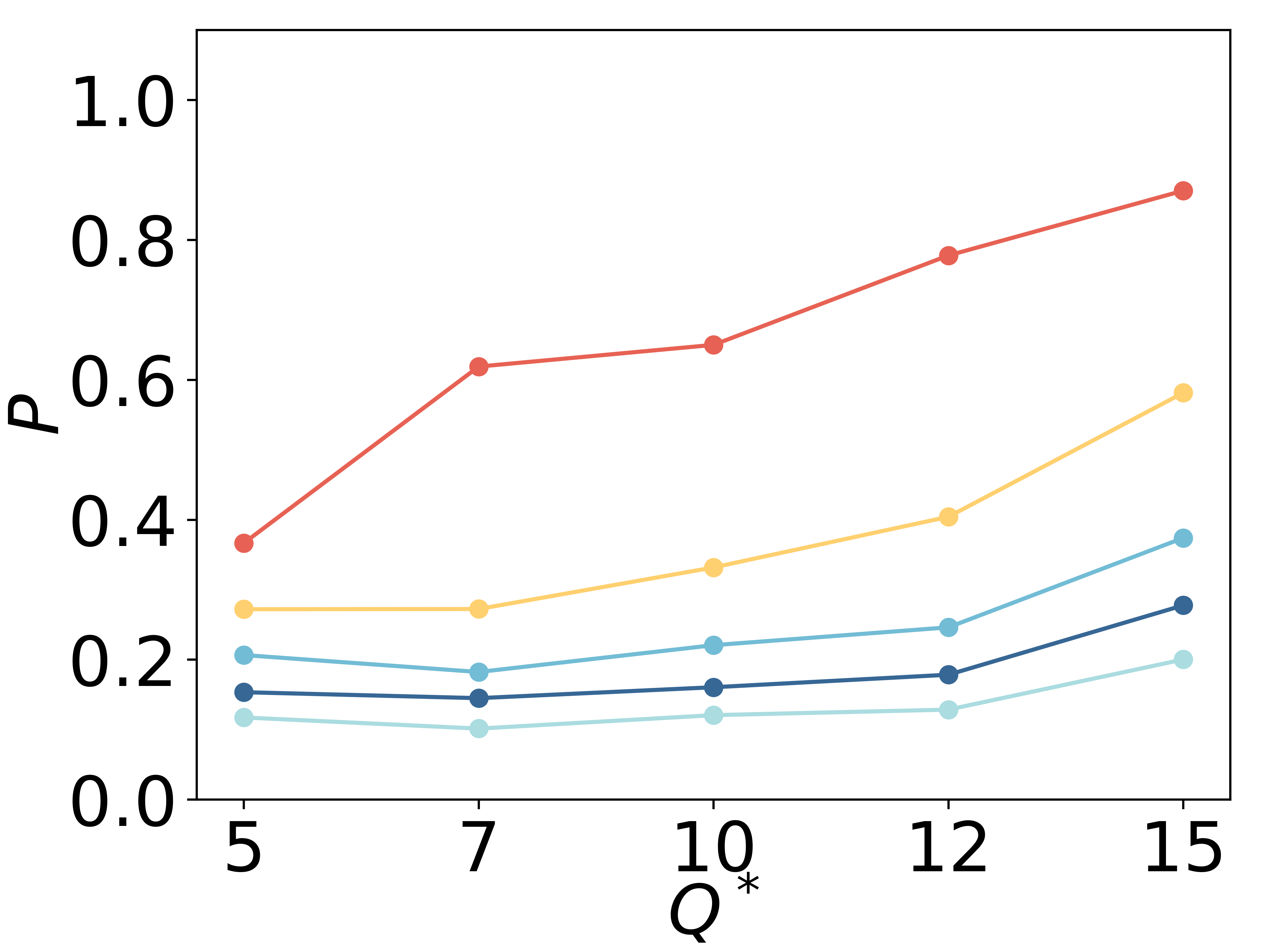}
    \end{minipage}
    \label{fig:precision_urban_node}
    }
    \subfigure[Attack within suburban region grouped by $K$]{
    \begin{minipage}[t]{0.22\textwidth}
    \centering
    \includegraphics[width=1.7in]{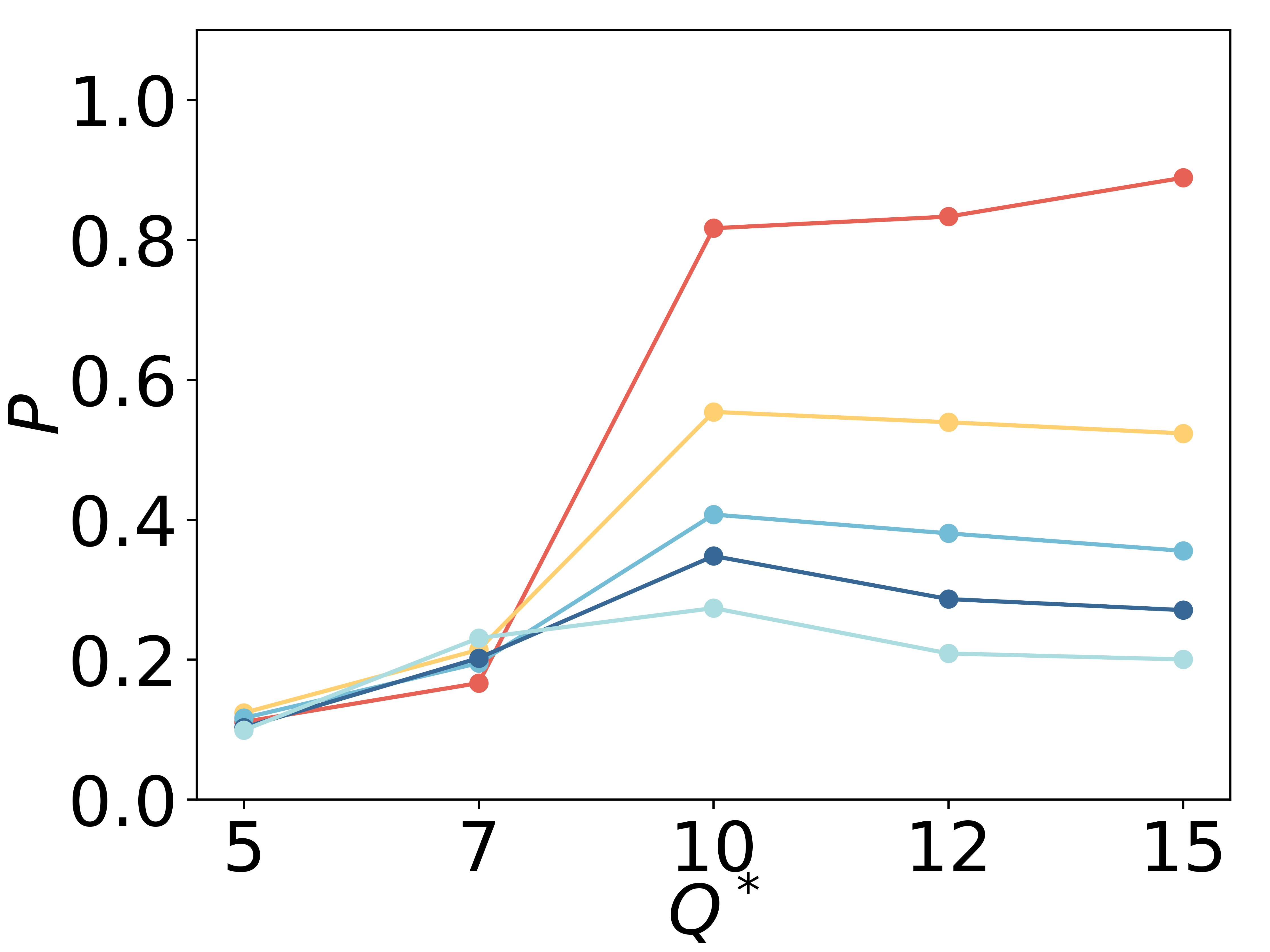}
    \end{minipage}
    \label{fig:precision_suburban_node}
    }
\caption{Comparison of attack precision $P$ within urban and suburban road networks: $y$-axes are the average values of $P$ among different $Q^*$ in (a) (b) and different road network areas in (c) (d).
The bar colors in all subplots represent the values of Top-$K$ as follows: \textcolor{red}{$\bullet$}~$K = 1$, \textcolor{yellow}{$\bullet$}~$K = 3$, \textcolor{blue}{$\bullet$}~$K = 5$, \textcolor{black}{$\bullet$}~$K = 7$, and \textcolor{cyan}{$\bullet$}~$K = 10$.}
\label{fig:precision}
\end{figure*}

\begin{figure*}[!htbp]
    \centering
    \subfigure[Attack within urban region when $K=1$ grouped by $Q^*$]{
    \begin{minipage}[t]{0.22\textwidth}
    \centering
    \includegraphics[width=1.7in]{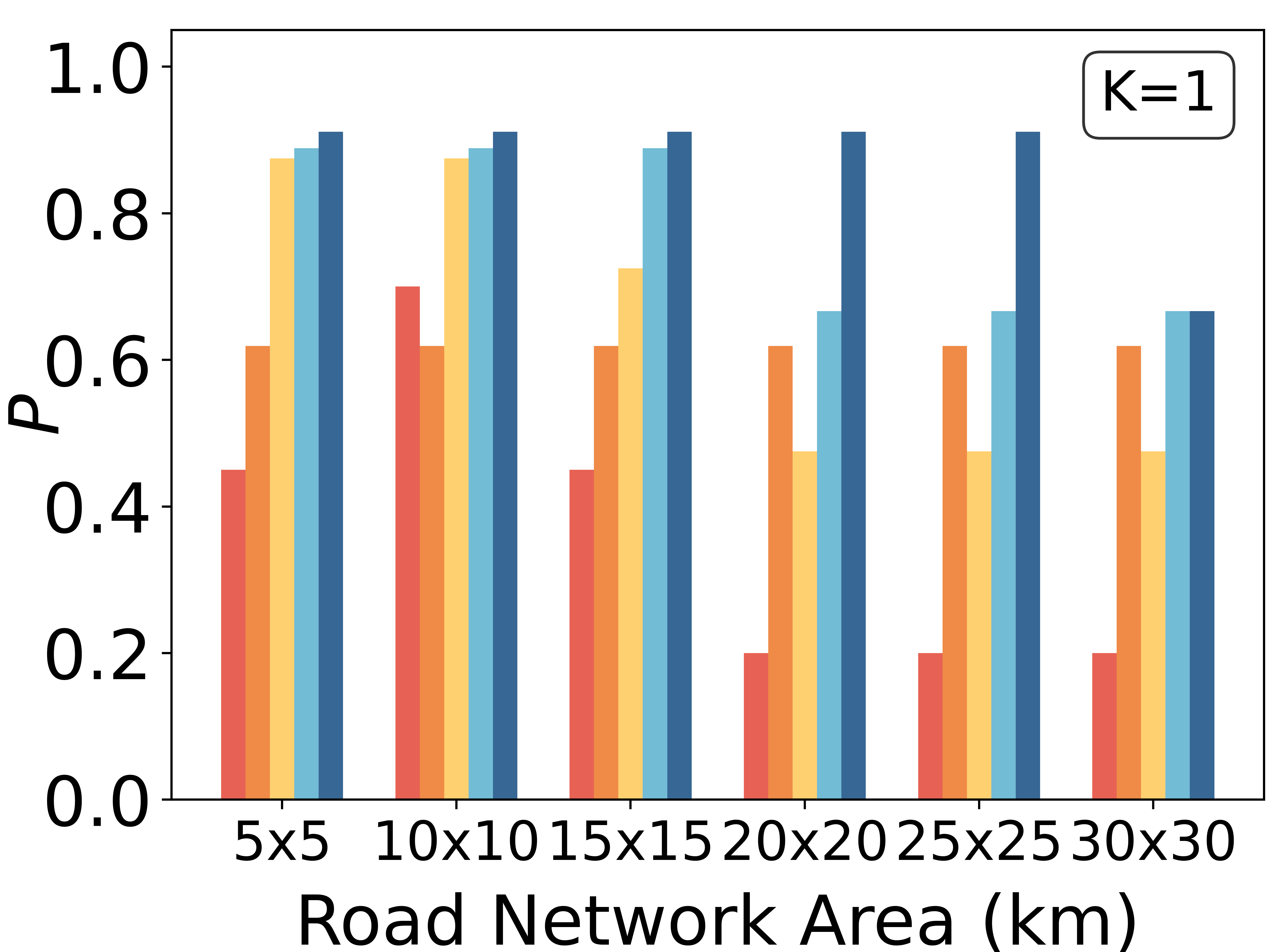}
    \end{minipage}
    \label{fig:precision_urban_area_k1}
    }
    \subfigure[Attack within suburban region when $K=1$ grouped by $Q^*$]{
    \begin{minipage}[t]{0.22\textwidth}
    \centering
    \includegraphics[width=1.7in]{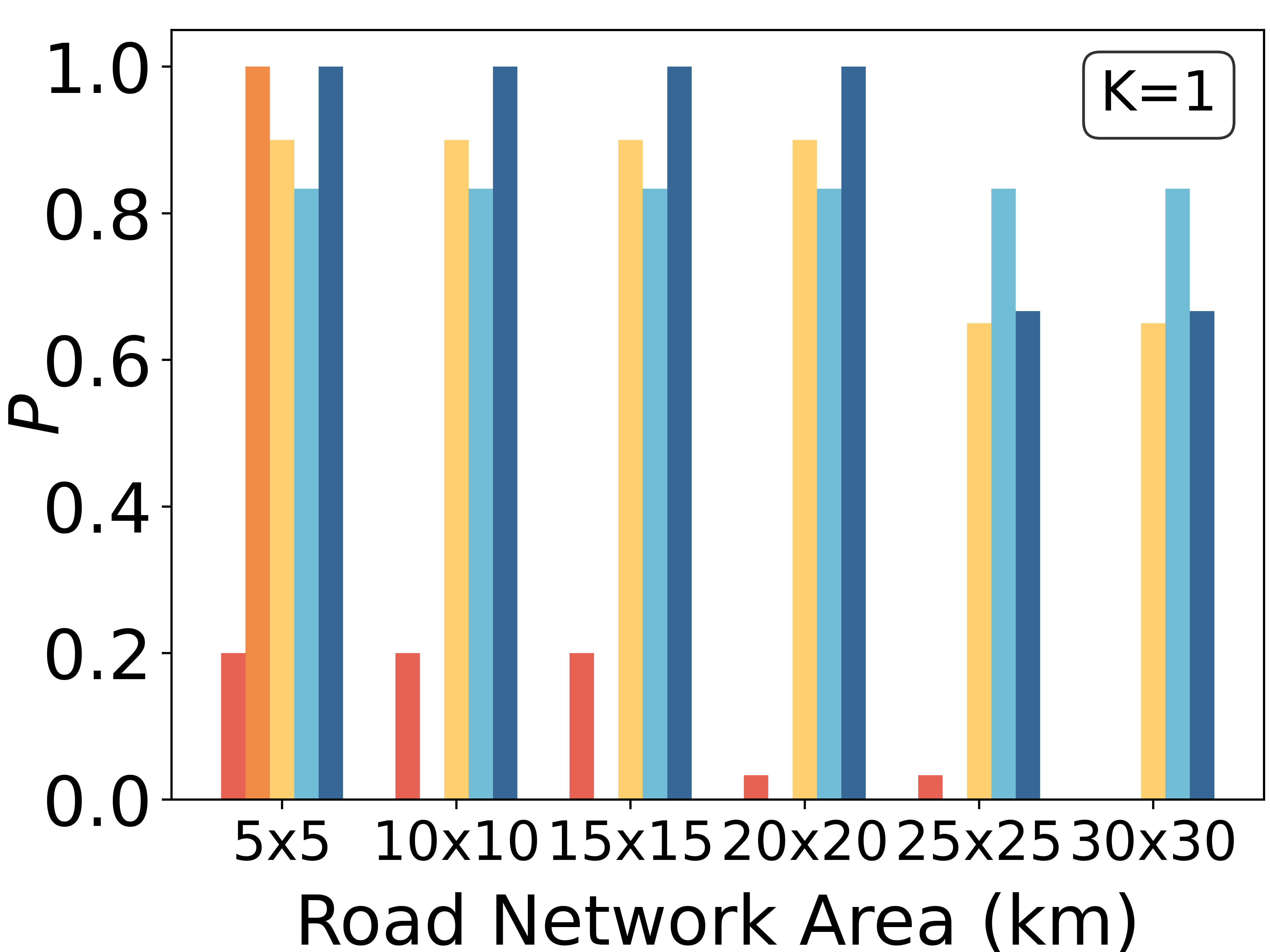}
    \end{minipage}
    \label{fig:precision_suburban_area_k1}
    }
    \subfigure[Attack within urban region when $K=3$ grouped by $Q^*$]{
    \begin{minipage}[t]{0.22\textwidth}
    \centering
    \includegraphics[width=1.7in]{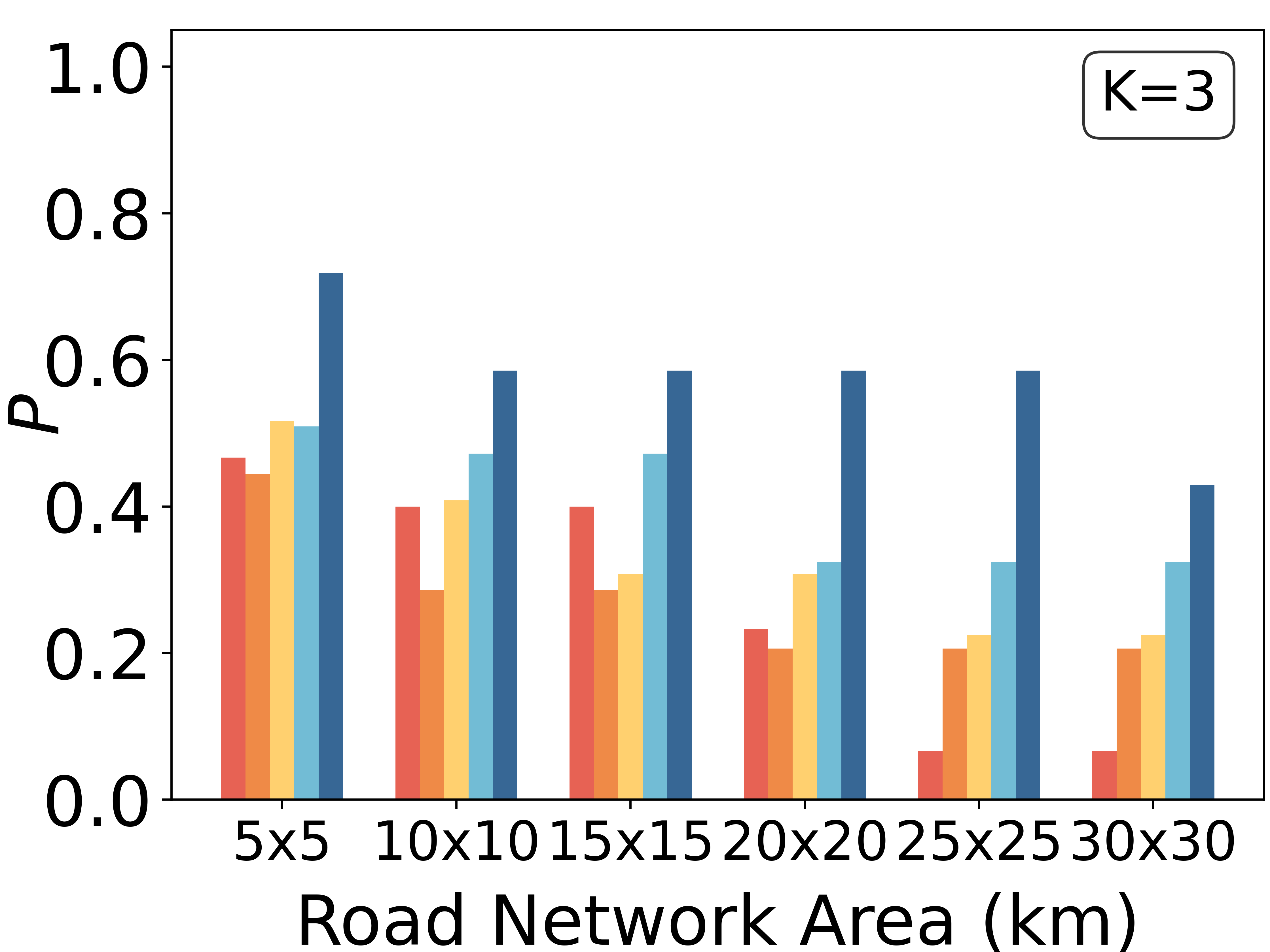}
    \end{minipage}
    \label{fig:precision_urban_area_k3}
    }
    \subfigure[Attack within suburban region when $K=3$ grouped by $Q^*$]{
    \begin{minipage}[t]{0.22\textwidth}
    \centering
    \includegraphics[width=1.7in]{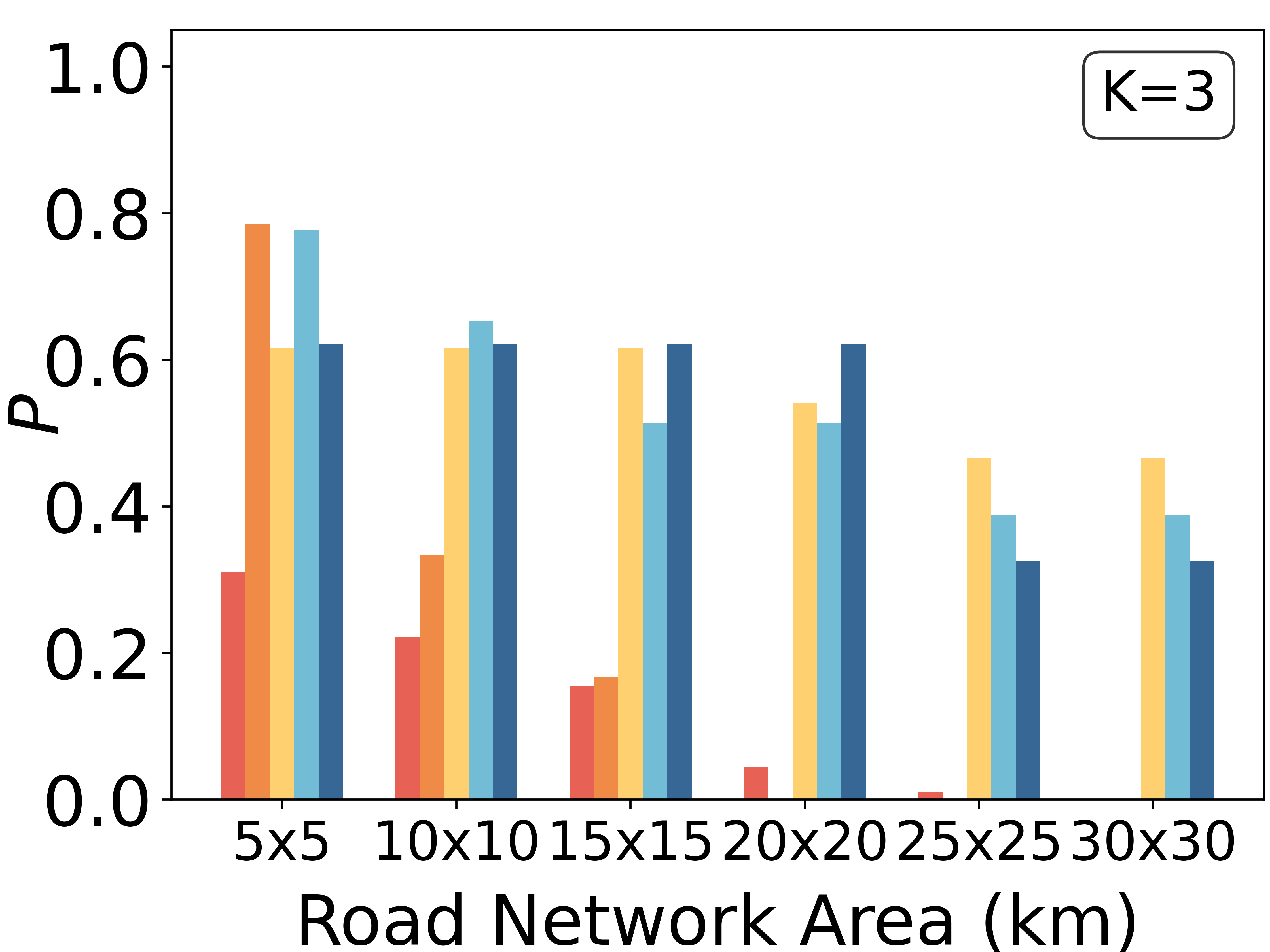}
    \end{minipage}
    \label{fig:precision_suburban_area_k3}
    }
\caption{Comparison of attack precision $P$ of $K \in \{1, 3\}$ within urban and suburban road networks: $y$-axes are the average values of $P$.
The bar colors in all subplots represent the numbers of nodes in the graph of actual driving trajectory as follows: \textcolor{red}{\rule{8pt}{4pt}}~$Q^* = 5$, \textcolor{orange}{\rule{8pt}{4pt}}~$Q^* = 7$, \textcolor{yellow}{\rule{8pt}{4pt}}~$Q^* = 10$, \textcolor{blue}{\rule{8pt}{4pt}}~$Q^* = 12$, and \textcolor{black}{\rule{8pt}{4pt}}~$Q^* = 15$.}
\label{fig:precision_k13}
\end{figure*}

\begin{figure*}[!htbp]
    \centering
    \subfigure[Attack within urban region grouped by $K$]{
    \begin{minipage}[t]{0.22\textwidth}
    \centering
    \includegraphics[width=1.7in]{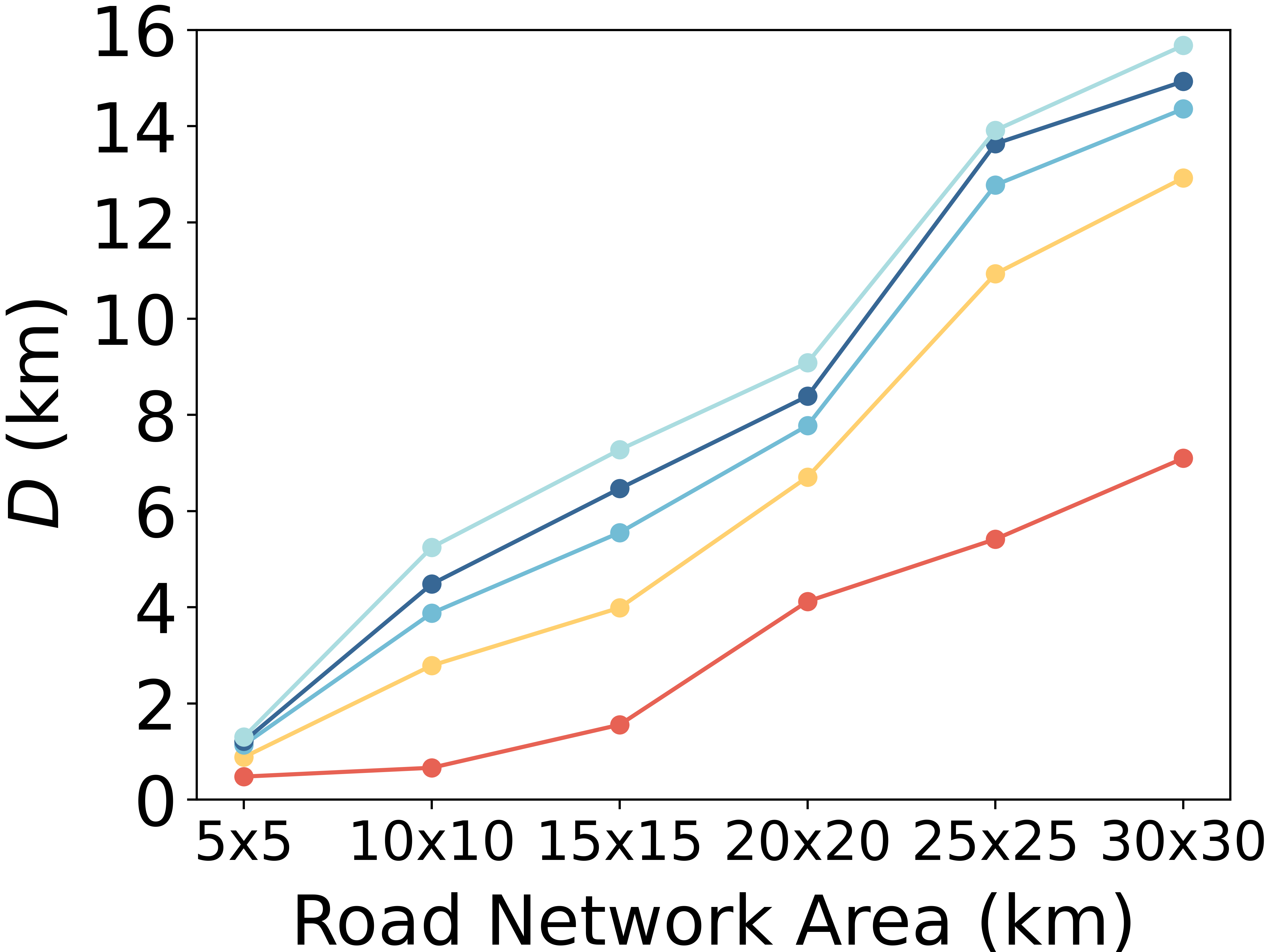}
    \end{minipage}
    \label{fig:distance_urban_area}
    }
    \subfigure[Attack within suburban region grouped by $K$]{
    \begin{minipage}[t]{0.22\textwidth}
    \centering
    \includegraphics[width=1.7in]{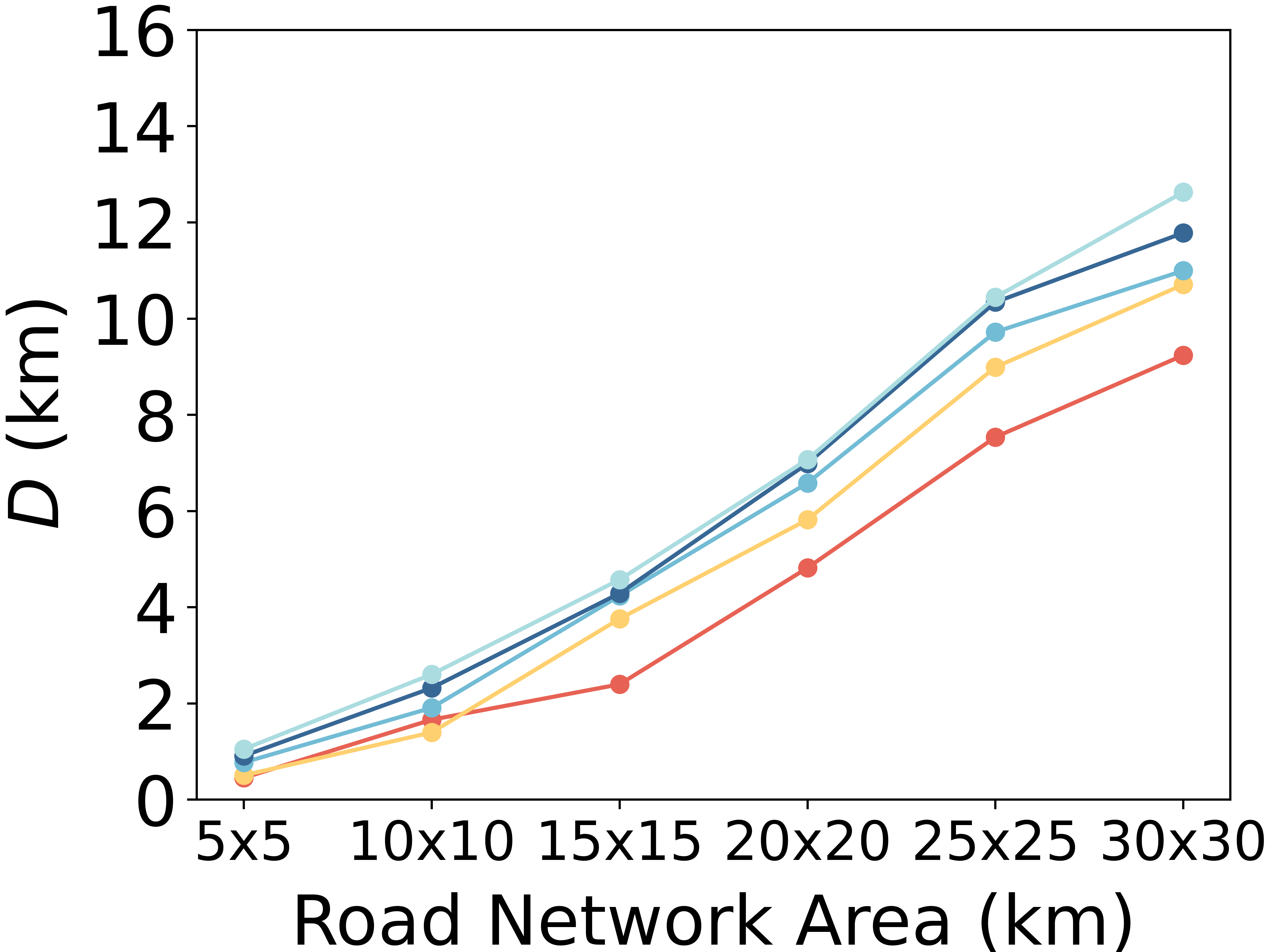}
    \end{minipage}
    \label{fig:distance_suburban_area}
    }
    \subfigure[Attack within urban region grouped by $K$]{
    \begin{minipage}[t]{0.22\textwidth}
    \centering
    \includegraphics[width=1.7in]{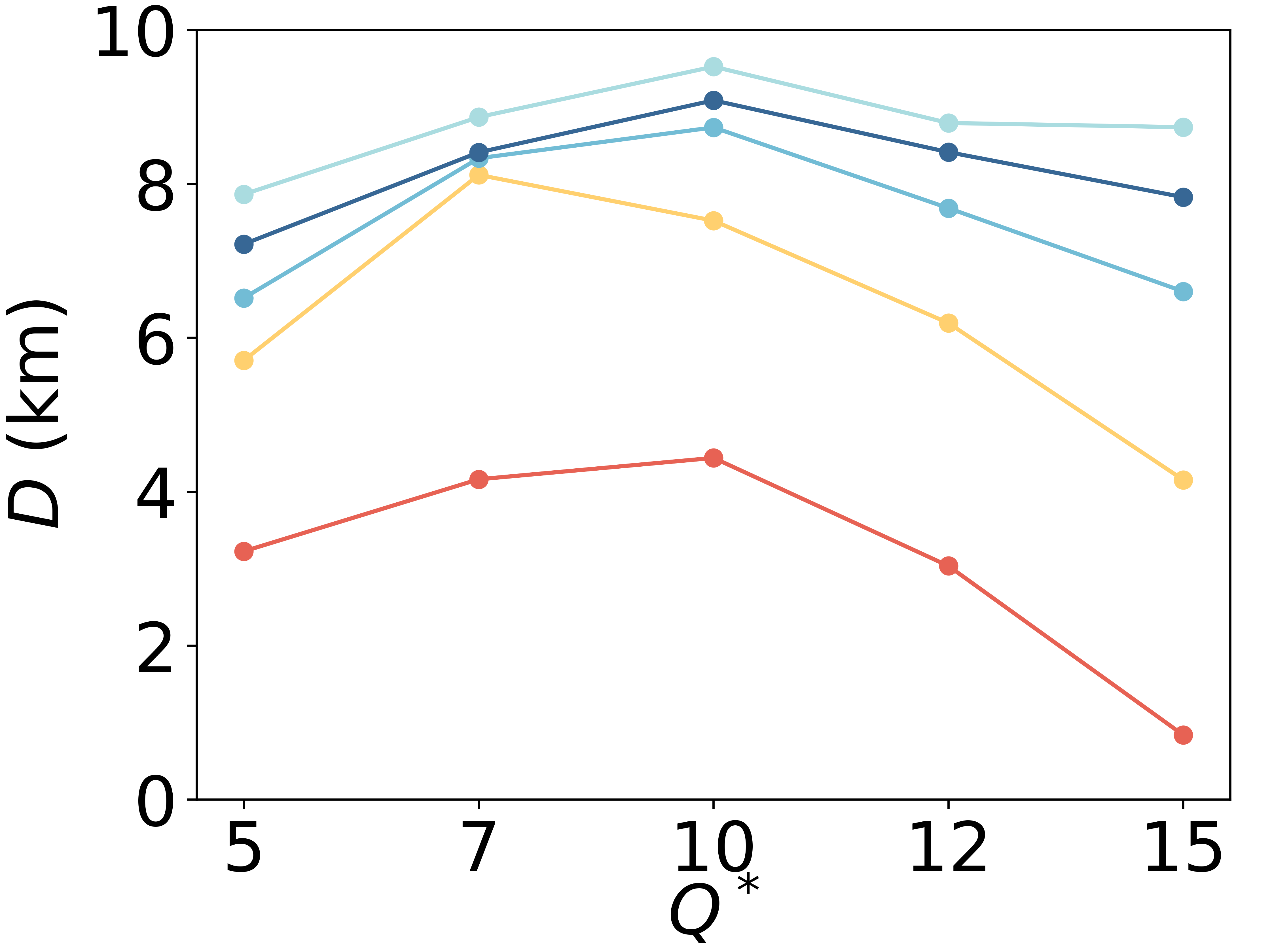}
    \end{minipage}
    \label{fig:distance_urban_node}
    }
    \subfigure[Attack within suburban region grouped by $K$]{
    \begin{minipage}[t]{0.22\textwidth}
    \centering
    \includegraphics[width=1.7in]{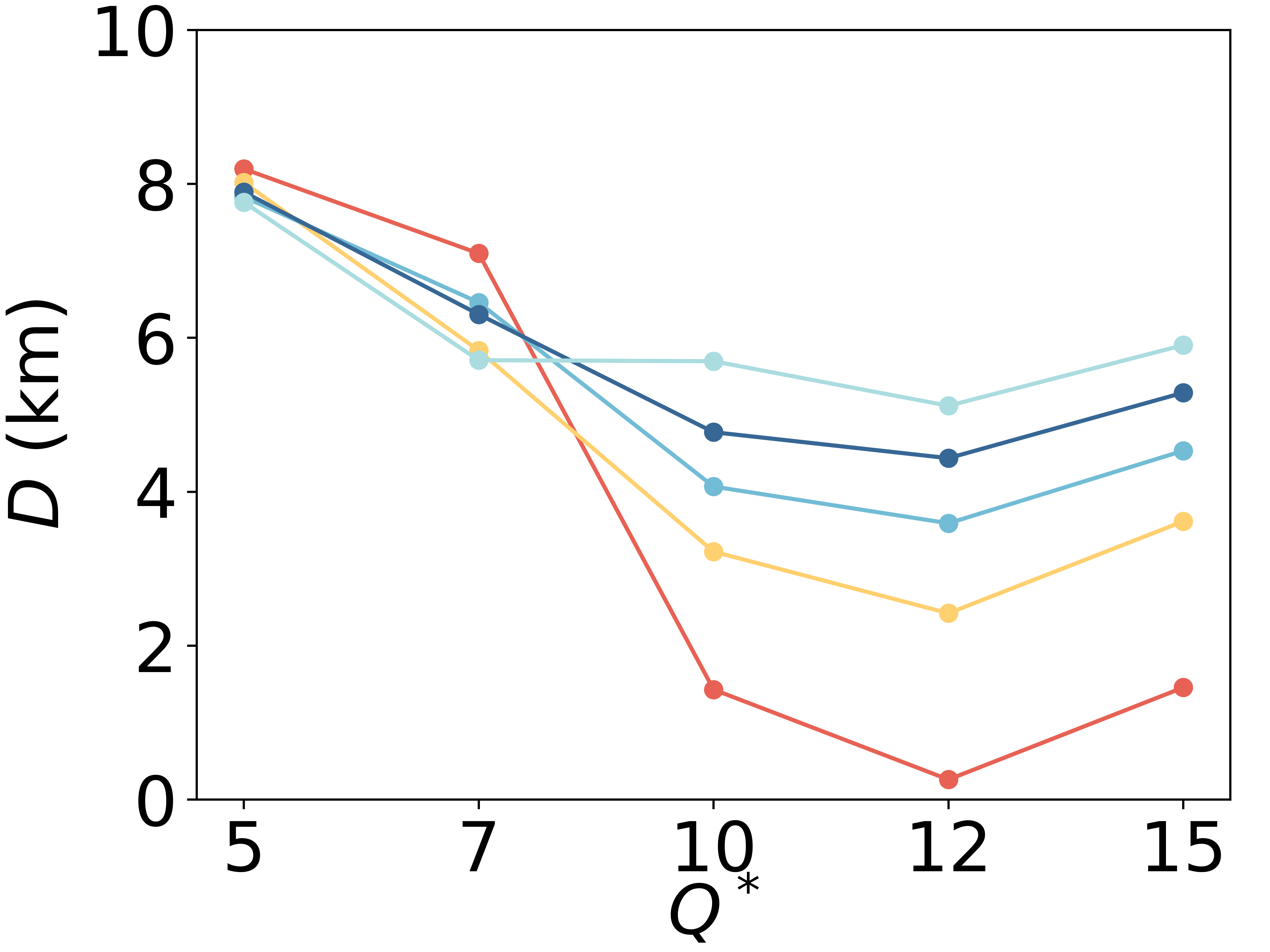}
    \end{minipage}
    \label{fig:distance_suburban_node}
    }
\caption{Comparison of spatial distance offset $D$ within urban and suburban road networks: $y$-axes are the average values of $D$ among different $Q^*$ in (a)(b) and different road network areas in (c)(d).
The bar colors in all subplots represent the values of Top-$K$ as follows: \textcolor{red}{$\bullet$}~$K = 1$, \textcolor{yellow}{$\bullet$}~$K = 3$, \textcolor{blue}{$\bullet$}~$K = 5$, \textcolor{black}{$\bullet$}~$K = 7$, and \textcolor{cyan}{$\bullet$}~$K = 10$.}
\label{fig:Distance}
\end{figure*}

The impact of different $K$ values is indicated in Fig.~\ref{fig:asr} where we compare $\Psi$ of different $K$ and discuss the related outcome under different road network areas or the driving trajectory length $Q^*$.
The $y$-axes in Figs.~\ref{fig:asr_urban_area} and~\ref{fig:asr_suburban_area} are the average values of $\Psi$ among different driving trajectory lengths $Q^*$, while the $y$-axes in Figs.~\ref{fig:asr_urban_node} and~\ref{fig:asr_suburban_node} are the average values of $\Psi$ among different road network areas.
In urban and suburban road networks, $\Psi$ progressively converges as the value of $K$ increases and stabilizes for $K \geq 5$.
A greater value of $K$ comes with a bigger tolerance $\sigma$ to accept the relative error of the edge weight.
Therefore, the correct nodes and edges can be found by enlarging the relative error in the matching process.
As shown in Figs.~\ref{fig:asr_urban_area} and~\ref{fig:asr_suburban_area}, $\Psi$ can reach \qty{90.59}{\%} in urban and \qty{99.41}{\%} in suburban at $K=10$ within the road network area of \qtyproduct{5 x 5}{\km}.
The result indicates that the CAN-Trace attack can deduce the majority of target driving trajectories within the top $10$ subgraph candidates, showing a high attack efficiency.
The CAN-Trace attack has a better attack success rate in the urban region.
As shown in Figs.~\ref{fig:asr_urban_area} and~\ref{fig:asr_suburban_area}, the average values of $\Psi$ are significantly higher in urban road networks, particularly in those with the road network area of \qtyproduct{20 x 20}{\km}, \qtyproduct{25 x 25}{\km} and \qtyproduct{30 x 30}{\km}.
The upward trend is obvious when $K=1$ with a \qtyproduct{30 x 30}{\km} road network area, where $\Psi$ increases from \qtyproduct{36.88}{\%} in suburban to \qtyproduct{50.35}{\%} in urban road networks.

The attack success rate $\Psi$ shows a positive correlation with the $Q^*$, as indicated in Figs.~\ref{fig:asr_urban_node} and~\ref{fig:asr_suburban_node}.
As shown in Figs.~\ref{fig:asr_urban_node} and~\ref{fig:asr_suburban_node}, the proposed attack can apply a small $K$ to achieve good performance on $\Psi$ with a longer driving trajectory length.
Notably, the CAN-Trace attack can gain a higher attack success rate $\Psi$ when $Q^* \geq 10$ in our experiments.
The success rate $\Psi$ in the urban region appears to be generally superior to that in the suburban region.
This could potentially be attributed to the more distinct features of road network patterns in the urban region.
In either urban or suburban regions, $\Psi$ rises to reach around \qtyproduct{90}{\%} with a greater~$Q^*$.

We proceed to compare the attack success rate of the proposed CAN-Trace method with the state-of-the-art trajectory detection methods\footnote{Strict comparisons are challenging due to differences in experimental setups, such as data sources, cities, and trajectories. We have attempted to compare results under similar parameters.}, i.e.,~\cite{li2018location} and~\cite{roth2021daroute}.
In comparison with~\cite{li2018location}, the proposed CAN-Trace achieves the attack success rates of 79.76\% and 50.35\% for \qtyproduct{10 x 10}{\km} and \qtyproduct{30 x 30}{\km} areas, respectively, when $K=1$. By contrast, the corresponding success rates in~\cite{li2018location} for \qty{100}{\km\squared} and \qty{900}{\km\squared} areas are 61.1\% and 33.5\%, respectively.
Compared to~\cite{roth2021daroute}, CAN-Trace achieves similar success rates on \qty{400}{\km\squared} areas. Specifically, CAN-Trace reaches a 78.18\% attack success rate for \qtyproduct{20 x 20}{\km} areas when $K=10$, as shown in Fig.~\ref{fig:asr_urban_area}. This success rate corresponds to a 97.73\% partial match rate (which is obtained from $\frac{78.18\%}{80\%}$ based on the 80\% matching requirement for partial match success rate in~\cite{roth2021daroute}), which is similar to the partial match success rate for the 400 km² area (i.e., Q1 results given in Fig. 8(b) of~\cite{roth2021daroute}).

\begin{figure*}[!htbp]
    \centering
    \subfigure[Attack within urban region grouped by $K$]{
    \begin{minipage}[t]{0.22\textwidth}
    \centering
    \includegraphics[width=1.7in]{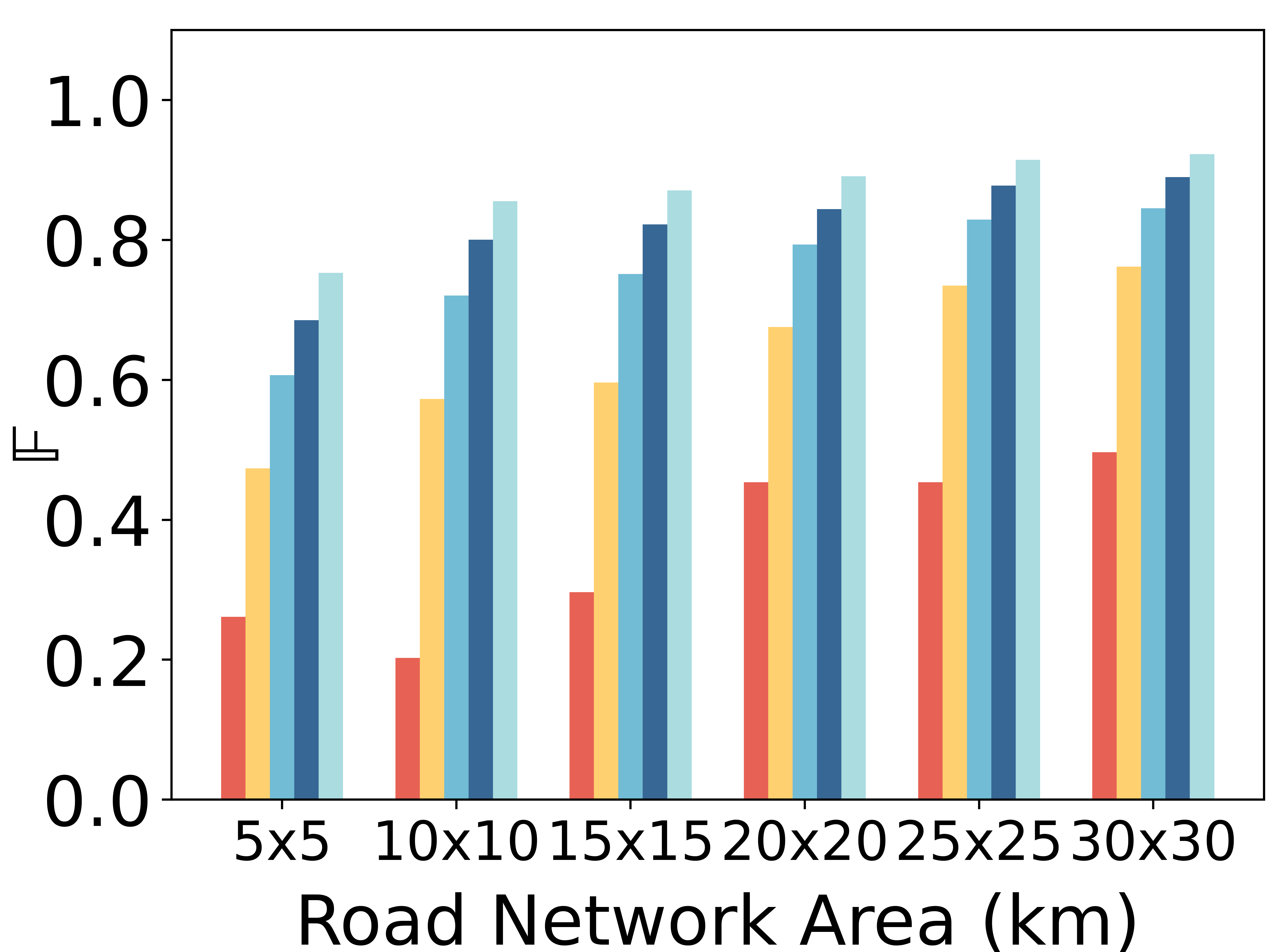}
    \end{minipage}
    \label{fig:precision_urban_area_fp}
    }
    \subfigure[Attack within suburban region grouped by $K$]{
    \begin{minipage}[t]{0.22\textwidth}
    \centering
    \includegraphics[width=1.7in]{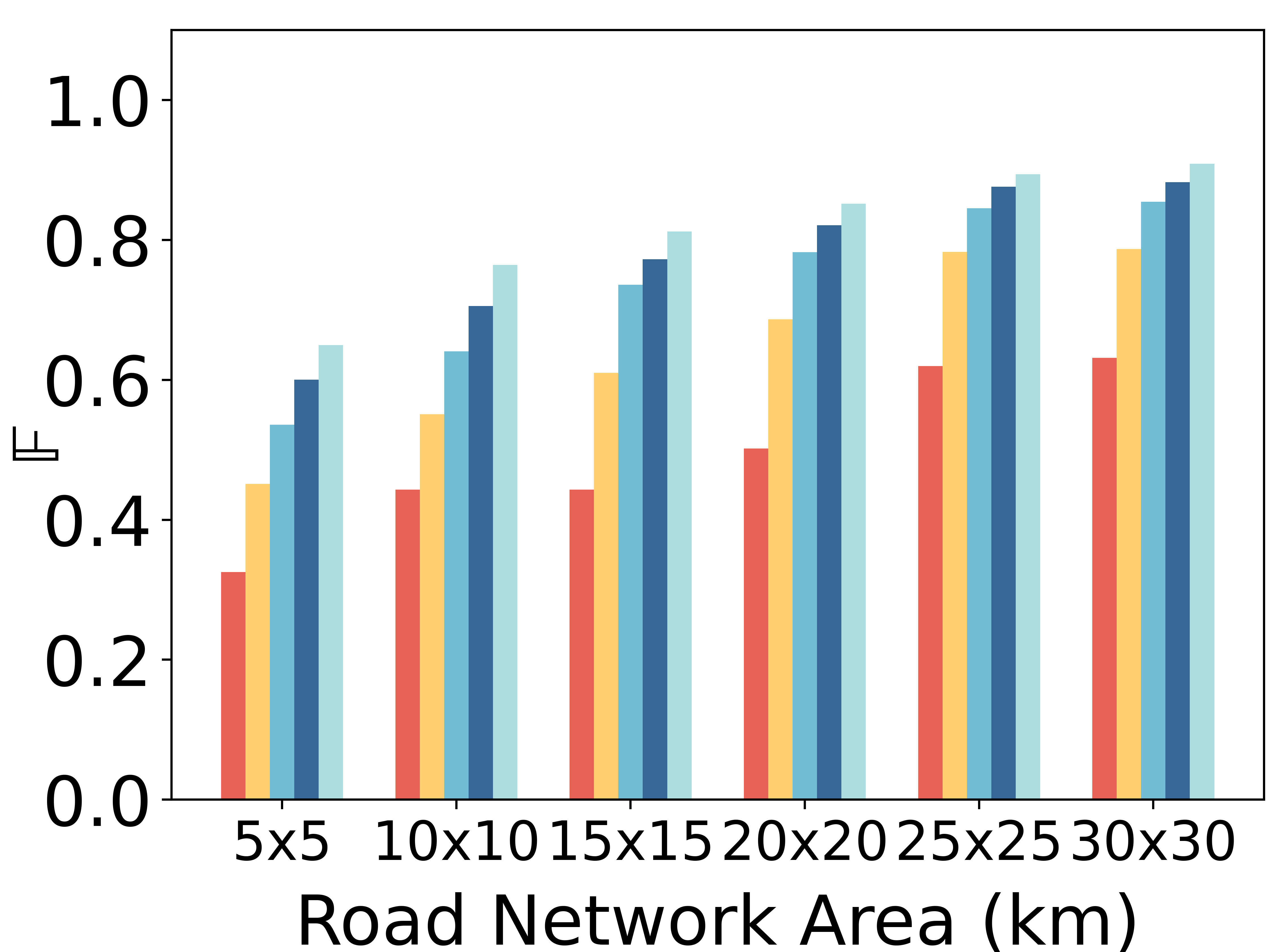}
    \end{minipage}
    \label{fig:precision_suburban_area_fp}
    }
    \subfigure[Attack within urban region grouped by $K$]{
    \begin{minipage}[t]{0.22\textwidth}
    \centering
    \includegraphics[width=1.7in]{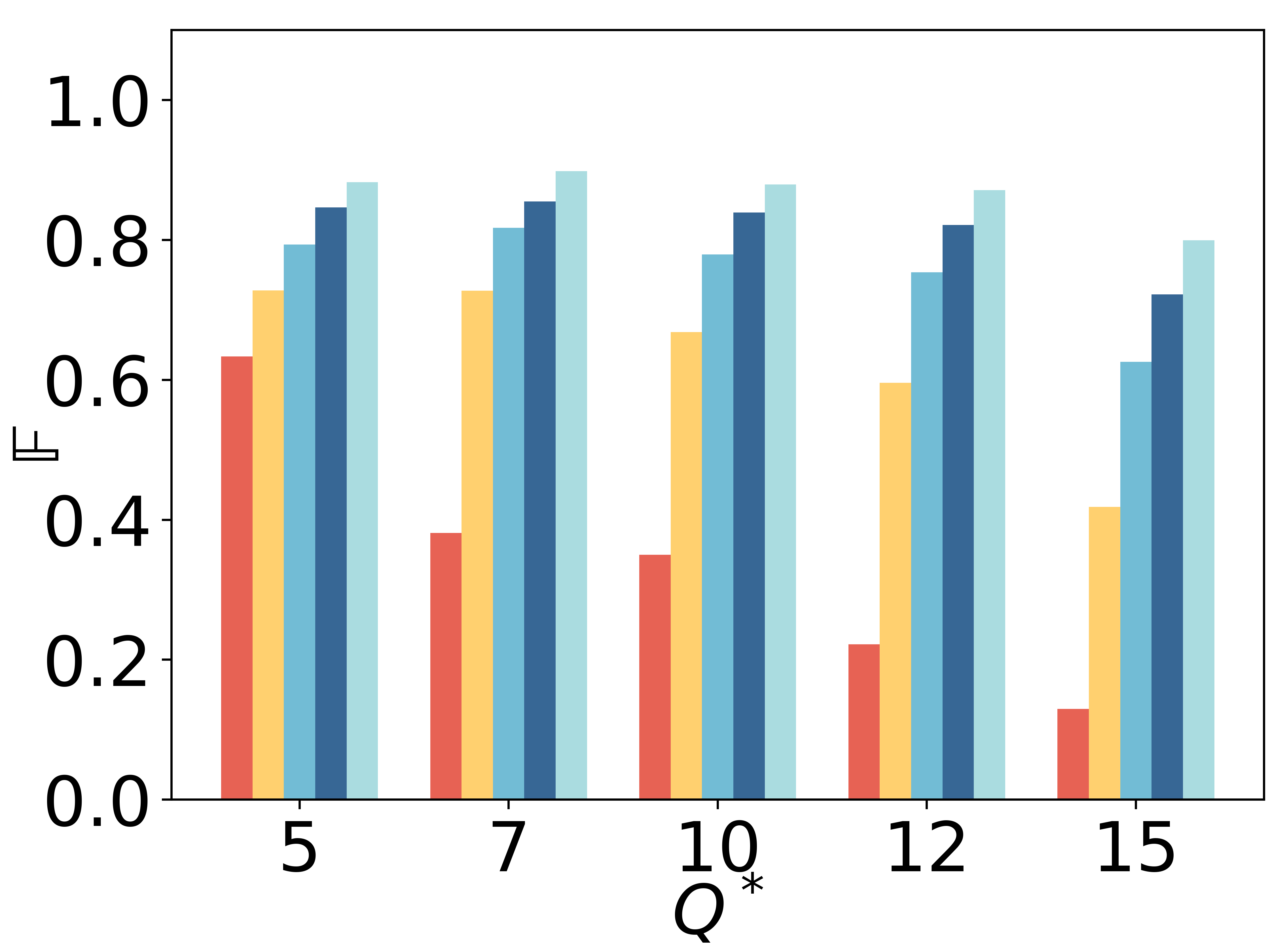}
    \end{minipage}
    \label{fig:precision_urban_node_fp}
    }
    \subfigure[Attack within suburban region grouped by $K$]{
    \begin{minipage}[t]{0.22\textwidth}
    \centering
    \includegraphics[width=1.7in]{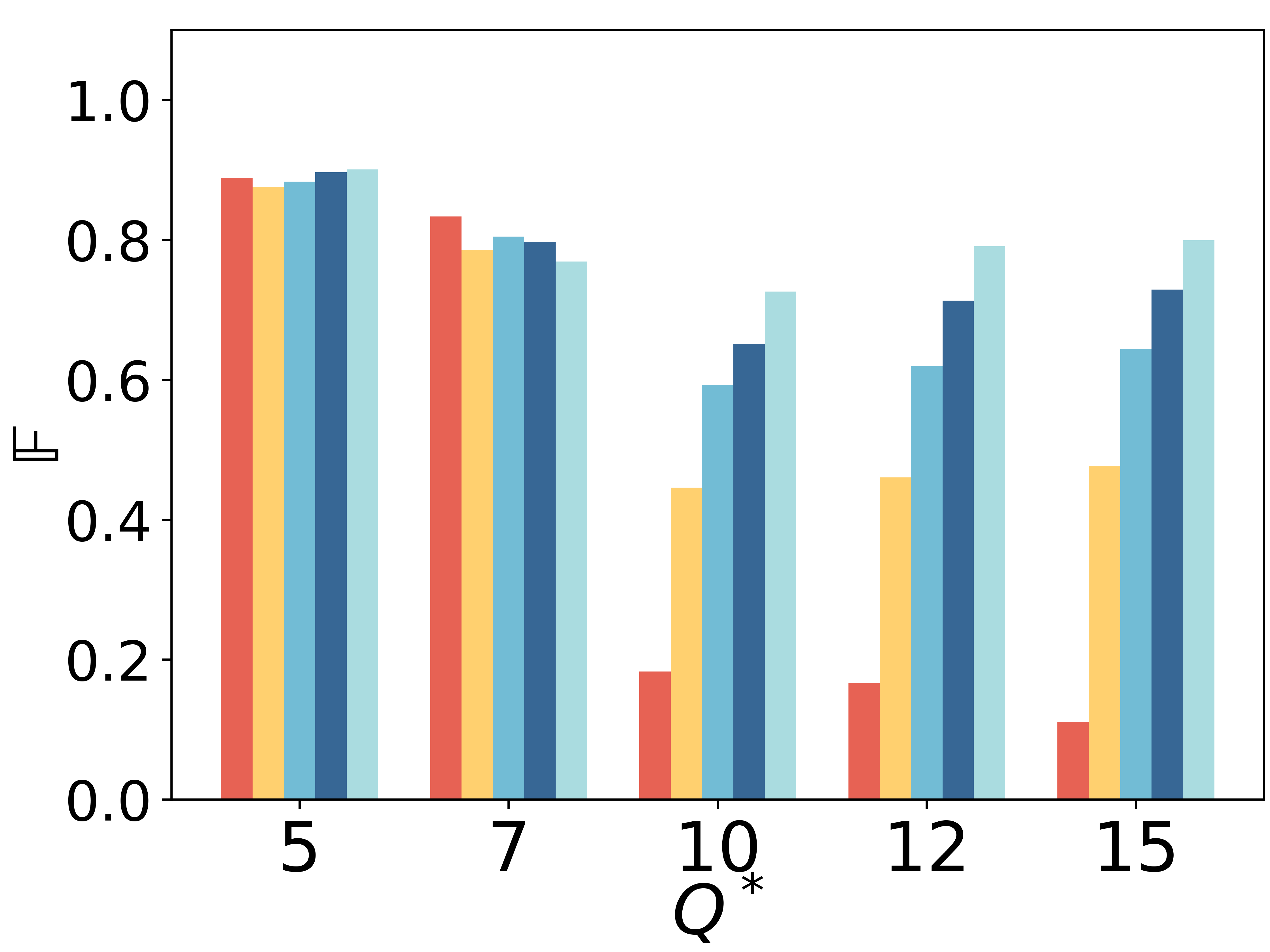}
    \end{minipage}
    \label{fig:precision_suburban_node_fp}
    }
\caption{Comparison of false negative rate $\mathbb{F}$ within urban and suburban road networks: $y$-axes are the average values of $\mathbb{F}$ among different road network areas in (a) (b) and different $Q^*$ in (c) (d). The bar colors in all subplots represent the values of Top-$K$ as follows: \textcolor{red}{\rule{8pt}{4pt}}~$K = 1$, \textcolor{yellow}{\rule{8pt}{4pt}}~$K = 3$, \textcolor{blue}{\rule{8pt}{4pt}}~$K = 5$, \textcolor{black}{\rule{8pt}{4pt}}~$K = 7$, and \textcolor{cyan}{\rule{8pt}{4pt}}~$K = 10$.}
\label{fig:FPFN}
\end{figure*}

\begin{figure*}[!htbp]
    \centering
    \subfigure[$\Psi$ within remote region grouped by $K$]{
    \begin{minipage}[t]{0.22\textwidth}
    \centering
    \includegraphics[width=1.7in]{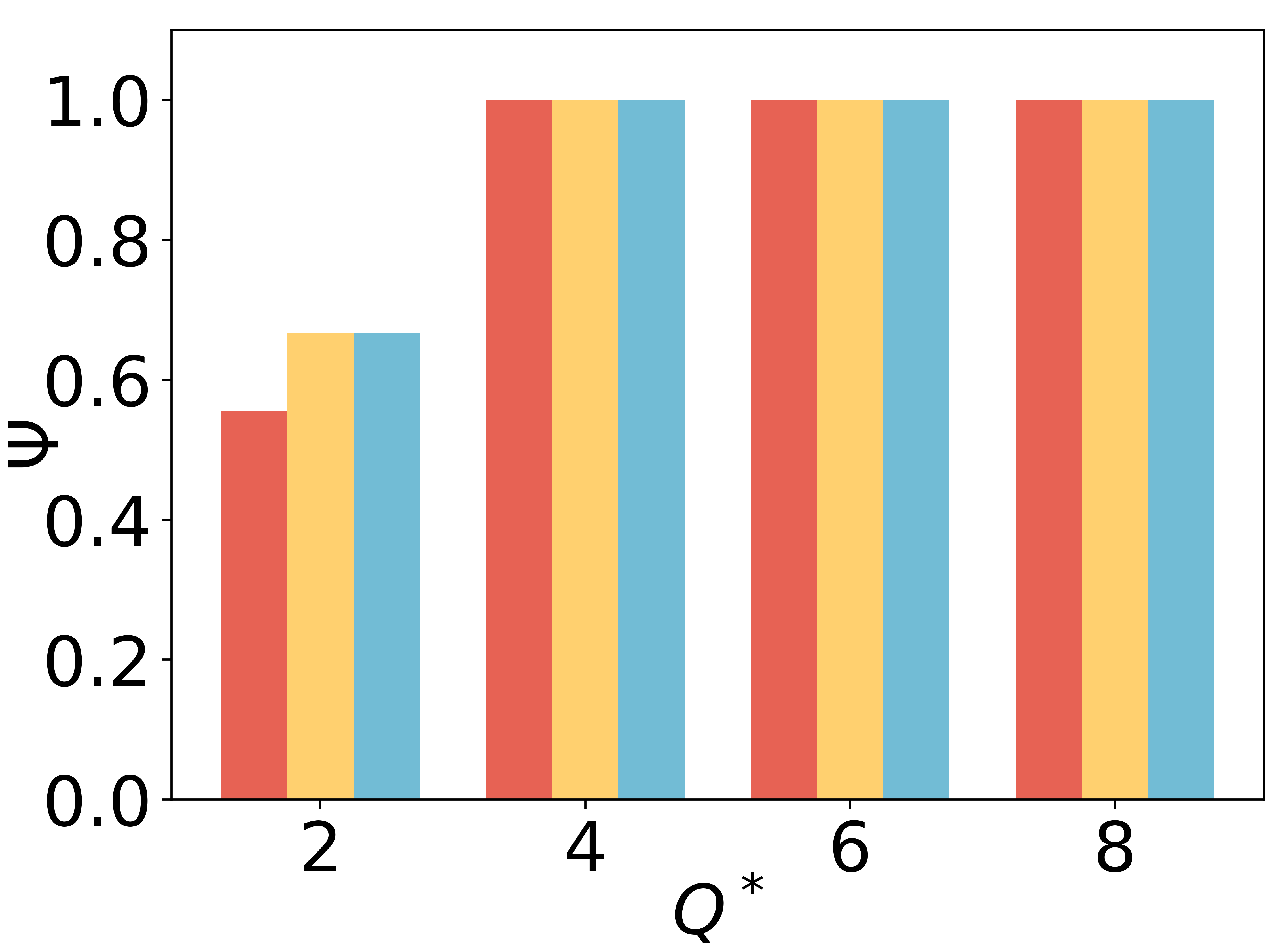}
    \end{minipage}
    \label{fig:remote_area_ask}
    }
    \subfigure[$P$ within remote region grouped by $K$]{
    \begin{minipage}[t]{0.22\textwidth}
    \centering
    \includegraphics[width=1.7in]{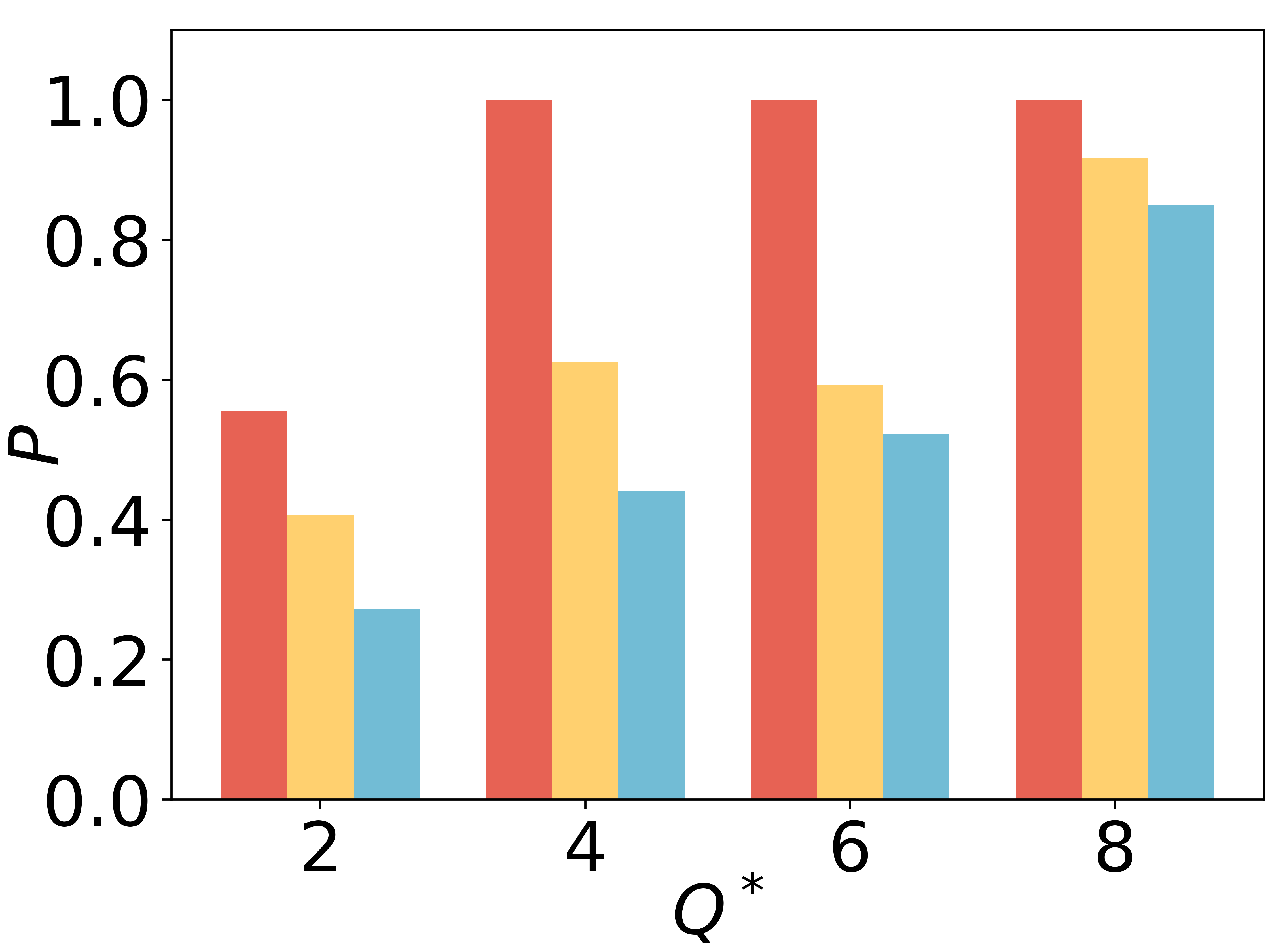}
    \end{minipage}
    \label{fig:remote_area_precision}
    }
    \subfigure[$D$ within remote region grouped by $K$]{
    \begin{minipage}[t]{0.22\textwidth}
    \centering
    \includegraphics[width=1.7in]{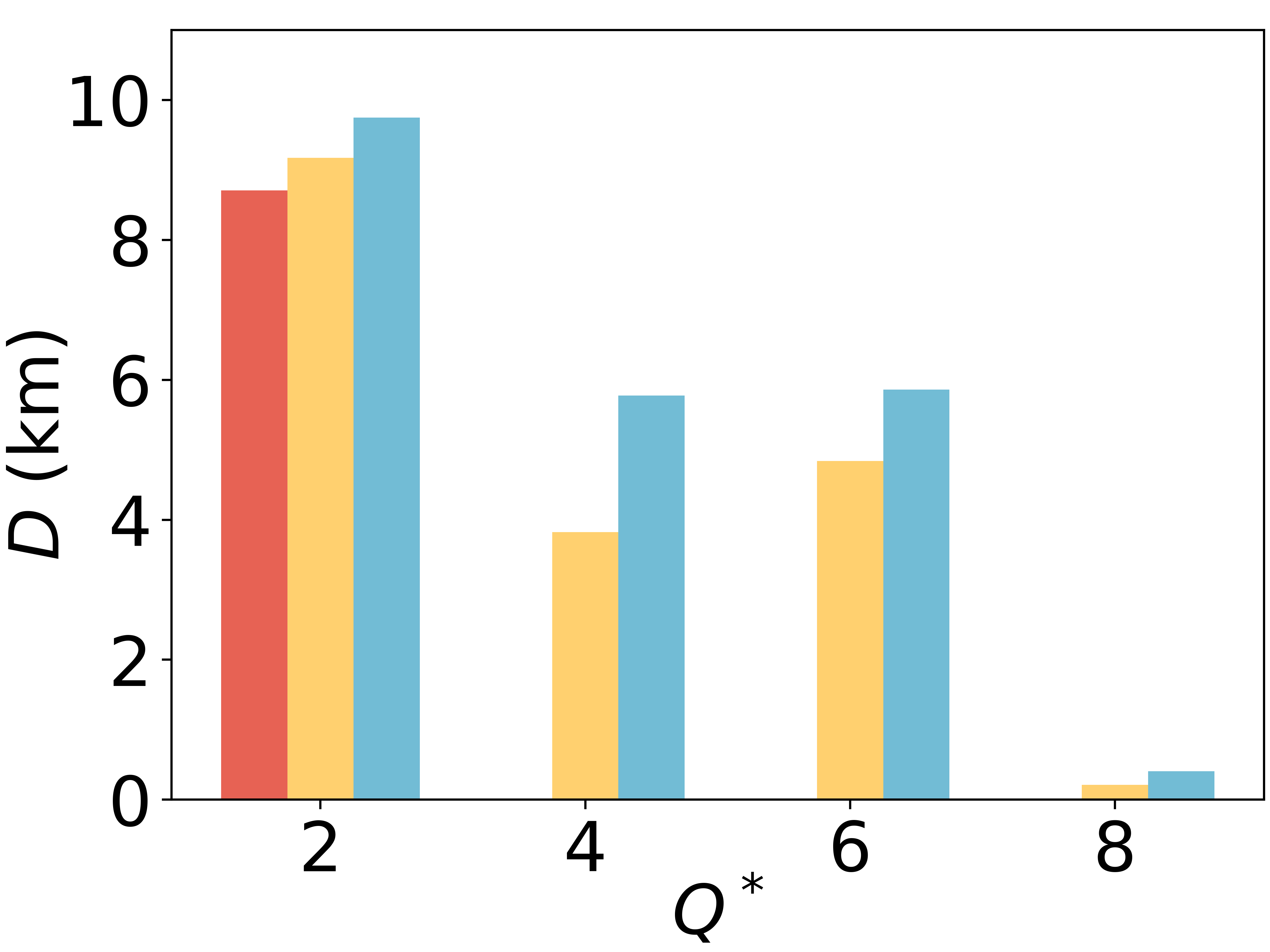}
    \end{minipage}
    \label{fig:remote_area_distance}
    }
    \subfigure[$\mathbb{F}$ within remote region grouped by $K$]{
    \begin{minipage}[t]{0.22\textwidth}
    \centering
    \includegraphics[width=1.7in]{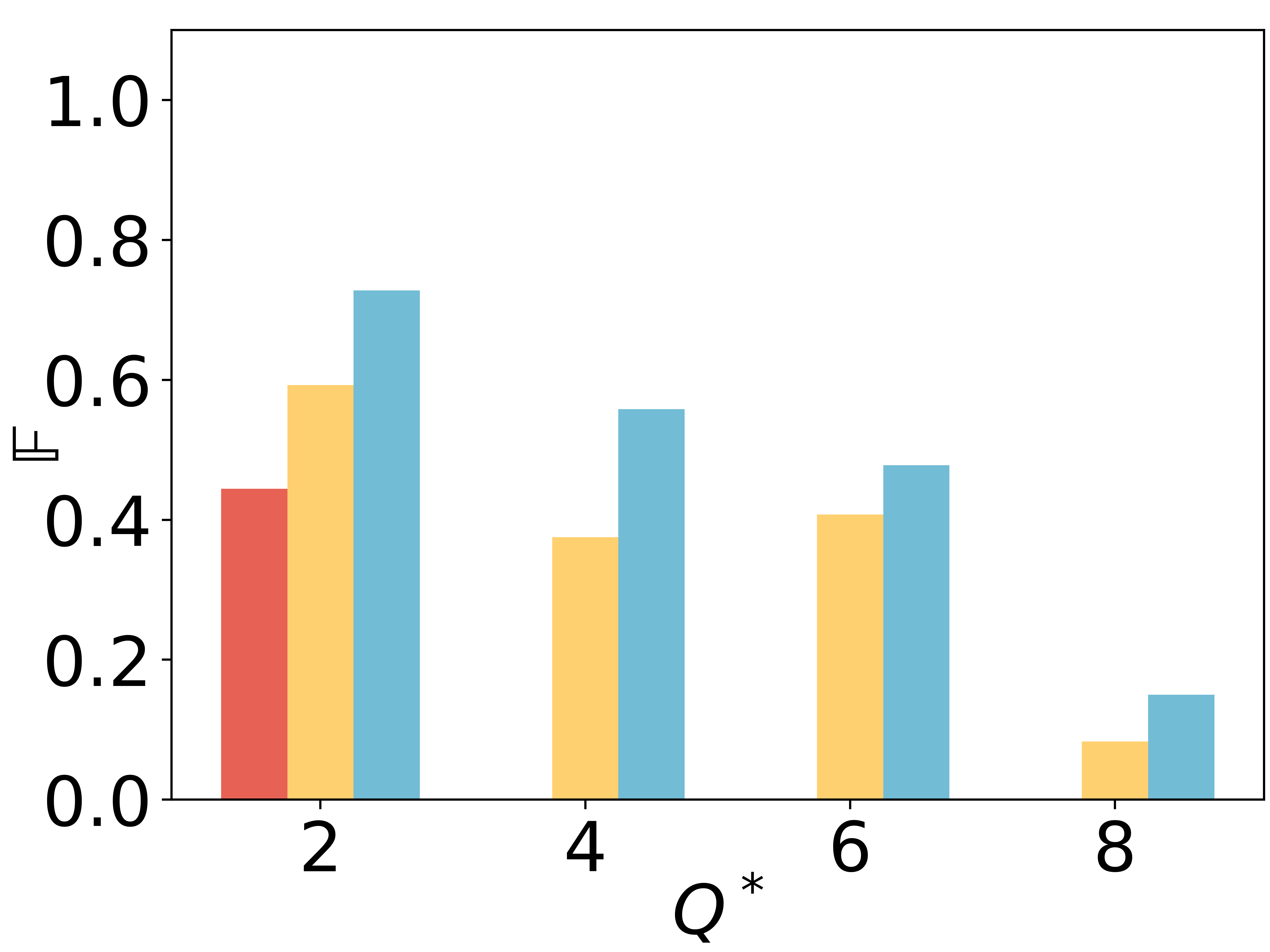}
    \end{minipage}
    \label{fig:remote_area_fnfp}
    }
\caption{Evaluation in remote regions: $y$-axes are the average values of attack success rate $\Psi$ in (a), the average values of the attack precision $P$ in (b), the average values of spatial distance offset $D$ in (c), and the average values of false negative $\mathbb{F}$ in (d). The $x$-axes are the numbers of nodes in the graph of actual driving trajectories, i.e., $Q^*$. The road network area is 30 km$\times$ 30 km. The bar colors in all subplots represent the values of Top-$K$ as follows: \textcolor{red}{\rule{8pt}{4pt}}~$K = 1$, \textcolor{yellow}{\rule{8pt}{4pt}}~$K = 3$, and \textcolor{blue}{\rule{8pt}{4pt}}~$K = 5$.}
\label{fig:remoteArea}
\end{figure*}

\smallskip
\subsubsection{\textbf{Attack Precision}}
The experiments of attack precision $P$ are shown in Figs.~\ref{fig:precision} and~\ref{fig:precision_k13}, where $P$ is discussed under different road network areas, $Q^*$ and $K$.
The $y$-axes in Figs.~\ref{fig:precision_urban_area} and~\ref{fig:precision_suburban_area} are the average values of $P$ among different driving trajectory lengths $Q^*$, while the $y$-axes in Figs.~\ref{fig:precision_urban_node} and~\ref{fig:precision_suburban_node} are the average values of $P$ among different road network areas.
Unlike the attack success rate $\Psi$ that increases with $K$, the CAN-Trace attack has the higher values of $P$ with a smaller $K$ as shown in Figs.~\ref{fig:precision} and~\ref{fig:precision_k13}.
As evidenced in Fig.~\ref{fig:precision}, $P$ exhibits superior values for $K=1$ than for $K \in \{3, 5, 7, 10\}$.
This is because the first ranking inferred subgraph candidate is extreme and has either a large or zero number of correctly deduced nodes.
As demonstrated in Figs.~\ref{fig:precision_urban_area} and~\ref{fig:precision_suburban_area}, $P$ attains a peak value of \qtyproduct{79.75}{\%} in urban when $K=1$ within a \qtyproduct{10 x 10}{\km} area and a peak value of \qtyproduct{67.45}{\%} in suburban $K=1$ within a \qtyproduct{5 x 5}{\km} road network area.
This also proves that the design of the Top-$K$ ranking rule leads to a better attack result and enhances the attack efficiency.

The attack performance is enhanced when the attack performs on a smaller road network size as indicated in Figs.~\ref{fig:precision_urban_area},~\ref{fig:precision_suburban_area},~\ref{fig:precision_urban_area_k1} and~\ref{fig:precision_suburban_area_k1}.
The observed increase of $P$ can be found at the \qtyproduct{5 x 5}{\km} road network area compared to the \qtyproduct{10 x 10}{\km} in urban road network, as indicated in Fig.~\ref{fig:precision_urban_area}.
As shown in Figs.~\ref{fig:precision_urban_area} and~\ref{fig:precision_suburban_area}, the attack precision $P$ stabilizes after surpassing a road network area value of \qtyproduct{20 x 20}{\km} in the urban region and that of \qtyproduct{25 x 25}{\km} in the suburban region, respectively.
Consistent with the trend of $\Psi$ in Figs.~\ref{fig:asr_urban_node} and~\ref{fig:asr_suburban_node}, the attack precision $P$ increases with a longer driving trajectory $Q^*$ as shown in Figs.~\ref{fig:precision_urban_node} and~\ref{fig:precision_suburban_node}.
$P$ increases from \qtyproduct{36.67}{\%} to \qtyproduct{87.04}{\%} in urban region and raises from \qtyproduct{11.11}{\%} to \qtyproduct{88.89}{\%} in suburban region when $Q^*$ grows from \qtyproduct{5}{} to \qtyproduct{15}{} with $K=1$.
The attack precision $P$ has a significant increase when $Q^* \geq 10$, especially in suburban road networks, as demonstrated in Fig.~\ref{fig:precision_suburban_node}.

As shown in Fig.~\ref{fig:precision_k13}, the overall attack performance of $P$ when $K=1$ is better than that of $K=3$.
Compared the Figs.~\ref{fig:precision_urban_area_k1} and~\ref{fig:precision_urban_area_k3}, the average value of $P$ is smaller when $K=3$ than $K=1$ in urban road networks.
The same trend can be found in Figs.~\ref{fig:precision_suburban_area_k1} and~\ref{fig:precision_suburban_area_k3}.
The reason is that the size of $\hat{R}$ is bigger with a larger $K$ value which decreases the average value of $P$.
However, a larger $K$ can be used to avoid extreme cases as shown in Figs.~\ref{fig:precision_suburban_area_k1} and~\ref{fig:precision_suburban_area_k3} where the CAN-Trace attack has $P=0$ when $K=1$ but $P>0$ when $K=3$ for the driving trajectory length $Q^*=7$ in the road network areas of \qtyproduct{10 x 10}{\km} and \qtyproduct{15 x 15}{\km}.
In terms of the $Q^*$, the attack precision tends to increase while the $Q^*$ grows.
As indicated in Fig.~\ref{fig:precision_suburban_area_k1}, CAN-Track may have no output in some extreme cases, i.e., when $K=1$ and $Q^* =7$ for the road network area larger than \qtyproduct{5 x 5}{\km} in the suburban region.
The same case happens when $K=3$ and $Q^* =7$ for the areas larger than \qtyproduct{15 x 15}{\km}, as shown in Fig.~\ref{fig:precision_suburban_area_k3}.
This indicates that the extreme cases can be avoided by enlarging the length of the driving trajectory, i.e., a greater $Q^*$.

\smallskip
\subsubsection{\textbf{Spatial Distance Offset}}

The spatial distance offset is compared under different road network areas and $Q^*$ in Fig~\ref{fig:Distance}.
The attack results filtered by different $K$ are compared among different settings.
The $y$-axes in Figs.~\ref{fig:distance_urban_area} and~\ref{fig:distance_suburban_area} are the average values of $D$ among different $Q^*$, and the $y$-axes in Figs.~\ref{fig:distance_urban_node} and~\ref{fig:distance_suburban_node} are the average values of $D$ among different road network areas.
As shown in Fig.~\ref{fig:Distance}, the CAN-Trace attack wins fewer distances offset $D$ in suburban rather than urban road network areas.
This is because the suburban region can have a similar road network pattern across the area and far away from each other.
The average spatial distance offset $D$ ranges from \qtyproduct{0.48}{\km} to \qtyproduct{1.30}{\km} in \qtyproduct{5 x 5}{\km} and from \qtyproduct{7.10}{\km} to \qtyproduct{15.68}{\km} in \qtyproduct{30 x 30}{\km} within the urban road network.
In the suburban region, $D$ varies from \qtyproduct{0.45}{\km} to \qtyproduct{1.05}{\km} and from \qtyproduct{9.24}{\km} to \qtyproduct{12.63}{\km} in the areas size of \qtyproduct{5 x 5}{\km} and \qtyproduct{30 x 30}{\km}, respectively.
As indicated in Figs.~\ref{fig:distance_urban_area} and~\ref{fig:distance_suburban_area}, the spatial distance offset $D$ has a robust positive correlation with the size of the road network. 
There is a notable decrease when the road network area becomes smaller, which means that CAN-Trace attack tends to locate the deduced driving trajectory close to the actual driving trajectory.
The observed trend can also be attributed to the fact that a smaller road network area yields a smaller number of subgraph candidates.
As demonstrated in Figs.~\ref{fig:distance_urban_node} and~\ref{fig:distance_suburban_node}, $D$ decreases with a bigger $Q^*$.
Similar to the attack precision $P$, the spatial distance offset $D$ gains better attack results when $Q^* \geq 10$, especially in the suburban road networks indicated in Fig.~\ref{fig:distance_suburban_node}.

Like the attack precision $P$, the overall attack performance of $D$ is enhanced with a smaller $K$.
Notably, $D$ for $K=1$ is about half that for $K=3$ as demonstrated in Fig.~\ref{fig:Distance}.
The best performance of $D$ comes with the lowest value \qtyproduct{0.8394}{\km} with $K=1$ and is \qtyproduct{4.15}{\km} when $K=3$ in urban road networks, as shown in Fig.~\ref{fig:distance_urban_node}.
In the suburban road network, $D$ drops to \qtyproduct{0.26}{\km} when $K=1$ but is \textcolor{black}{greater} than \qtyproduct{2.42}{\km} when $K \in \{3, 5, 7, 10\}$ as demonstrated in Fig.~\ref{fig:distance_suburban_node}.
The difference of $D$ between $K=1$ and $K \in \{3, 5, 7, 10\}$ is extremely large, as shown in Figs.~\ref{fig:distance_urban_node} and~\ref{fig:distance_suburban_node}.
The reason is that a larger $K$ value brings in candidates with different node matches, and some are less aligned.

\subsubsection{\textbf{False Negative Rate}}
The attack false negative rate $\mathbb{F}$ is examined in Fig.~\ref{fig:FPFN} across different road network areas, the number of actual driving trajectories, and $K$ for Top-$K$, in urban and suburban regions. The false negative rate can be high, especially when $K$ is large, e.g., $K=7$. This is because the Top-$K$ selection mechanism includes multiple trajectories even though only one is correct, increasing the numbers of incorrect trajectories and nodes selected. For example, in the case of $Q^*=15$, the false negative rate $\mathbb{F}$ is 11.11\% when $K=1$ in the suburban region, as shown in Fig.~\ref{fig:precision_suburban_node_fp}, indicating that CAN-Trace has identified the majority of nodes in the actual trajectory. However, as $K$ increases, incorrect trajectories are included in the Top-$K$ selection, leading to a higher $\mathbb{F}$, e.g., $\mathbb{F}=79.96\%$ for $K=10$, as the second through tenth results barely cover any right nodes.

The false negative rate $\mathbb{F}$ gradually increases as the road network size expands. For example, when $K=1$, $\mathbb{F}$ is 32.55\% for the 5~km $\times$ 5~km road networks in suburban regions and 53.14\% for the 30~km $\times$ 30~km road networks in suburban regions, as shown in  Fig.~\ref{fig:precision_suburban_area_fp}.
The false negative rate~$\mathbb{F}$ decreases as the number of nodes in the actual driving trajectories increases. According to Fig.~\ref{fig:precision_urban_node_fp}, when $K=1$, $\mathbb{F}$ is 63.33\% on average for all trajectory graphs with five nodes and drops to only 12.96\% for all trajectory graphs with 15 nodes. The experiment results reveal that CAN-Trace requires a certain number of nodes to accurately identify the actual trajectories, and $K$ should be tuned to balance the attack success rate and the false negative rate.

\subsubsection{\textbf{Performance in Remote Regions}}
The proposed CAN-Trace attack is also validated in remote regions. The attack performance metrics, including success rate ($\Psi$), precision ($P$), spatial distance offset ($D$), and false negative rate ($\mathbb{F}$), are illustrated in Fig.~\ref{fig:remoteArea} Testing results confirm the effectiveness of the CAN-Trace attack in remote regions. Similar to the results in urban and suburban regions, the attack performance improves with increasing $Q^*$, leading to higher $\Psi$ and $P$ and lower $D$ and $\mathbb{F}$. Notably, the results in the remote regions demonstrate better performance of the proposed attack compared to urban and suburban regions, with the attack success rate ($\Psi$) reaching the upper bound with a small $K$ and few nodes in the driving trajectory graph. This is due to the strong heterogeneity of road segments in remote regions, which allows for exclusive matching.

\subsection{Discussion}
\subsubsection{Lesson Learned}
In summary, the proposed CAN-Trace attack has a good performance on the attack success rate $\Psi$, attack precision $P$, and spatial distance offset $D$ in the real-world environment.
In our experiments, the CAN-Trace attack infers the vehicle driving trajectory with a higher value of $\Psi$ and $P$ within a smaller road network area, which means the attack is quite efficient in a small size of the road network area.
CAN-Trace attack performance in the suburban region is slightly superior to those in the urban region, especially when the driving trajectory length $Q^*$ is greater than $10$.
The experiments highlight the impact of the Top-$K$ value, which wins the highest attack precision $P$ when $K=1$ and gains a good attack success rate $\Psi$ representing the coverage when $K$ is greater than $5$.
Thus, a $K$ value between $1$ and $5$ is recommended to balance the attack efficiency and coverage.

\subsubsection{Privacy Concern}
The proposed CAN-Trace attack can persist because it utilizes basic vehicle motion data, i.e., speed and pedal data, from the standard OBD-II protocol that is widely used in the automotive industry. The attack could compromise drivers' privacy on personal addresses, locations, and driving trajectories by analyzing CAN messages alone. In our experiments, all driving data are collected with consent. In practice, the application of the proposed CAN-Trace should follow privacy regulations such as the General Data Protection Regulation (GDPR) in Europe and the California Consumer Privacy Act (CCPA). Organizations like car rental or logistics companies could leverage CAN-Trace to track vehicles, provided there is full disclosure and compliance with these regulations.

\subsubsection{Mitigation}
To safeguard against the proposed CAN-Trace attack, we recommend actions from car manufacturers, drivers, and CAN-based service providers. Car manufacturers should design secure, privacy-preserving CAN networks with secure access controls. Drivers should regularly inspect the OBD-II port and other potential CAN sniffers to prevent unauthorized physical access. CAN-based service providers should recognize the privacy risks of disclosing CAN messages and implement encryption, data anonymization (e.g., k-anonymity), and privacy-preserving technologies (e.g., differential privacy) during the storage, processing, and sharing of CAN data. In summary, all stakeholders must recognize the privacy risks of trajectory leakage from CAN messages, ensure that only trusted parties have access, and apply anonymity, data privacy, and other privacy-preserving technologies to protect CAN data.

\begin{table*}[!htbp]
\renewcommand{\arraystretch}{1.2}
  \centering
  \caption{Comparison with trajectory detection studies}
\begin{tabular}{lllll}
\hline
\textbf{Reference}             & \textbf{Data Sources}                                                               & \textbf{Approaches}                                                  & \textbf{Advantages}                                                                                      & \textbf{Limitations}                                                                                      \\ \hline
L2MM~\cite{jiang2023l2mm}      & GPS data                                                                            & Deep Learning                                                        & Handel poor-quality GPS data                                                                             & \begin{tabular}[c]{@{}l@{}}GPS coverage and access permission\\ Data quality still matters\end{tabular}   \\ \hline
GOI~\cite{xiao2018goi}         & \begin{tabular}[c]{@{}l@{}}GPS data\\ Vehicle motion data\end{tabular}              & SVR                                                                  & Data fusion for robustness                                                                               & \begin{tabular}[c]{@{}l@{}}GPS coverage and access permission\\ Specific vehicle motion data\end{tabular} \\ \hline
DMM~\cite{shen2024dmm}         & \begin{tabular}[c]{@{}l@{}}Cell tower location data\\ Cell access data\end{tabular} & RNN                                                                  &  \begin{tabular}[c]{@{}l@{}}GPS-agnostic\\ Passive interference by operators \end{tabular}                                                                         & \begin{tabular}[c]{@{}l@{}}Cellular network coverage\\ Cell towers locations\end{tabular}                 \\ \hline
MBT~\cite{li2018location}      & Magnetometer data                                                                   & Turn angle matching                                                  & GPS-agnostic                                                                                             & \begin{tabular}[c]{@{}l@{}}Need external fixed IMU\\ High cost to calculate turn angles\end{tabular}      \\ \hline
DaRoute~\cite{roth2021daroute} & IMU data                                                                            & Route ranking                                                        & \begin{tabular}[c]{@{}l@{}}GPS-agnostic\\ No wiring to vehicles\end{tabular}                             & \begin{tabular}[c]{@{}l@{}}Need external fixed IMU\\ Limited IMU data accuracy\end{tabular}               \\ \hline
Invasion~\cite{kim2023invasion}      & Mobile motion data                                                                  & \begin{tabular}[c]{@{}l@{}}Noise removing\\ DTW matching\end{tabular} & \begin{tabular}[c]{@{}l@{}}GPS-agnostic\\ Tolerate human-generate noise\end{tabular}                     & \begin{tabular}[c]{@{}l@{}}Need external mobile/IMU\\ Restrict to subway trajectories\end{tabular}        \\ \hline
\textbf{CAN-Trace}             & \begin{tabular}[c]{@{}l@{}}CAN data \\ (speed and pedal)\end{tabular}                                                                            & Subgraph matching                                                    & \begin{tabular}[c]{@{}l@{}}GPS-agnostic\\ No attached motion sensors\\ Generic vehicle motion data\end{tabular} & Need access to CAN messages                                                                               \\ \hline
\end{tabular}
 \begin{tablenotes}
        \scriptsize
        \item 
        GPS, SVR, RNN, IMU, DTW and CAN stand for Global Positioning System, Support Vector Regression, Recurrent Neural Network, Inertial Measurement Unit, \\Dynamic Time Warping, and Controller Area Network, respectively.
    \end{tablenotes}
\end{table*}

\section{Related Work}\label{related}
\subsection{Map-Matching: Methods and Sensor Utilization}
The existing map-matching methods~\cite{quddus2007current} that detect vehicle location and trajectory are discussed in this section, focusing on the comparison of data sources. Map-matching is to find the road segments on which the vehicle drives and the location of the vehicles to finally map out the driving trajectory. Since the driving trajectory contains personal information~\cite{xu2022preserving}, sensitive information such as the driver's habits and identity can be inferred from the trajectory data~\cite{ma2021trajectory, xin2017privacy}, which leads to the driver's privacy leakage. To infer the locations and trajectories of the vehicle, adversaries can launch side-channel attacks with the Global Positioning System (GPS) data~\cite{iqbal2010privacy, guha2012autowitness}, passive sensor data~\cite{ho2015pressure, rouf2010security, han2012accomplice, li2018location, chen2019trajcompressor}, and the vehicular network data~\cite{raya2007securing, othmane2015survey, bloessl2015scrambler}.

In~\cite{han2012accomplice}, the vehicle motion data from mobile magnetometer sensors is introduced as additional data in map-matching. In~\cite{li2018location}, Li et al. revealed the driving trajectory by matching intersection angles with the magnetometer sensor data. Li et al. developed the first attack model~\cite{li2018location} to match car turn angles with the road network intersection angles to construct the driving trajectory. The GPS data is not required in the developed attack model, but the compass data from the mobile phone. Unfortunately, the attack model degrades in the case of disorientation due to the perturbed phone position by drivers or passengers.

\subsection{GPS Data Utilization}
The most used data to deduce the vehicle trajectories is the GPS data. In~\cite{iqbal2010privacy}, the authors utilized a passive GPS device to collect GPS data and vehicle state by inferring the victims' home and working addresses from a large number of guesses. The map-matching process using GPS data is to convert a sequence of GPS data into a sequence of road segments~\cite{cao2009gps}. However, the GPS data is not reliable due to the difficult access and data noise and loss. The missing data caused by the GPS outage problem requires additional data such as the odometer, Light Detection And Ranging (LiDAR) or camera data~\cite{roncella2005photogrammetric, liu2021vehicle, georgy2011enhanced, sasani2016improving, chen2019trajcompressor,aftatah2016fusion, huang2019visual,  zhao2021super,liu2022deep} to enhance the inference accuracy.

\subsection{Innovative Approaches to Trajectory Reconstruction}
Unlike identifying only a few road points, Guhu et al. analyzed the sequence of vehicle motions (i.e., movement, stop, turn) collected by a small wireless tag to reconstruct the complete driving path with knowing the initial and final GPS position~\cite{guha2012autowitness}. Xiao et al. classified the road section types with the motion information from the OBD data and integrated the road section types with GPS data to reconstruct the driving trajectory~\cite{chen2019trajcompressor}. The Gated Recurrent Unit (GRU) model is used to identify the candidate path during GPS outages. Both approaches solve the GPS outage problem by only constructing partial driving trajectories that require GPS data.

\subsection{Challenges and Novel Attack Surfaces}
The GPS data and sensor data from external devices (e.g., smartphones) are unreliable data sources due to issues like restricted access and data noise. The GPS outage problem can be partially addressed by incorporating additional data, such as vehicle motion data from compass sensors.
The vehicle motion data collected from the smartphone is easily perturbed by unexpected user movement and error positioning~\cite{chen2019trajcompressor}. 
In order to enhance attack efficiency, various attack surfaces are being investigated, including tire pressure sensors~\cite{rouf2010security} and the hardware-embedded scrambling algorithm~\cite{bloessl2015scrambler}.
However, these attack surfaces either demand sophisticated techniques or involve complex access requirements.

\subsection{Comparison Highlights}
Various data sources have been studied for driving trajectory detection. GPS data enables accurate position and trajectory detection~\cite{xiao2018goi, jiang2023l2mm}, although it suffers from limited coverage and requires sensitive permissions to access. In the absence of GPS, DMM~\cite{shen2024dmm} leverages cell connectivity data and cell tower locations to infer the trajectory of mobile users; however, this approach is suitable only for attackers with extensive knowledge of mobile networks. Motion data has also gained interest for trajectory detection~\cite{li2018location, roth2021daroute}, but it requires an external Inertial Measurement Units (IMUs) to be mounted in vehicles, with the IMU remaining fixed during driving to minimize human-generated noise. In~\cite{kim2023invasion}, the authors designed a new algorithm to remove human-generated noise but only managed to match subway trajectories of passengers.

This paper proposes a new vector for driving trajectory detection by leveraging the vehicle motion data from in-vehicle CAN, specifically speed and pedal data. These vehicle-generated CAN data can be more accurate than the data from external IMUs and do not require external IMUs. The CAN data can be collected via the OBD-II ports of vehicles or obtained from CAN data-based service providers.

\textcolor{black}{
Evolving vehicular communication technologies, such as Automotive Ethernet~\cite{de2024systematic} for in-vehicle network communication and Cellular Vehicle-to-Everything (C-V2X)~\cite{rammohan2023revolutionizing} for V2X communication, heighten privacy risks. These technologies enable the exchange of large volumes of sensitive driving data, including vehicle motion, environmental surroundings, and other contextual information, to support advanced applications like autonomous driving. While these technologies enhance vehicle capabilities and connectivity, they also create new attack surfaces for privacy breaches~\cite{watney2022addressing}. The increased connectivity and cloud processing creates additional opportunities for attackers to intercept or exploit sensitive data from vehicles. Consequently, privacy threats akin to the CAN-Trace attack may exploit new data types to compromise personal and vehicular privacy on a broader scale~\cite{lu2018survey, khan2020cyber}.
}

\section{Conclusion and Future Work}
\label{conclusion}

In this paper, we proposed the CAN-Trace attack to deduce vehicle driving trajectories mapped into real-world road networks.
This work is the first to infer driving trajectories using CAN messages and to apply subgraph matching to align the trajectories with road networks.
The proposed CAN-Trace attack obtains the vehicle kinematic data by accessing the CAN bus via the OBD-II port, ensuring the integrity of the data and the stealthiness of the proposed attack.
The proposed attack uses the Top-$K$ ranking rule to improve the attack efficiency.
The proposed CAN-Trace attack is evaluated with real-world driving data.
As demonstrated by the experimental results, the CAN-Trace attack performs effectively in both urban and suburban regions.

In future work, we plan to consider weather conditions, traffic patterns, city layouts, vehicle-specific factors, and driving behaviors to further refine and enhance the method. Additionally, we aim to develop new algorithms capable of handling pure CAN messages or integrating CAN data with other sources to effectively and accurately reconstruct driving trajectories.

\section*{Acknowledgment}

We thank our industry partner IAG for providing access to test vehicles. We thank Thanh Phuoc Nguyen for assisting with technical issues and experiments. This work is funded in part by the University of Technology Sydney, Insurance Australia Group (IAG), and the iMOVE CRC under Grant 5-028. This work is also supported by the Cooperative Research Centres program, an Australian Government initiative.

% \bibliographystyle{IEEEtran}
% % \bibliography{IEEEabrv}
% \bibliography{IEEEabrv}

\bibliographystyle{IEEEtran}

\vskip -2\baselineskip plus -1fil 

\begin{IEEEbiography}[{\includegraphics[width=1in,height=1.25in,clip,keepaspectratio]{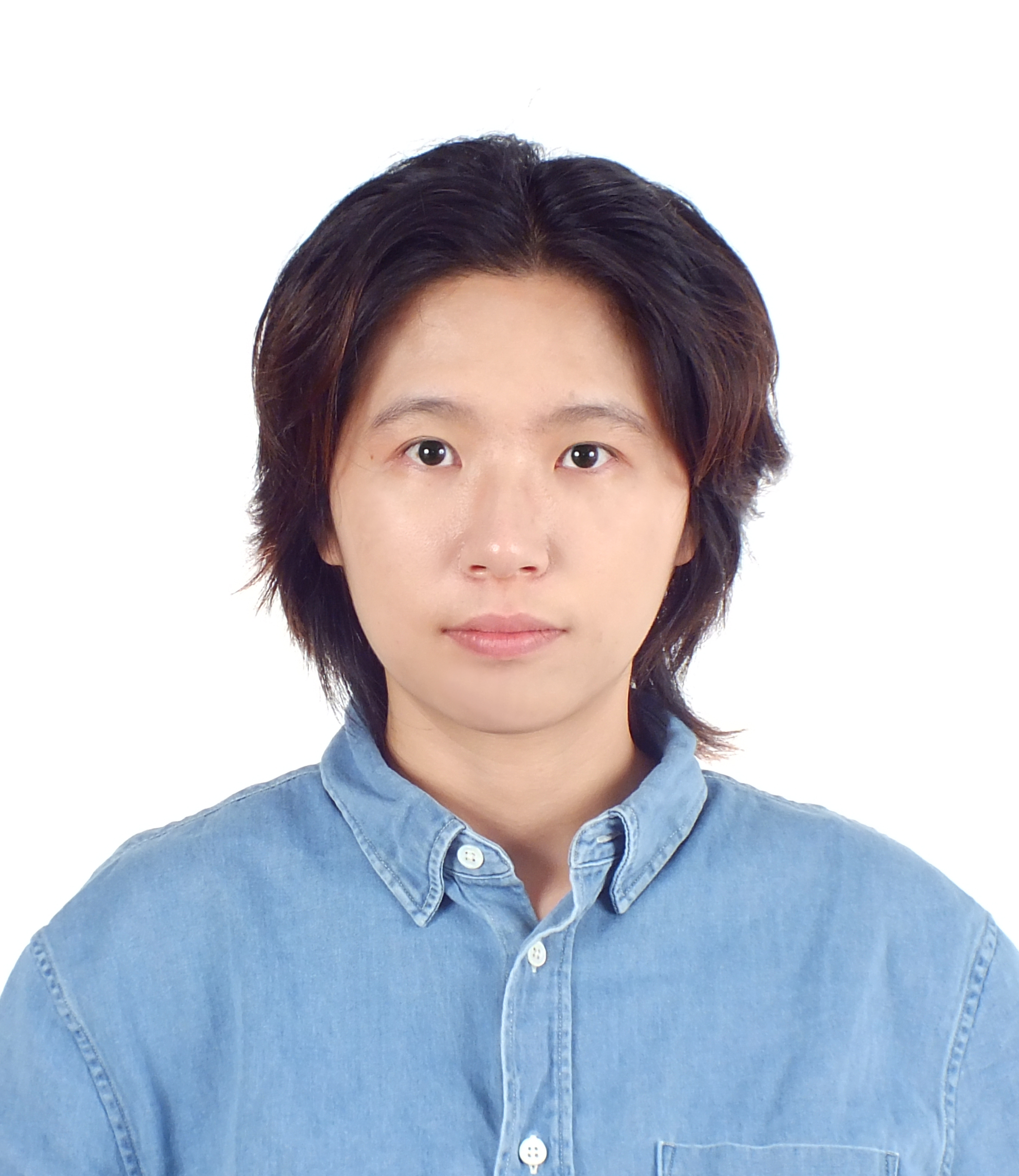}}]{Xiaojie Lin} is currently a Ph.D. student at University of Technology Sydney, Australia. She received the M.I.T. degree in 2020 from the Faculty of Engineering and Information Technology, University of Technology Sydney, Australia. Her main research interests include cybersecurity, privacy, and machine learning.
\end{IEEEbiography}

\vskip -2\baselineskip plus -1fil

\begin{IEEEbiography}[{\includegraphics[width=1in,height=1.25in,clip,keepaspectratio]{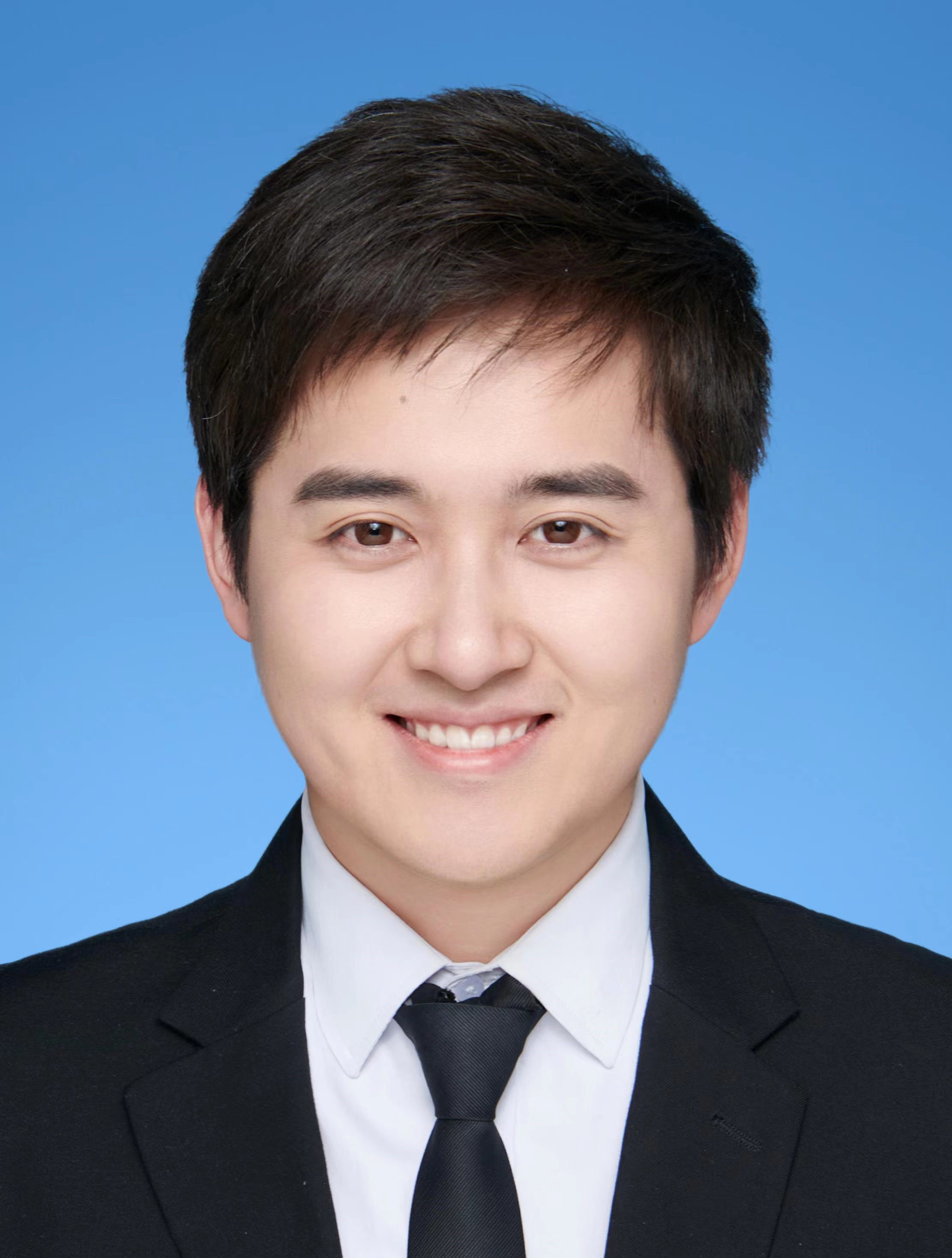}}]{Baihe Ma} received his B.E. and M.E. degrees from Xidian University, China, in 2016 and 2019, and Ph.D. degree from University of Technology Sydney, Australia, in 2024. He is currently a Lecturer with the School of Electrical and Data Engineering, University of Technology Sydney, Australia. His main research interests include Differential Privacy, Data Privacy, IoV Privacy, Cybersecurity, and Machine Learning Privacy.
\end{IEEEbiography}

\vskip -2\baselineskip plus -1fil

\begin{IEEEbiography}[{\includegraphics[width=1in,height=1.25in,clip,keepaspectratio]{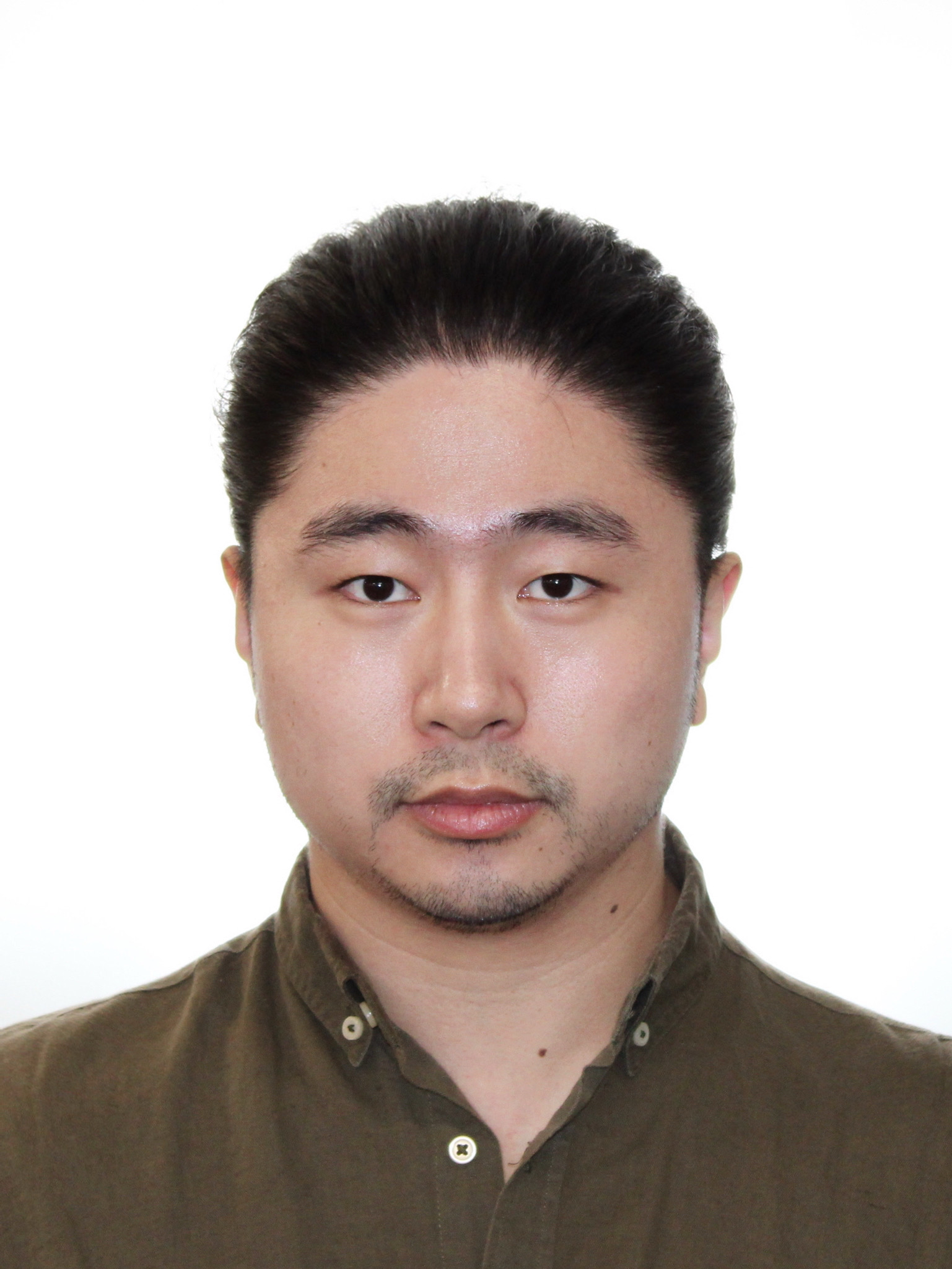}}]{Xu Wang} received his B.E. from Beijing Information Science and Technology University, China, in 2010, and dual Ph.D. degrees from Beijing University of Posts and Telecommunications, China, in 2019 and University of Technology Sydney, Australia, in 2020. He is currently a Senior Lecturer with the School of Electrical and Data Engineering, University of Technology Sydney, Australia. His research interests include cybersecurity, blockchain, privacy, and network dynamics.
\end{IEEEbiography}

\vskip -2\baselineskip plus -1fil

\begin{IEEEbiography}[{\includegraphics[width=1in,height=1.25in,clip,keepaspectratio]{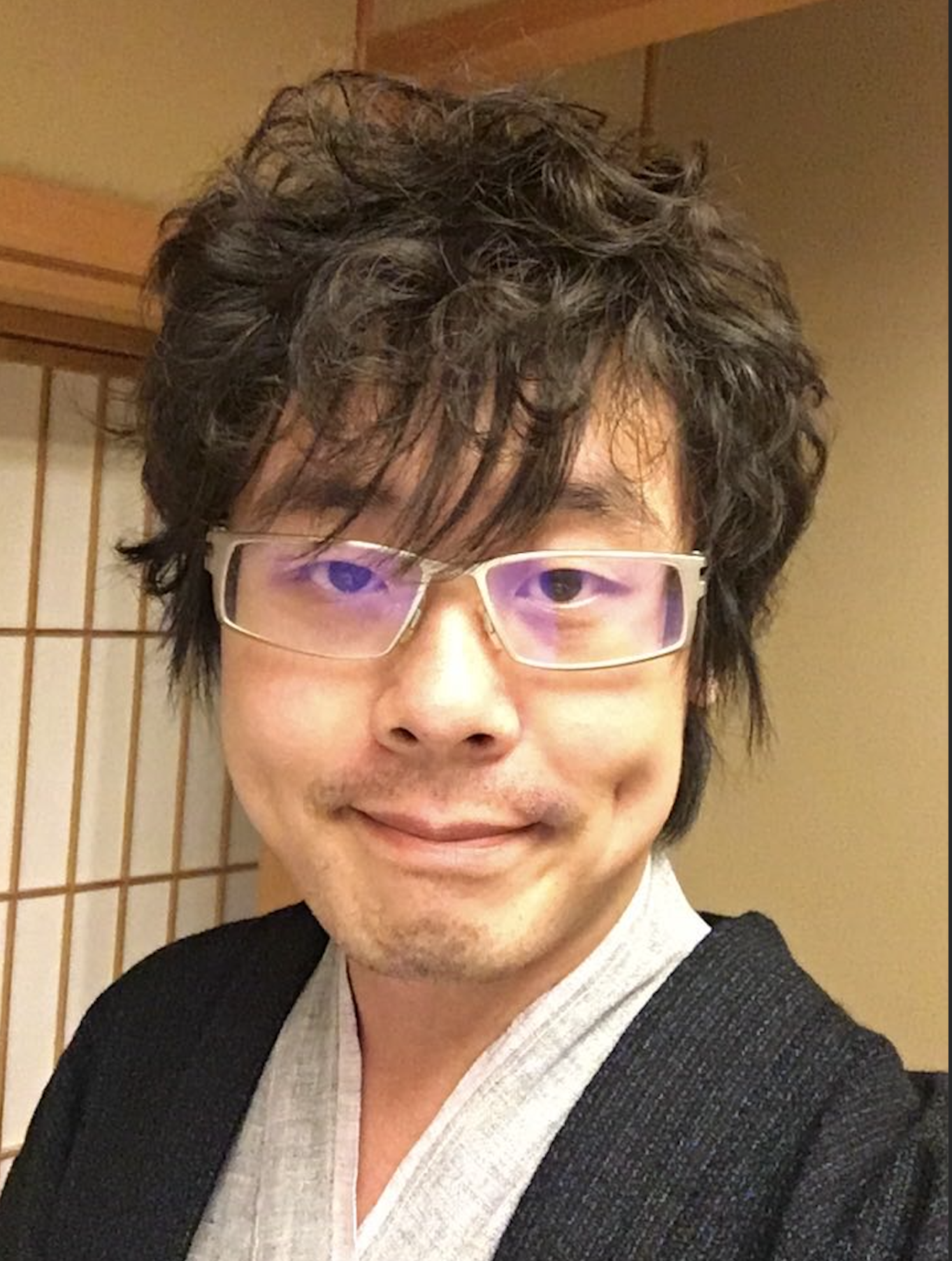}}]{Guangsheng Yu} is currently a Lecturer at University of Technology Sydney (UTS). He was a Research Fellow with CSIRO Data61, Sydney from 2021 to 2024. He received the Ph.D. degree in 2021 with the Faculty of Engineering and Information Technology, UTS. He received the B.Sc. degree and M.Sc degree from the University of New South Wales, Sydney, Australia, from 2011 to 2015. His main research interests lie in cybersecurity, blockchain, and distributed learning.
\end{IEEEbiography}

\vskip -2\baselineskip plus -1fil

\begin{IEEEbiography}[{\includegraphics[width=1in,height=1.25in,clip,keepaspectratio]{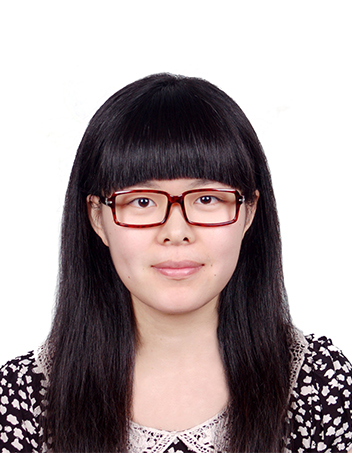}}]{Ying He} received the B.Eng. degree in telecommunications engineering from Beijing University of Posts and Telecommunications, Beijing, China, in 2009, and the Ph.D. degree in telecommunications engineering from the University of Technology Sydney, Australia, in 2017. She is currently a Senior Lecturer with the School of Electrical and Data Engineering, University of Technology Sydney. Her research interests are physical layer algorithms in wireless communication with machine learning, vehicular communication, spectrum sharing and satellite communication.
\end{IEEEbiography}

\vskip -2\baselineskip plus -1fil

\begin{IEEEbiography}[{\includegraphics[width=1in,height=1.25in,clip,keepaspectratio]{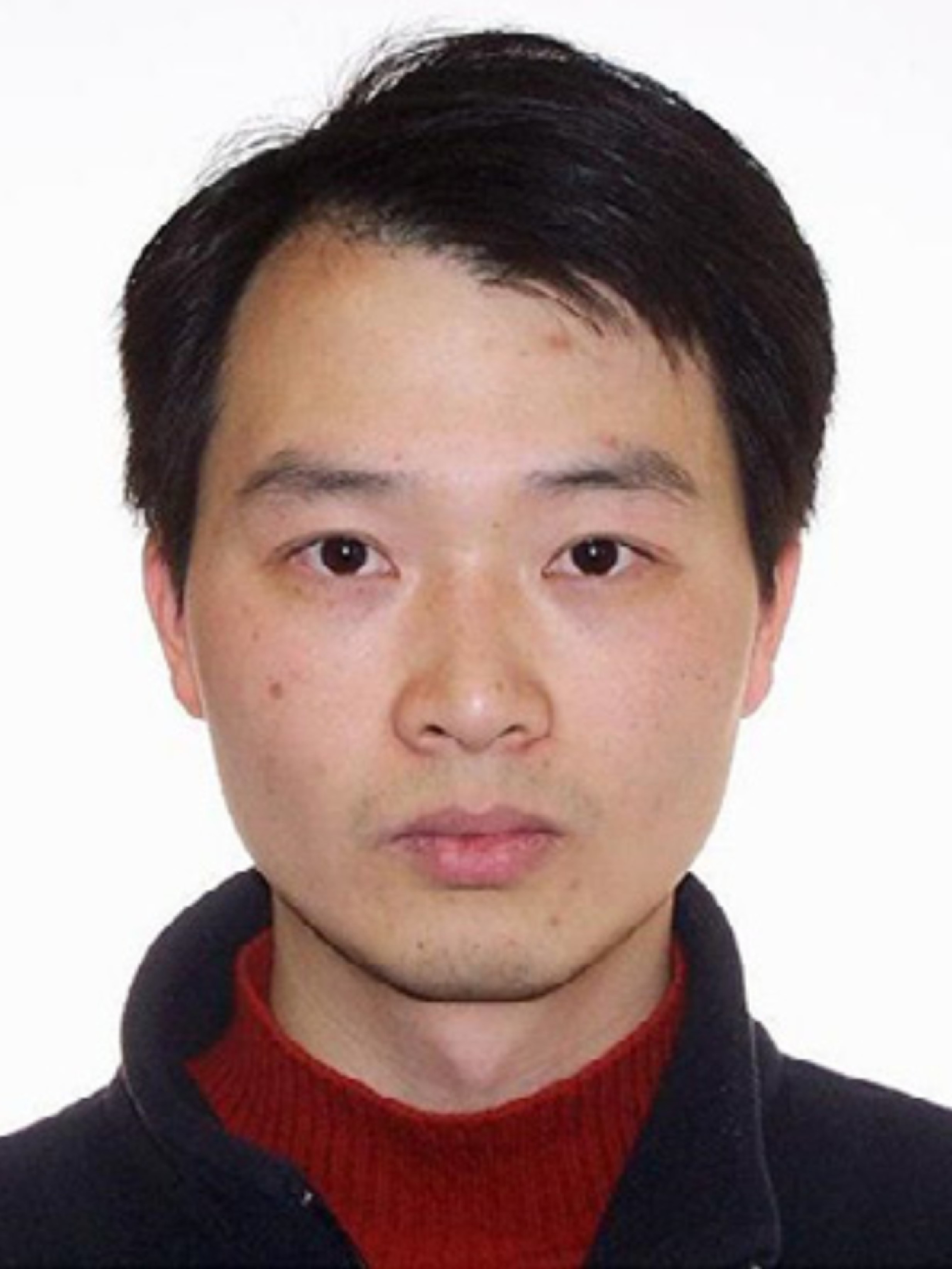}}]{Wei Ni}(Fellow, IEEE) received the B.E. and Ph.D. degrees in communication science and engineering from Fudan University, Shanghai, China, in 2000 and 2005, respectively. He is a Senior Principal Research Scientist of the Commonwealth Scientific and Industrial Research Organisation and a Conjoint Full Professor with the University of Technology Sydney, Sydney, Australia. He has coauthored three books, eleven book chapters, more than 300 journal articles, 120 conference papers, 27 patents, and 10 standard proposals accepted by IEEE. His research interests include machine learning, online learning, stochastic optimization, and their applications to system efficiency and integrity. He has won several research awards, including the 2022 IEEE IWCMC Best Paper Award, the 2022 Elsevier Best Review Paper Award, and the 2021 Elsevier YJNCA Best Review Paper Award, as well as the 2021 IEEE Vehicular Technology Society (VTS) Chapter of the Year Award.

Dr. Ni has served as an Editor for IEEE Transactions on Wireless Communications since 2018, an Editor for IEEE Transactions on Vehicular Technology since 2022, an Editor for IEEE Communications Surveys and Tutorials and IEEE Transactions on Information Forensics and Security since 2024, and an Editor for IEEE Transactions Network Science and Engineering since 2025. He served first as the Secretary, Vice-Chair, and then Chair for IEEE New South Wales VTS Chapter from 2015 to 2023, the Track Chair for VTC-Spring 2017, the Track Co-Chair for IEEE VTC-Spring 2016, the Publication Chair for BodyNet 2015, and the Student Travel Grant Chair for WPMC 2014. 
\end{IEEEbiography}

\vskip -2\baselineskip plus -1fil

\begin{IEEEbiography}[{\includegraphics[width=1in,height=1.25in,clip,keepaspectratio]{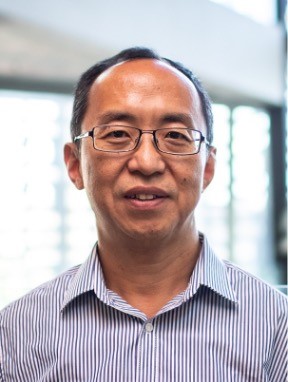}}]{Ren Ping Liu} (M’09-SM’14) received his B.E. degree from Beijing University of Posts and Telecommunications, China, and the Ph.D. degree from the University of Newcastle, Australia, in 1985 and 1996, respectively. He is a Professor and Head of Discipline of Network \& Cybersecurity at University of Technology Sydney (UTS). As a research leader, a certified network professional, and a full stack web developer, he has delivered networking and cybersecurity solutions to government agencies and industry customers. His research interests include wireless networking, 5G, IoT, Vehicular Networks, 6G, Cybersecurity, and Blockchain. He has supervised over 30 PhD students and has over 200 research publications. Professor Liu was the winner of NSW iAwards 2020 for leading the BeFAQT (Blockchain-enabled Fish provenance And Quality Tracking) project. He was awarded the Australian Engineering Innovation Award 2012 and the CSIRO Chairman’s Medal for contributing to the Wireless Backhaul project. Professor Liu was the founding chair of the IEEE NSW VTS Chapter and a Senior Member of IEEE.
\end{IEEEbiography}

\end{document}